\documentclass[paper]{JHEP3}
\usepackage{booktabs}
\usepackage{psfig}
\usepackage{epsfig,bm,amsmath}
\preprint{Cavendish--HEP--07/07}
{\title{{\tt Herwig++} Monte Carlo At Next-To-Leading Order \\
          for \boldmath{$e^+e^-$} annihilation and lepton pair production}}
\author{Oluseyi Latunde-Dada\\
  Cavendish Laboratory, University of Cambridge,\\
  JJ Thomson Avenue, Cambridge CB3 0HE, U.K.\\
  E-mail: \email{seyi@hep.phy.cam.ac.uk}}
\abstract{This paper describes the MC@NLO method for matching next-to-leading order (NLO) perturbative
QCD with the parton shower and hadronization model of the Monte Carlo (MC) event
generator {\tt Herwig++}, for $e^+e^-$ annihilation and Drell-Yan lepton pair production. Details of the event generation method as well as spin, flavour,
momentum and colour assignments are presented. We obtain predictions for various distributions which are
compared with experimental data.}


\begin{document}

\section{Introduction}

\begin{figure}[h!!]
\[
\psfig{figure=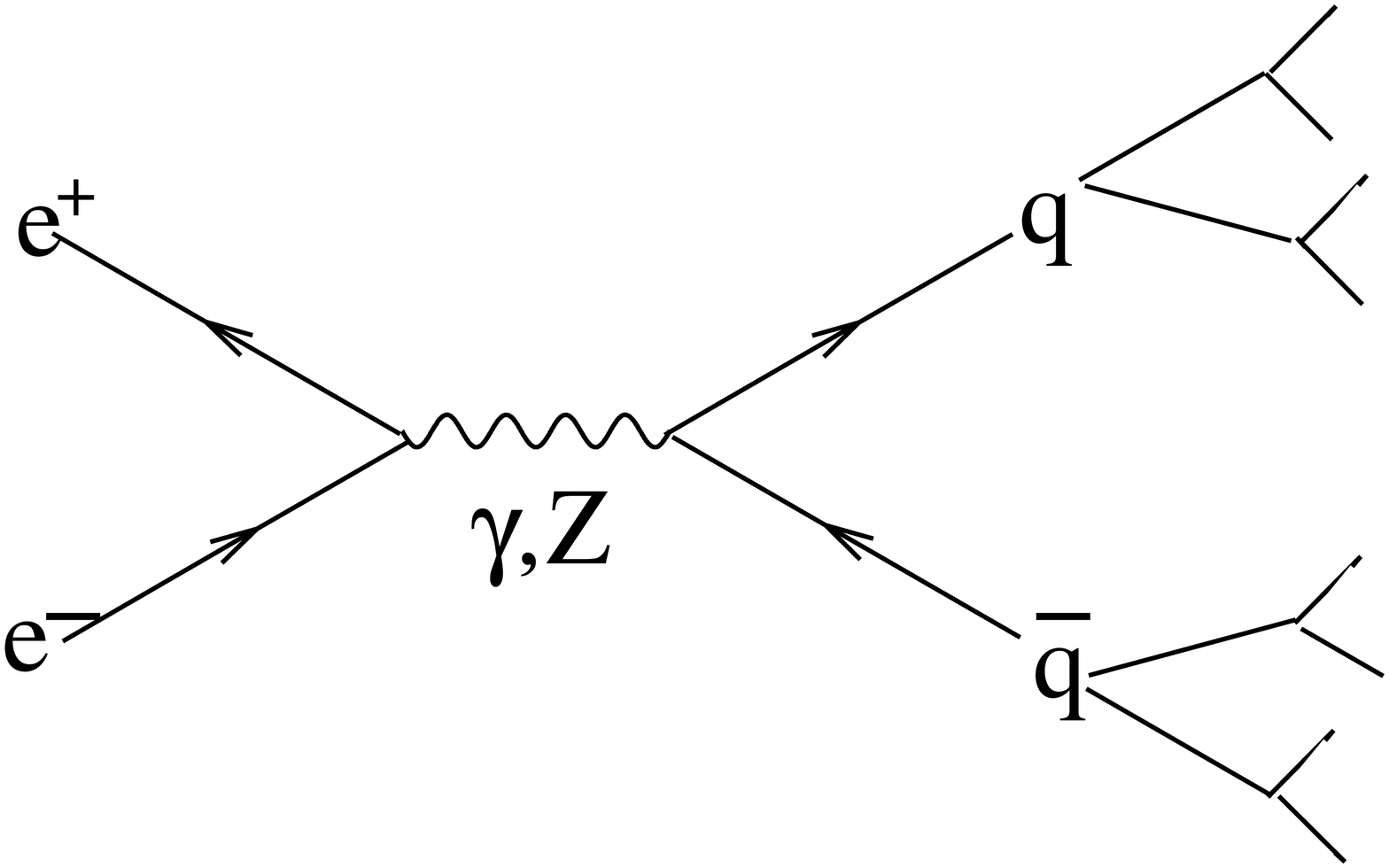,%
width=2in,height=1.7in,angle=0}
\]
\caption{2 jet formation for $e^+e^-$ annihilation }
\label{2jet}
\end{figure}

\begin{figure}[!ht]
\[
\psfig{figure=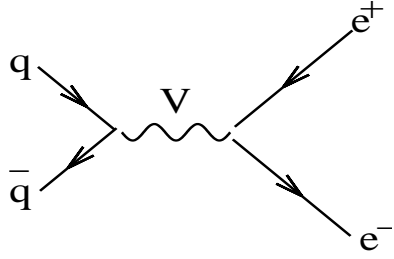,%
width=2in,height=1.3in,angle=0} 
\]
\caption{LO diagram for Drell-Yan lepton pair production.}
\label{fig:LOO}    
\end{figure}

\begin{figure}[h!!]
\[
\psfig{figure=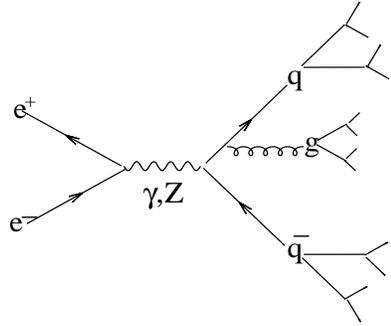,%
width=2in,height=1.7in,angle=0}
\]
\caption{3 jet formation}
\label{figee3}
\end{figure}

\begin{figure}[h!!]
\[
\psfig{figure=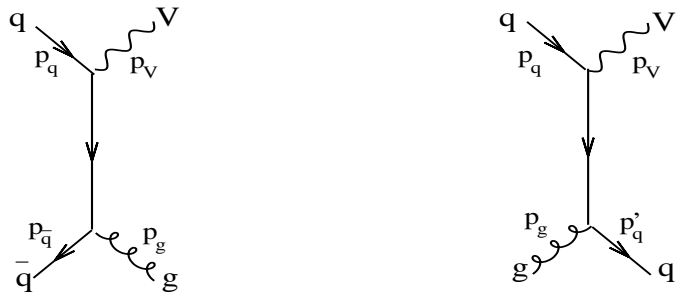,%
width=3.5in,height=1.5in,angle=0}
\]
\caption{NLO diagrams for lepton pair production.}
\label{fig:NLOO}    
\end{figure}

{\tt Herwig++} \cite{Gieseke:2006ga}  is a general purpose
Monte Carlo event generator used for simulating hard lepton-lepton, lepton-hadron and
hadron-hadron collisions. It uses the parton shower approach for initial and final state
parton branching processes including colour coherence effects and azimuthal
correlations. One example of a process modelled by {\tt Herwig++} is $ e^+$$e^-$
annihilation to $q\bar{q}$ to form two jets (Figure \ref{2jet}).
The jet topology (the  number of jets) is determined by the hard
cross-section of the process whilst the jet structure is determined by {\tt Herwig++} by
simulating soft and collinear branching from the primary partons and the conversion of the
partonic final states into hadrons (hadronization). 
Another process modelled by {\tt Herwig++} is Drell-Yan lepton pair production from
hadron-hadron collisions which is illustrated at leading order in Figure
\ref{fig:LOO}.
Different methods of matching next-to-leading order calculations to parton
shower generators have been proposed and implemented
\cite{Seymour:1994we,Nason:2004rx,Nason:2006hf,LatundeDada:2006gx,Frixione:2007nw,Frixione:2007nu}.
The aim of this paper is to extend the parton shower simulation to next-to-leading order  using the
MC@NLO method to include the formation of an extra jet and NLO virtual
 corrections without any double counting of events. This is illustrated for $e^+e^-$
 annihilation in Figure \ref{figee3}.
For Drell-Yan lepton pair production, there are 2 real emission
contributions at next-to-leading order. They are the emission of a gluon,
$q + \bar{q} \rightarrow V + g$
and the QCD Compton subprocess, $q + g \rightarrow V + q$. Both are illustrated in Figure \ref{fig:NLOO}.

The generic MC@NLO method is described in \cite{Frixione:2002ik} and has previously been successfully applied to the
hadroproduction of gauge boson pairs \cite{Frixione:2002ik, Frixione:2002bd}, heavy quark-antiquark pairs
\cite{Frixione:2003ei} and single-top production \cite{Frixione:2005vw}. In these applications, the Fortran Monte Carlo event generator {\tt HERWIG} \cite{Corcella:2000bw} was used to simulate
the parton shower and hadronization. In
this paper however, the MC@NLO method is applied to the $e^+e^-$ annihilation and Drell-Yan processes  using {\tt Herwig++}, a
redeveloped  version which implements new shower variables and an improved
hadronization model \cite{Gieseke:2003hm}.

\section{\boldmath{$e^+e^-$} annihilation }
\label{sigma}
\begin{figure}[h]
\[
\psfig{figure=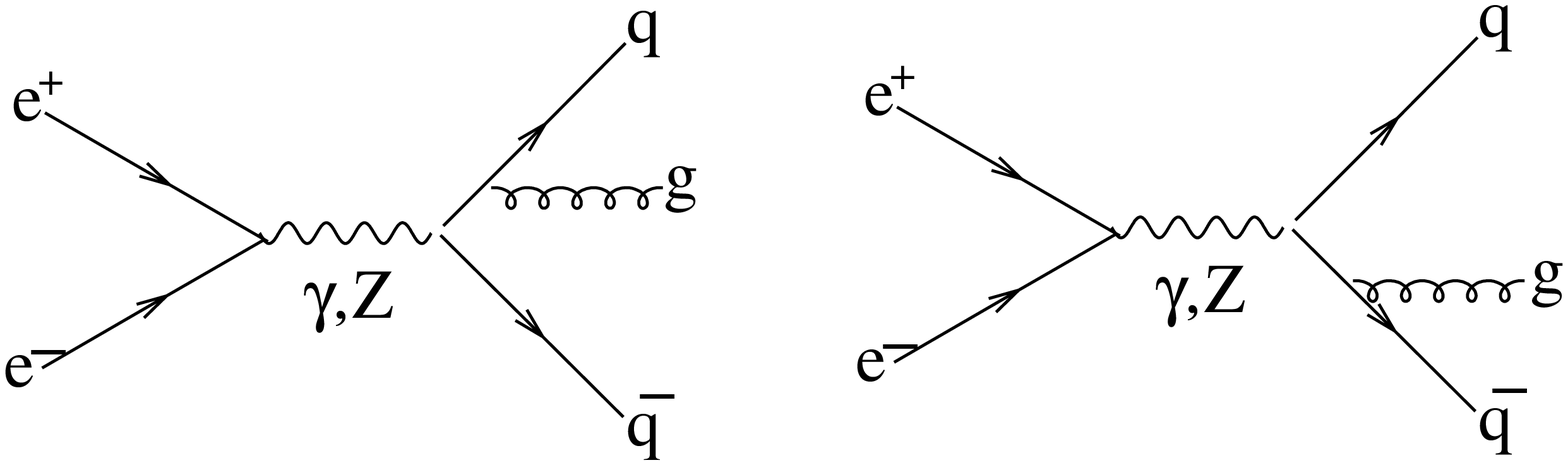,%
width=4in,height=2in,angle=0}
\]
\caption{Feynman diagrams for $e^{+}e^{-} \rightarrow q{\bar{q}}g$}
\label{figeee}
\end{figure}

In the massless limit, the 3-particle cross-section for the process, $ e^+e^- \rightarrow
\gamma^* \rightarrow q\bar{q}g $ shown in Figure~\ref{figeee}  
\\
is given by (neglecting Z boson exchange contributions for the moment)
\begin{eqnarray}
\sigma^{q\bar{q}g}=\sigma_0 \int dx_q
dx_{\bar{q}}\frac{\alpha_S}{2\pi}C_FM(x_q,x_{\bar{q}})
\label{eqqqg}
\end{eqnarray}
where
\begin{eqnarray}
M(x_q,x_{\bar{q}})&=&\frac{x_q^2+x_{\bar q}^2}{(1-x_q)(1-x_{\bar{q}})} \,,\nonumber \\
x_q&=&\frac{2E_{q}}{\sqrt{s}} \,,\nonumber \\
x_{\bar{q}}&=&\frac{2E_{\bar{q}}}{\sqrt{s}}\,,\nonumber \\
\sigma_0&=&3\sum_q{Q_q}^2\frac{4\pi\alpha^2}{3s}\,,
\end{eqnarray}
$C_{F}=4/3$ (for the case of SU(3)colour representations), and the integration region is: $ 0\leq x_q, x_{\bar{q}}\leq1, x_q+x_{\bar{q}}\geq1$
\cite{Webber}.\\ 

The integrand in (\ref{eqqqg}) is divergent at $x_q, x_{\bar{q}}=1$
where the gluon is collinear with the quark or antiquark or where the gluon is soft. As we shall see shortly, these singularities are cancelled out in
the total cross-section to next-to-leading order in $\alpha_S$. 
Using dimensional regularization, (\ref{eqqqg}) can be evaluated to give,
\begin{equation}
\sigma^{q\bar{q}g}(\epsilon)=\sigma_0\frac{C_F\alpha_S}{2\pi}H(\epsilon)\left[\frac{2}{\epsilon^2}+\frac{3}{\epsilon}+\frac{19}{2}-\pi^2+O(\epsilon)\right]
\label{eqqqe}
\end{equation}
where
\begin{eqnarray}
\label{eqe}
\epsilon&=&\frac{1}{2}(4-n)\,, \nonumber \\
H(\epsilon)&=&\frac{3(1-\epsilon)}{(3-2\epsilon)\Gamma(2-2\epsilon)}(4\pi)^{2\epsilon}\,,
\nonumber \\
           &=&1 +O(\epsilon)
\end{eqnarray}
and $n=$ number of dimensions. 

\begin{figure}[h]
\[
\psfig{figure=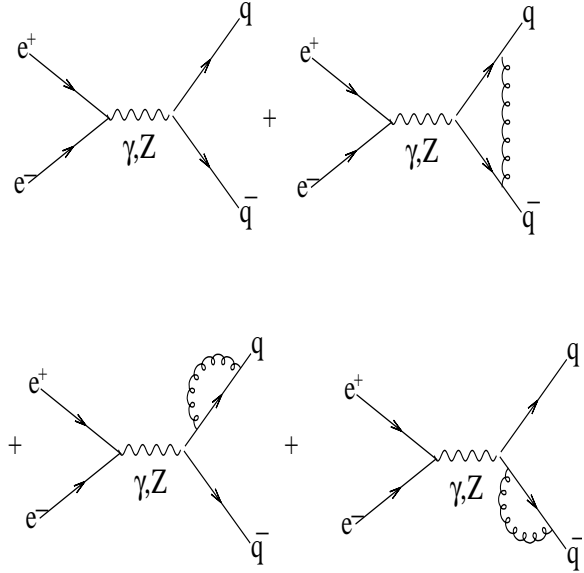,%
width=3in,height=3in,angle=0}
\]
\caption{Leading order and virtual gluon Feynman diagrams}
\label{figqq}
\end{figure}

The total cross-section to order $\alpha_S$ is obtained by adding the contributions from
the leading order and virtual gluon Feynman diagrams in Figure \ref{figqq} to
(\ref{eqqqe}). 
This contribution is
\begin{equation}
\sigma_0\left[1+\frac{C_F\alpha_S}{2\pi}H(\epsilon)\left\{-\frac{2}{\epsilon^2}-\frac{3}{\epsilon}-8+\pi^2+O(\epsilon)\right\}\right]\;.
\label{eqqq(g)}
\end{equation}
Taking $C_{F}=4/3$ (for the SU(3) colour group), the total cross-section is found to be \cite{Webber},
\begin{equation}
\sigma_{\rm total}=\sigma_0\left[1+\frac{\alpha_S}{\pi}+O({\alpha_S}^2)\right] \;.
\label{eqtotal}
\end{equation}
For massless partons, the QCD correction  at O($\alpha_S$) is independent of the
nature of the exchanged boson. Hence at this order for Z boson exchange,
\begin{equation}
\sigma_{ \rm total}=\sigma_0^Z[1+\frac{\alpha_S}{\pi}]
\end{equation}
where 
\begin{eqnarray}
\sigma_0^Z&=&3\sum_q({A_q}^2+{V_q}^2)({A_e}^2+{V_e}^2)\frac{4 \pi \alpha^2
  \kappa^2}{3\Gamma^2} \,,\nonumber \\
\kappa&=&\frac{\sqrt{2}G_FM_{z}^2}{16 \pi \alpha} \,,
\end{eqnarray}
 $A_q,V_q,A_e$ and $V_e$ are the axial and vector coupling constants for
the quarks  and leptons respectively, 
$G_F$ is the Fermi constant, $M_Z$ is the pole mass of the Z boson and $\Gamma$
is the total decay width of the Z boson. Note that this is the cross-section at $\sqrt{s}=M_Z$.

As was mentioned earlier, it can be seen that the collinear and soft singularities in
(\ref{eqqqe}) have been cancelled in $\sigma_{ \rm total}$. 
However, writing $\sigma_{\rm total}$ as a sum of two separate parts (\ref{eqqqe}) and
(\ref{eqqq(g)}), makes it possible to generate  a Monte-Carlo events in
$x_q,x_{\bar{q}}$ space which can be fed into {\tt Herwig++} to
simulate 2 and 3-jet processes. This is the subject of section~\ref{section3}. 
\section{MC@NLO method}
\label{section3}

Writing $\sigma_{\rm total}$ explicitly as the sum of equations (\ref{eqqqe}) and
(\ref{eqqq(g)}) gives 
\begin{eqnarray}
\sigma_{\rm total}&=&\sigma_0\left[1+\frac{C_F\alpha_S}{2\pi}H(\epsilon)\left(-\frac{2}{\epsilon^2}-\frac{3}{\epsilon}-8+\pi^2+O(\epsilon)\right)
 \right. \nonumber \\
&+&\left. \frac{C_F\alpha_S}{2\pi} H(\epsilon)\left(\frac{2}{\epsilon^2}+\frac{3}{\epsilon}+\frac{19}{2}-\pi^2+O(\epsilon)\right)\right]
\label{eq2+3jet}
\end{eqnarray}
This can be re-written in integral form as 
\begin{equation}
\sigma_{\rm total}=\sigma_0\int dx_q dx_{\bar{q}}\left[2-\frac{\alpha_S}{2\pi}C_F\left\{M(x_q,x_{\bar{q}},\epsilon)-3\right\}+\frac{\alpha_S}{2\pi}C_FM(x_q,x_{\bar{q}},\epsilon)\right]
\label{eqnewsigma}
\end{equation}
where 
\begin{equation}
M(x_q,x_{\bar{q}},\epsilon)=\frac{H(\epsilon)}{{\left[(1-x_{q})(1-x_{\bar{q}})(1-x_g)\right]}^{\epsilon}}\left[\frac{(1-\epsilon)(x_q^2+x_{\bar{q}}^2)+2\epsilon(1-x_g)}{(1-x_q)(1-x_{\bar{q}})}-2\epsilon\right]
\end{equation}
 is the $e^+e^- \rightarrow q\bar{q}g$ hard matrix element and $x_g=2-x_q-x_{\bar{q}}$.
Now, if we define a functional $ F_i$  as the functional which represents hadronic final states
generated by the parton shower starting from a configuration {\em i}, a generating functional for the process
$e^+e^-\rightarrow$ hadrons can be written as  
\begin{equation}
F=\sigma_0\int dx_q dx_{\bar{q}}\left[F_{q\bar{q}} \left
      \{2-\frac{\alpha_S}{2\pi}C_F\left(M(x_q,x_{\bar{q}})-3\right)\right\}+F_{q\bar{q}g}\frac{\alpha_S}{2\pi}C_FM(x_q,x_{\bar{q}})\right]
\label{eqF}
\end{equation}
where $F_{q\bar{q}}$ is the functional representing the shower final states resulting from
the process $e^+e^-\rightarrow {q\bar{q}}$ and $F_{q\bar{q}g}$ represents the final
states from $e^+e^-\rightarrow {q\bar{q}g}$. We have set $\epsilon=0$ so that
$H(\epsilon)= 1$ and $M(x_q,x_{\bar{q}},\epsilon)=M(x_q,x_{\bar{q}})$ which is defined in
(\ref{eqqqg}).

\begin{figure}[h]
\[
 \psfig{figure=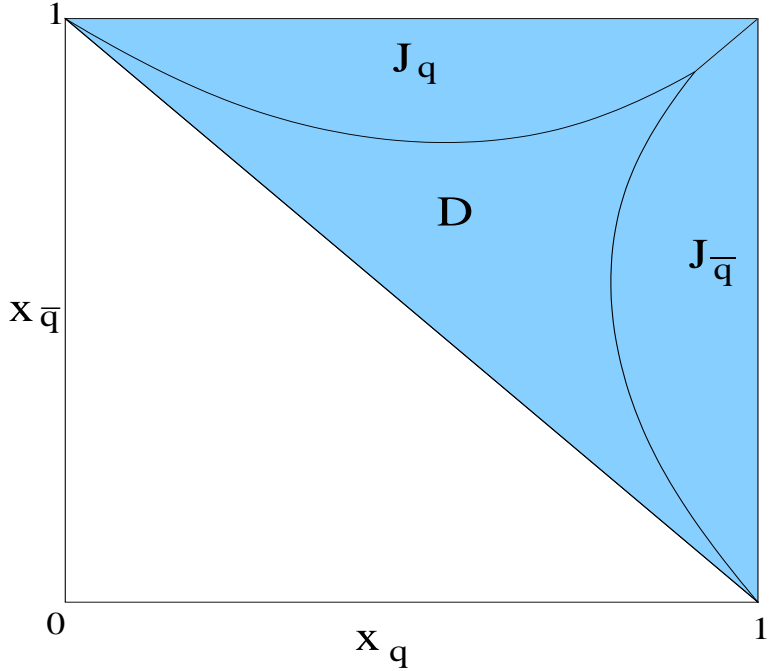,%
width=4in,height=3.5in,angle=0}
\]
\caption{Phase space for $x_q,x_{\bar{q}}$ showing hard ($D$) and soft/collinear
  ($J_{q},J_{\bar{q}}$) gluon emission regions.}
\label{figphase}
\end{figure}

This would be wrong however because configurations starting with $q\bar{q}$, would also
radiate quasi-collinear gluons, with a distribution, $M_{C}( x_q,x_{\bar{q}})$ given by
the parton shower. Likewise, configurations with $q\bar{q}g$ would generate $q
\bar{q}$-like configurations if the gluon is quasi-collinear to the quark or antiquark. $M_{C}(x_q,x_{\bar{q}})$
is the parton shower branching cross-section, which in the massless case is given in
{\tt Herwig++} by
\begin{equation}
\frac{\alpha_S}{2\pi}C_FM_{C}(x_q,x_{\bar{q}})= \frac{\alpha_S}{2\pi}C_F\frac{1+\left(\frac{x_q+x_{\bar{q}}-1}{x_{q}}\right)^2}{(1-x_{\bar{q}})(1-x_{q})}
\label{eqMC}
\end{equation}
for a gluon branching quasi-collinearly off the antiquark. (Interchange $x_q$ and $
x_{\bar{q}}$ for a gluon branching off the quark). 
This can be derived from the quasi-collinear splitting function defined terms of the {\tt
  Herwig++} evolution variables, $z$ and $\tilde{q}$ in (\ref{eqdP})
\cite{Gieseke:2003rz}: 
\begin{equation}
dP(q \rightarrow qg)= \frac{C_{F}}{2\pi}\alpha_S[z^{2}(1-z)^{2}\tilde{q}^{2}]\frac{d\tilde{q}^2}{\tilde{q}^2}\frac{dz}{1-z}[1+z^{2}-\frac{2m^{2}}{z\tilde{q}^{2}}]
\label{eqdP}
\end{equation}
where $z$ is the momentum fraction of the quark after gluon emission relative to the parent quark
and $\tilde{q}$ is an angular variable related to the relative transverse momentum, ${\bf p_{T}}$ of the
quark after gluon emission via:
\begin{equation}
\mid {\bf p_{T}} \mid=\sqrt{(1-z)^{2})(z^{2}\tilde{q}^{2}-\mu^{2})-z{Q_{g}}^{2}} \,,
\label{q}
\end{equation}
 $\mu =$ max $(m, Q_{g})$ and ${Q_{g}}^{2}$ is the minimum virtuality for the quarks and gluons
 which is required to define a resolvable emission.
 The Dalitz plot variables $x_{q}$ and $x_{\bar{q}}$ are related to the evolution
 variables $z$ and $\tilde{q}$ via;
\begin{eqnarray}
x_{q}&=&1-z(1-z)\kappa \,, \nonumber\\
x_{\bar{q}}&=&(2-x_q)r+(z-r)\sqrt{x_{q}^2-4\rho}
\end{eqnarray}
where 
\begin{eqnarray}
\rho&=&\frac{m^{2}}{s} \,,\nonumber \\
r&=&\frac{1}{2}\left(1+\frac{\rho}{1+\rho-x_q}\right)\,, \nonumber \\
z&=&r+\frac{x_{\bar{q}}-(2-x_q)r}{\sqrt{x_q^2-4\rho}}\,,\nonumber \\
\kappa&=&\frac{\tilde{q}^{2}}{s} \;.
\end{eqnarray}
By changing the evolution variables in (\ref{eqdP}) to the Dalitz plot variables in
the limit where $m=\rho=0$, (\ref{eqMC}) can be derived. The Jacobian factor for the
transformation is 
\begin{equation}
z(1-z)\sqrt{x_{q}^{2}-4\rho}\;.
\end{equation}
The equations given above apply to a radiating antiquark. For a radiating quark,
interchange $x_q$ and $x_{\bar{q}}$. Imposing the condition 
\begin{equation}
\kappa < \frac{1}{2}(1+\sqrt{1-4\rho})
\end{equation}
defines the regions of the
phase space covered by the parton showers i.e the quark and antiquark jets. In the massless limit this yields the function $\Theta_{PS}$ in (\ref{H}) which defines the
phase space regions $J_{q}$, $J_{\bar{q}}$ and $D$ in Figure \ref{figphase}.  
\begin{equation}
\Theta_{PS}=\Theta[(1-x_q)(x_q+x_{\bar{q}}-1)-x_{\bar{q}}^2(1-x_{\bar{q}})]+\Theta[(1-x_{\bar{q}})(x_q+x_{\bar{q}}-1)-x_q^2(1-x_q)]\;.
\label{H}
\end{equation}

The full integration region is shown shaded in the figure above. Regions $J_{q}$ and $J_{\bar{q}}$ include soft and quasi-collinear gluon emission events whilst region $D$
includes hard and non-collinear gluon emission events, giving rise to additional jets. 

Having defined $M_{C}$, we can now obtain the correct overall functional by subtracting the integral of
(\ref{eqMC}) from the second term in (\ref{eqF}) and adding it to the first term: 
\begin{equation}
F=\sigma_0\int dx_q dx_{\bar{q}}\left[F_{q\bar{q}}\left\{2-\frac{\alpha_S}{2\pi}C_F\left(M-M_{C}-3\right)\right\}+F_{q\bar{q}g}\frac{\alpha_S}{2\pi}C_F(M-M_{C})\right]\;.
\label{eqFF}
\end{equation}
This subtraction is relevant only to regions $J_{q}$ and $J_{\bar{q}}$ in Figure \ref{figphase} which
include soft and collinear emission events. In region $D$, which contains hard emission events, we simply integrate the hard matrix
element, $M(x_q,x_{\bar{q}})$ over the region. Therefore the overall generating functional can
be written as 
\begin{eqnarray}\label{eq11}
F&=&\sigma_0\left[\int_{J}dx_q
  dx_{\bar{q}}\left[F_{q\bar{q}}\left\{2-\frac{\alpha_S}{2\pi}C_F(M-M_{C}-3)\right\}+F_{q\bar{q}g}\frac{\alpha_S}{2\pi}C_F(M-M_{C})\right] \right.\nonumber \\
&+& \left. \int_Ddx_qdx_{\bar{q}}\left[F_{q\bar{q}}\left\{2-\frac{\alpha_S}{2\pi}C_F(M-3)\right\}+F_{q\bar{q}g}\frac{\alpha_S}{2\pi}C_FM\right]\right]
\end{eqnarray}
where $J$ denotes the region $J_{q} \cup J_{\bar{q}}$.
Note that the total cross-section can be retrieved by making the substitution
$F_{q\bar{q}}, F_{q\bar{q}g} = 1$ in (\ref{eq11}) as shown.
\begin{eqnarray}\label{eq12}
\sigma_{\rm total}&=&\sigma_0\left[\int_{J}dx_q dx_{\bar{q}}\left\{2-\frac{\alpha_S}{2\pi}C_F(M-M_{C}-3)+\frac{\alpha_S}{2\pi}C_F(M-M_{C})\right\}\right.\nonumber\\
&+& \left. \int_Ddx_qdx_{\bar{q}}\left\{2-\frac{\alpha_S}{2\pi}C_F(M-3)
+\frac{\alpha_S}{2\pi}C_FM\right\}\right]\;.
\end{eqnarray}
The `Hit or Miss' Monte Carlo method was used to evaluate the integrals and the events
were generated using the importance sampling method. The algorithm is described in Appendix \ref{HM}.

 \section{Heavy quarks}
\label{HP}
So far, we have discussed the limit in which the quark and anti-quark are massless. We
shall now discuss the case where heavy flavour quarks are produced i.e. charm quarks of mass
$1.6$ GeV and bottom quarks of mass $5$ GeV. There are two ways in which this can be
treated. Both methods are described below.

\subsection{Method 1: Using the heavy quark matrix element}
\label{heavyME}
The 3-particle differential cross-section
for the process $e^{+}e^{-} \rightarrow V \rightarrow Q{\bar{Q}}g$, where V represents a
vector current like the photon is given by \cite{Dokshitzer:1994jt}; 

\begin{equation}
\frac{1}{\sigma_{V}}\frac{d^{2}\sigma_{V}}{dx_{Q}dx_{\bar{Q}}}=\frac{\alpha_{S}}{2\pi}\frac{C_{F}}{v}\left[\frac{(x_Q^2+2\rho)^2+(x_{\bar {Q}}^2+2\rho)^2+\zeta_{V}}{(1+2\rho)(1-x_Q)(1-x_{\bar{Q}})}-\frac{2\rho}{(1-x_Q)^2}-\frac{2\rho}{(1-x_{\bar{Q}})^2}\right]
\label{eqmassvect}
\end{equation}
where
\begin{eqnarray}
\rho&=&\frac{m_Q^2}{s}\,,\nonumber \\
v&=&\sqrt{1-4\rho}\,,\nonumber \\
\zeta_{V}&=&-8\rho(1+2\rho)\,,\nonumber\\
\sigma_{V}&=&\sigma_0(1+2\rho)v\,.
\end{eqnarray}
For an axial current contribution $e^{+}e^{-} \rightarrow A \rightarrow Q{\bar{Q}}g$, we have
\begin{equation}
\frac{1}{\sigma_{A}}\frac{d^{2}\sigma_{A}}{dx_{Q}dx_{\bar{Q}}}=\frac{\alpha_{S}}{2\pi}\frac{C_{F}}{v}\left[\frac{(x_Q^2+2\rho)^2+(x_{\bar {Q}}^2+2\rho)^2+\zeta_{A}}{v^2(1-x_Q)(1-x_{\bar{Q}})}-\frac{2\rho}{(1-x_Q)^2}-\frac{2\rho}{(1-x_{\bar{Q}})^2}\right]
\label{eqmassax}
\end{equation}
where
\begin{eqnarray}
\zeta_{A}&=&2\rho[(3+x_g)^2-19+4\rho]\,,\nonumber \\
\sigma_{A}&=&\sigma_0v^3\,.
\end{eqnarray}
$\sigma_V$ and $\sigma_A$ are the Born cross-sections for heavy quark production by a
vector and axial current respectively whilst $\sigma_0$ is the massless quark Born
cross-section. 

Since the partons are massive, the phase space available for gluon emission is reduced. It
is determined by the triangle relation: 
\begin{equation}
\Delta(x_Q^2-\rho,x_{\bar{Q}}^2-\rho,x_g^2) \leq 0
\end{equation}
where $\Delta(a,b,c)=a^2+b^2+c^2-2ab-2ac-2bc$. This is equivalent to satisfying the condition
\begin{equation}
(1-x_Q)(1-x_{\bar{Q}})(x_Q+x_{\bar{Q}}-1) > \rho(2-x_Q-x_{\bar{Q}})^2
\end{equation}
in the phase space.

\begin{figure}[h]
\[
\psfig{figure=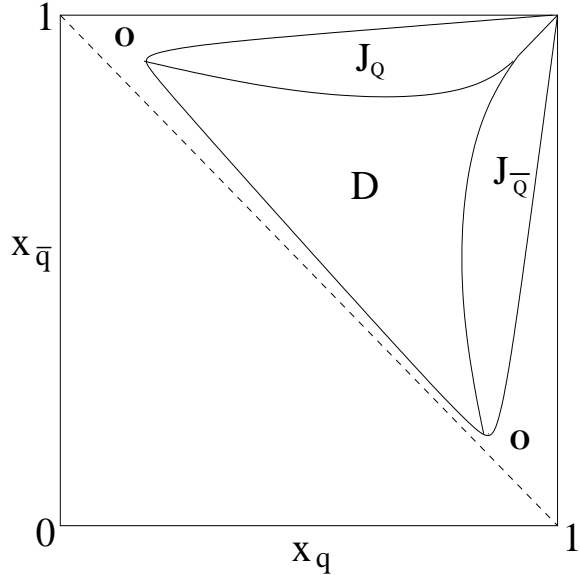,%
width=3in,height=3in,angle=0}
\]
\caption{Heavy flavour phase space for $x_Q,x_{\bar{Q}}$ showing hard ($D$) and soft
  ($J_{Q},J_{\bar{Q}}$) gluon emission regions. Not to scale}
\label{figmassphase}
\end{figure}

Just as in the massless limit, the phase space region can again be split into 2 regions $J_{Q}$ and $J_{\bar{Q}}$, containing soft and
quasi-collinear gluon emission events and a region $D$ containing hard and non-collinear
emission events as shown in Figure \ref{figmassphase}. There is also an additional region labeled $O$  
outside the phase space which as we shall see in (\ref{eqnsplit}) generates 2-jet events.

As in the massless case, the total 3-particle cross-section to $O(\alpha_S)$ is
calculated by adding leading order and virtual gluon contributions to the integrals of
(\ref{eqmassvect}) and (\ref{eqmassax}) over the phase space in Figure
\ref{figmassphase}. This yields for photon exchange \cite{Webber} 
\begin{equation}
\sigma_{\rm total}=\sigma_{V}\left[1+c_1\frac{\alpha_S}{\pi}\right]
\end{equation}
where  $c_1\approx1+12\rho$ \cite{Djouadi:1989uk}.

For Z boson exchange, to $O(\alpha_S)$, the cross-section is given by \cite{Webber}
\begin{equation}
\sigma_{\rm total}=\sigma_{V}\left[1+c_1\frac{\alpha_S}{\pi}\right]+\sigma_{A}\left[1+d_1\frac{\alpha_S}{\pi}\right]
\label{sig}
\end{equation}
where $d_1\approx1-22\rho$ \cite{Djouadi:1989uk}. 

Following the same procedure as for the massless case, Monte Carlo events can be generated
by writing $\sigma_{\rm total}$ explicitly. For example, the jet generating functional, $F$ can now be written as;
\begin{eqnarray}
F&=&\sigma_{V}\left[\int_{O}dx_{\bar{Q}}dx_QF_{Q\bar{Q}}\left\{2+3c_1\frac{\alpha_S}{2\pi}C_F\right\}\right.\nonumber\\
&+& \int_{D}dx_Qdx_{\bar{Q}}\left[F_{Q\bar{Q}}\left\{2-\frac{\alpha_S}{2\pi}C_F(M-3c_1)\right\}+F_{Q\bar{Q}g}\frac{\alpha_S}{2\pi}C_FM\right]\nonumber\\
&+& \left.\int_Jdx_Qdx_{\bar{Q}}\left[F_{Q\bar{Q}}\left\{2-\frac{\alpha_S}{2\pi}C_F(M-M_C-3c_1)\right\}+F_{Q\bar{Q}g}\frac{\alpha_S}{2\pi}C_F(M-M_{C})\right]\right]
\label{eqnsplit}
\end{eqnarray}
where $J=J_{Q} \cup J_{\bar{Q}}$, $M$ is the differential cross-section defined in
(\ref{eqmassvect}) and $M_C$ is the heavy quark
quasi-collinear branching probability given in (\ref{eqdP}).\\
Just as in the massless limit, setting $F_{Q\bar{Q}},F_{Q\bar{Q}g}=1$ in (\ref{eqnsplit})
recovers the vector exchange part of $\sigma_{\rm total}$ in (\ref{sig}). The only
difference is the integral over the region $O$ outside the phase space which is required
to recover the full cross-section. As before, the coefficients of
$F_{Q\bar{Q}}$ and $F_{Q\bar{Q}g}$ generate 2-jet and 3-jet events respectively.
Details of the evaluation of the integrals can be found in Appendix \ref{HQ}. Event
generation follows the same lines as described on Appendix \ref{HM}.

\subsection{Method 2: Using the massless quark matrix element}
\label{massless}
Since the masses of the charm and bottom quarks ($1.6$ and $5$ GeV respectively) are small compared to the
center of mass energy at the Large Electron-Positron (LEP) collider ($91.2$ GeV), they can
in the first approximation be assumed to be massless. Hence, the massless matrix element can be used to obtain
the 4-momentum distributions for charm and bottom quarks as described above for up, down
and strange quarks. This is less rigorous than the method described in section
\ref{heavyME} but it has the advantage of a smoother distribution of events due to the
unweighting procedure being more efficient. This is the method used in this paper.


\section{Results on \boldmath{$e^+e^-$} annihilation}
The methods described above were used to generate $e^+e^-$ events for comparison with LEP
1 data. Details of the assignment of partonic final-state properties are described in
Appendix \ref{dress}. Figures \ref{fig:tt}-\ref{fig:bsbd} show comparisons of event shape distributions
obtained from our results and LEP 1
data. The massless quark matrix element method described in section \ref{massless} was used for heavy
quark generation. Also compared are event shapes obtained from {\tt Herwig++} with the matrix element
correction switched on. This is the method whereby emissions are only accepted into the
dead region D of the phase space at a rate given by the matrix element. In both cases {\tt Herwig++} version 2.0.1 was used. The hadronization scale which is the scale at which the shower is
turned off was set
to the default value of $0.631$ GeV and the 2-loop $\alpha_S$ value was used.

\begin{figure}[!ht]
\vspace{2cm}
\hspace{0.5cm}
\psfig{figure=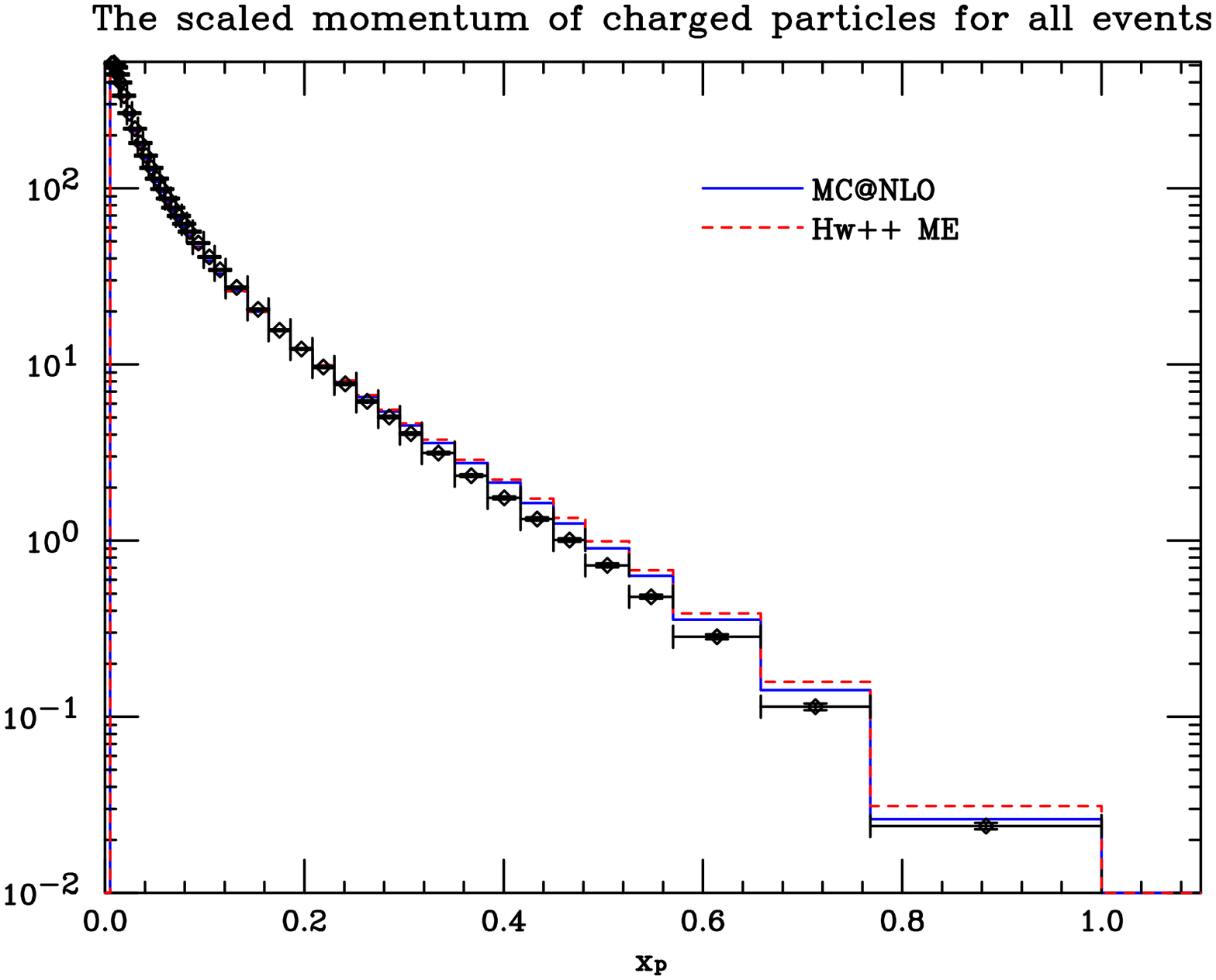,%
width=3.03in,height=2.93in,angle=90}
\psfig{figure=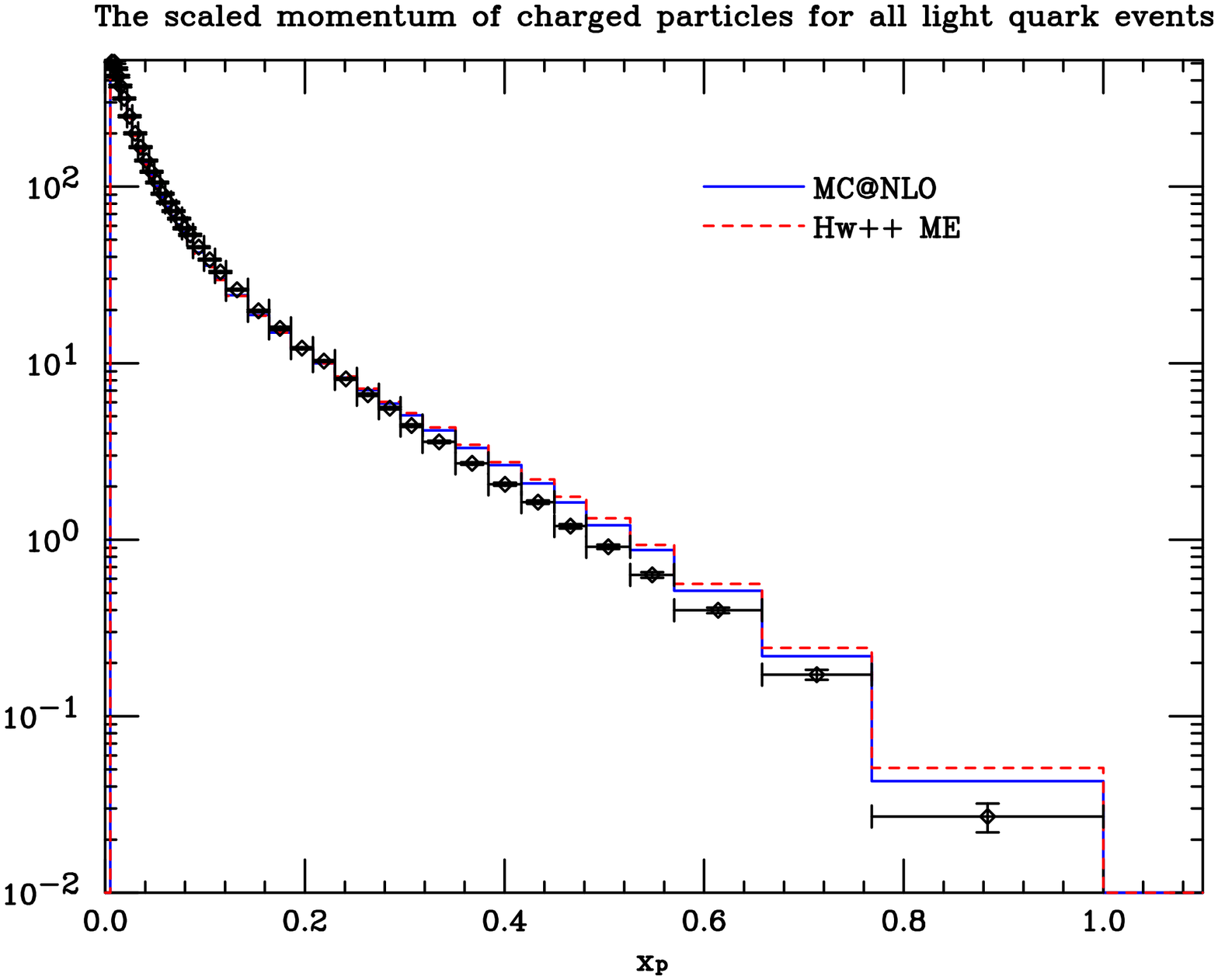,%
width=3.03in,height=2.93in,angle=90} 

\hspace{0.5cm}
\psfig{figure=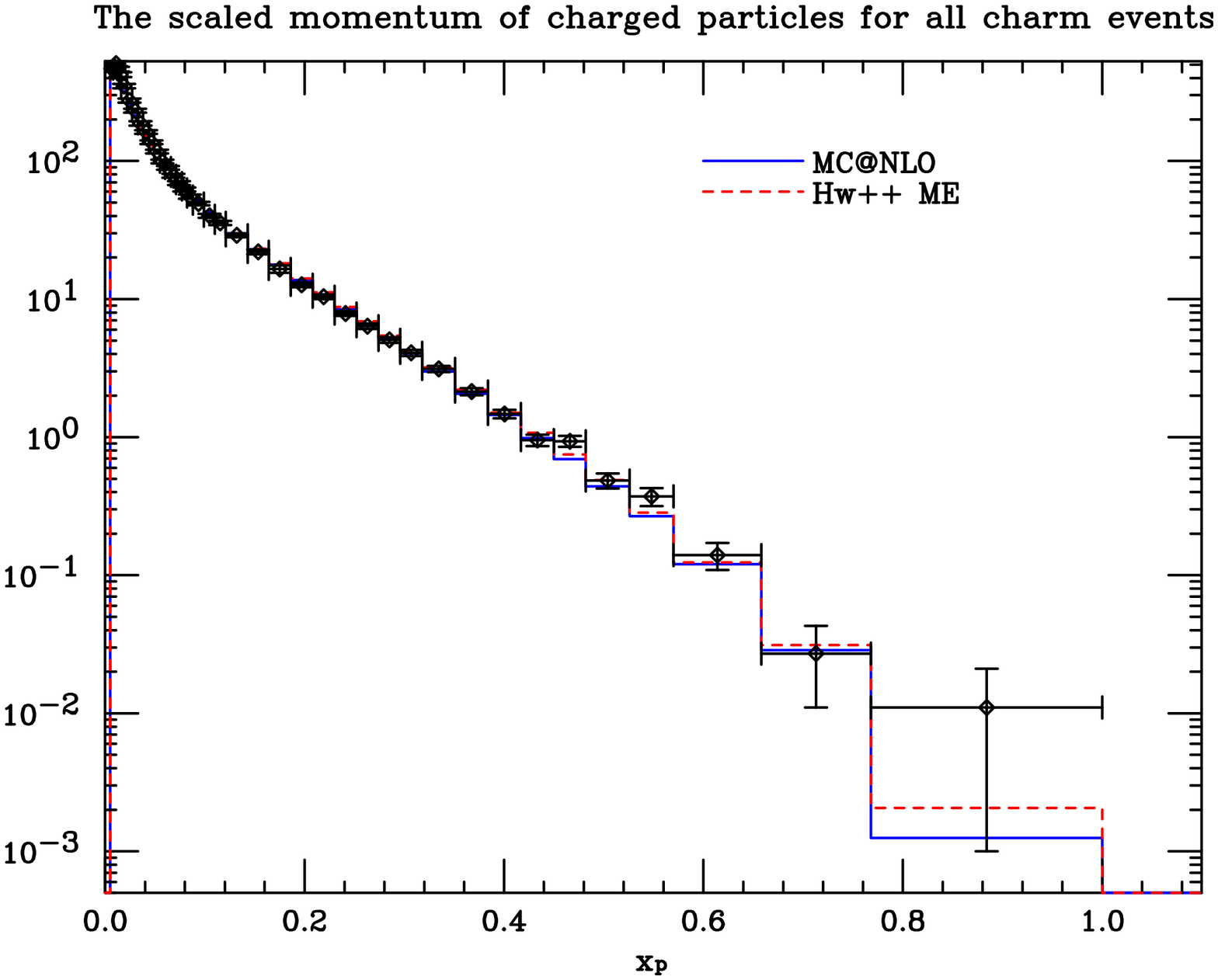,%
width=3.03in,height=2.93in,angle=90}
\psfig{figure=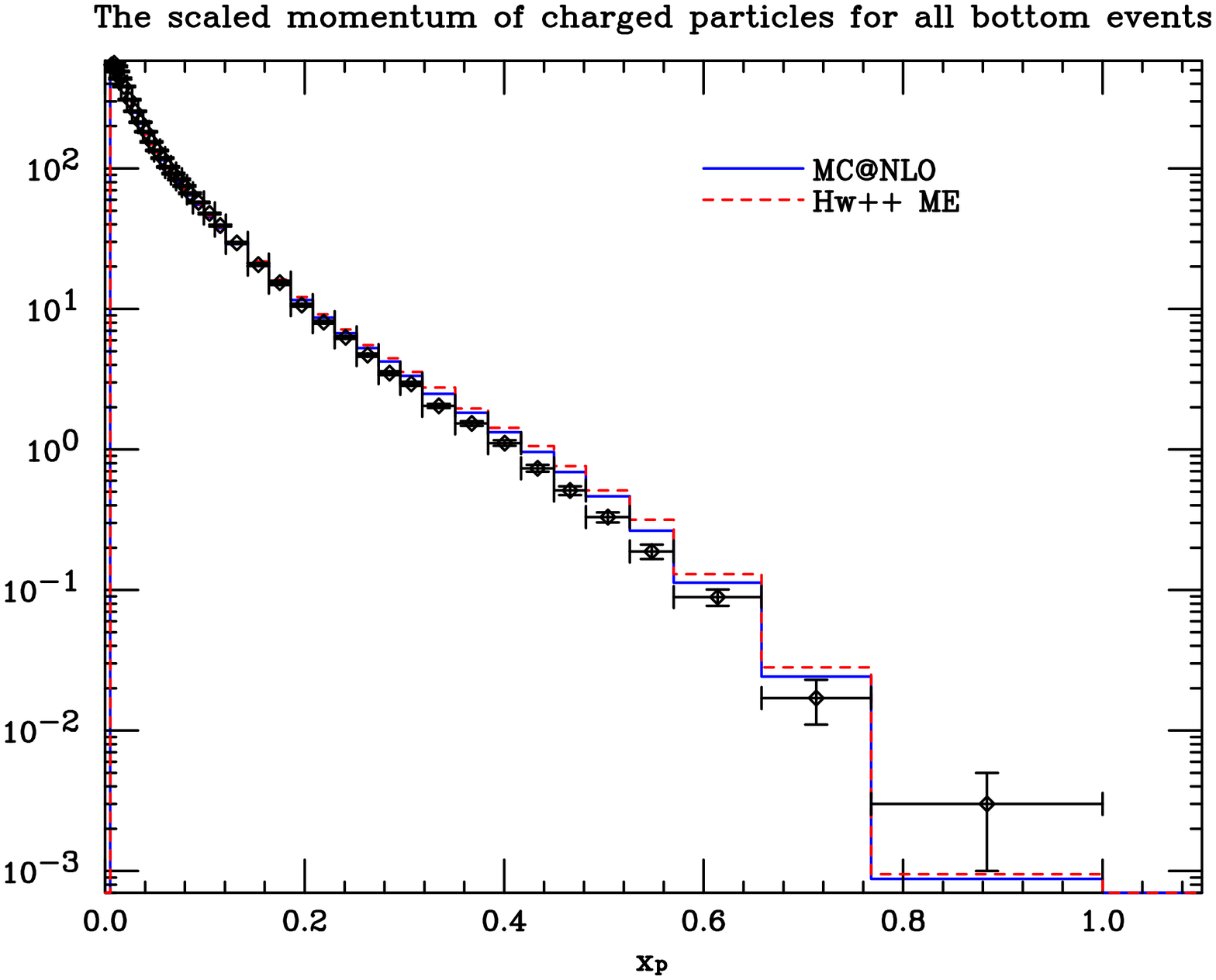,%
width=3.03in,height=2.93in,angle=90}
\caption{The scaled momentum distributions of all charged particles.}
\label{fig:cdlh2}    
\end{figure}

\begin{figure}[!ht]
\vspace{2cm}
\hspace{0.5cm}
\psfig{figure=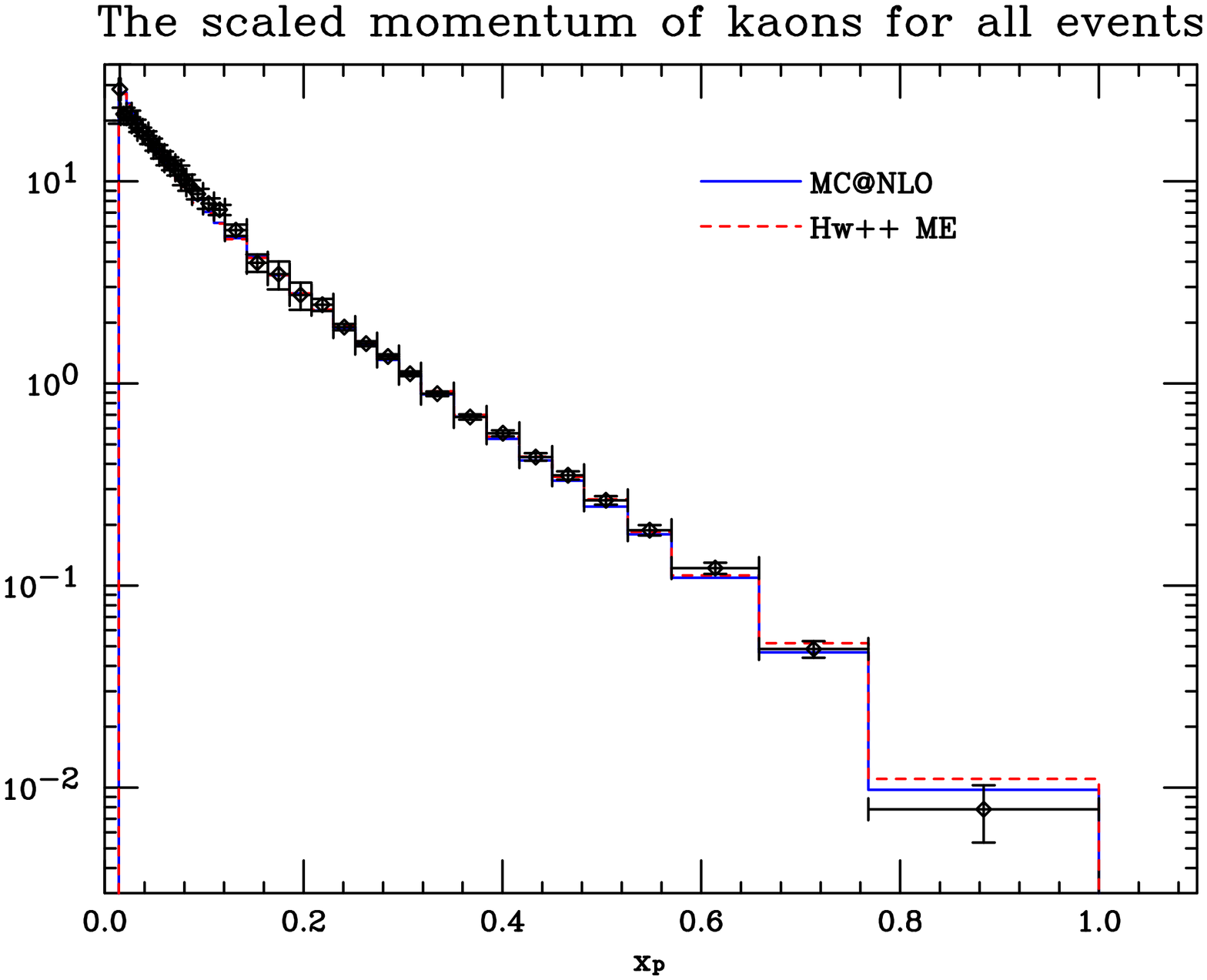,%
width=3.03in,height=2.93in,angle=90}
\psfig{figure=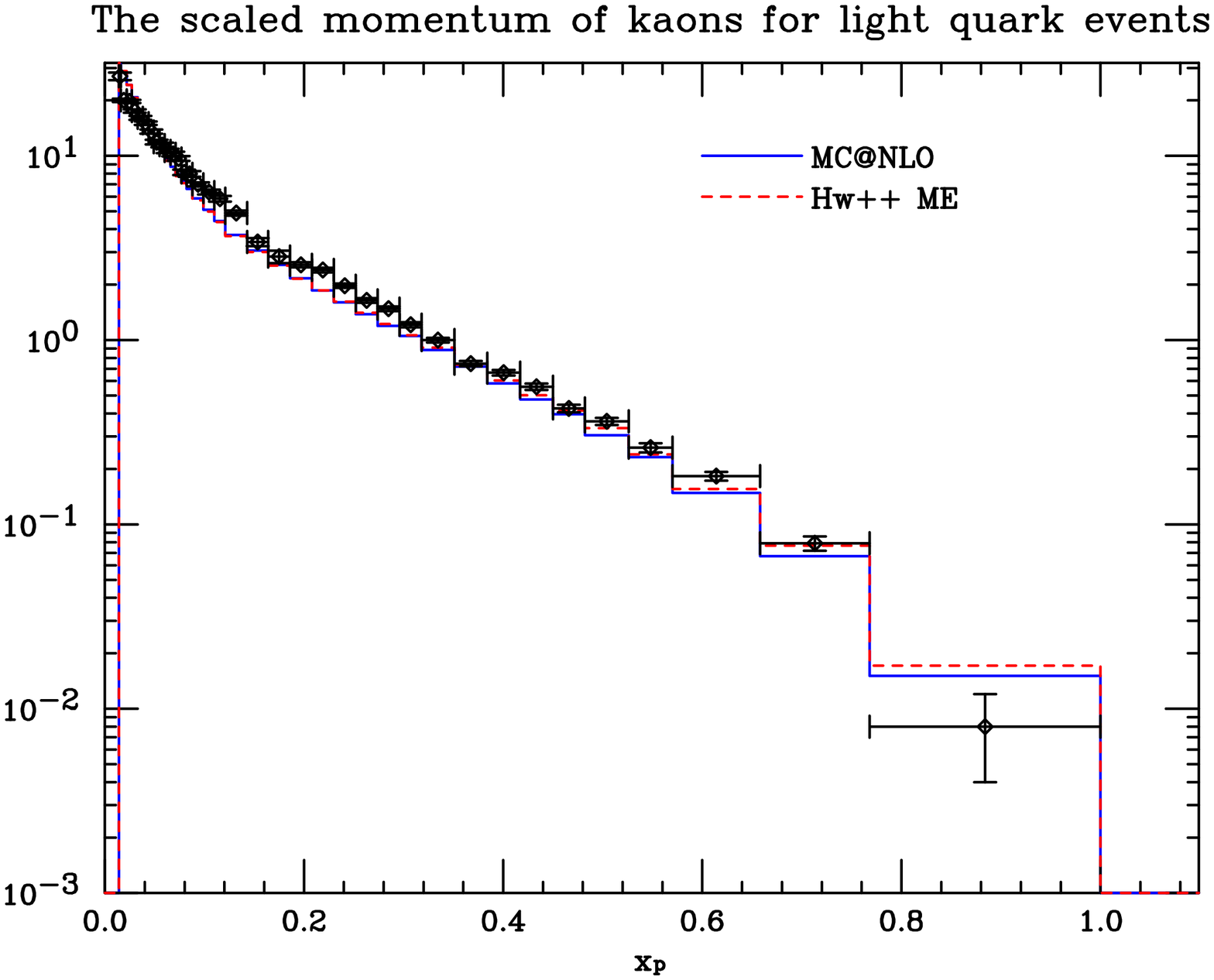,%
width=3.03in,height=2.93in,angle=90} 

\hspace{0.5cm}
\psfig{figure=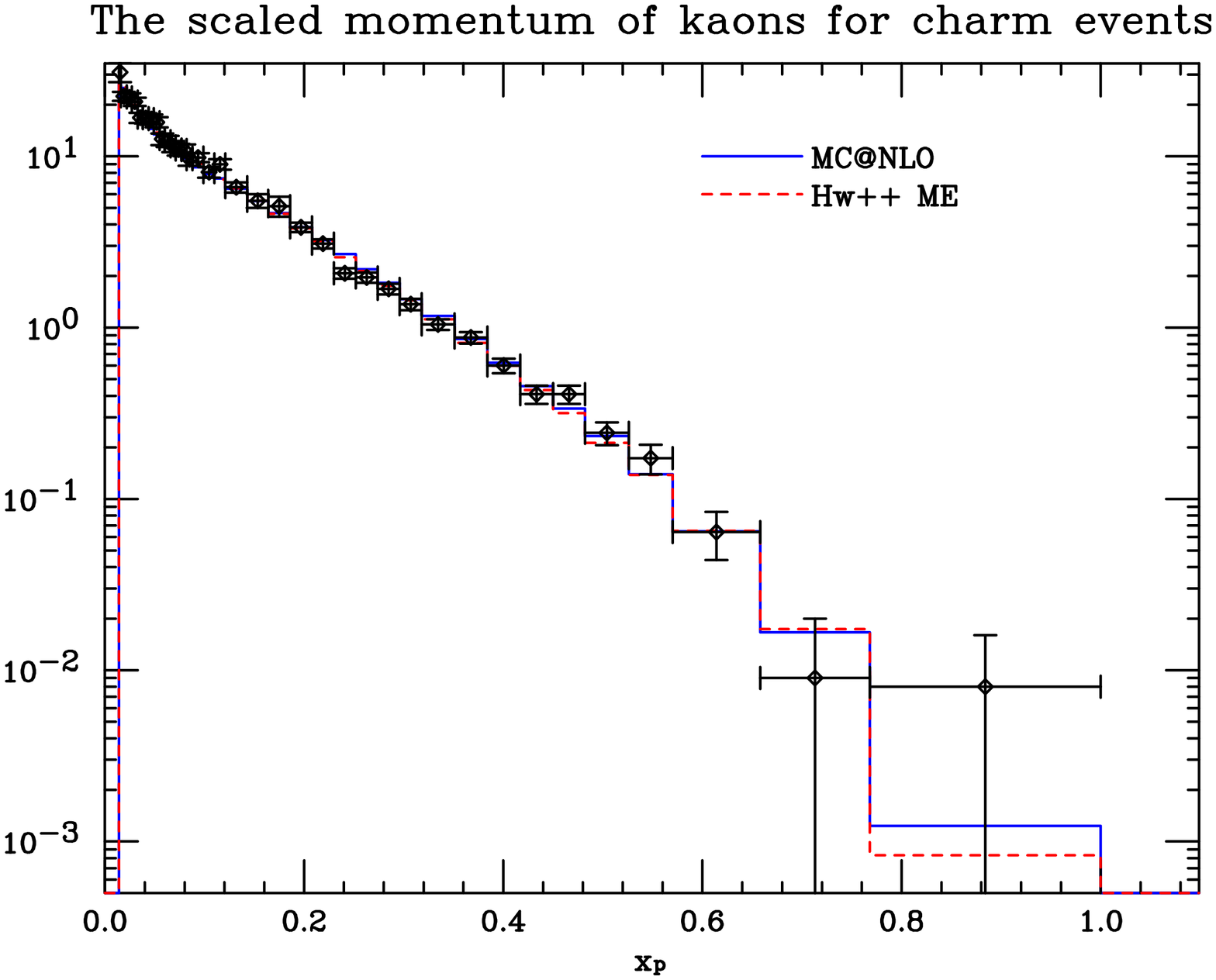,%
width=3.03in,height=2.93in,angle=90}
\psfig{figure=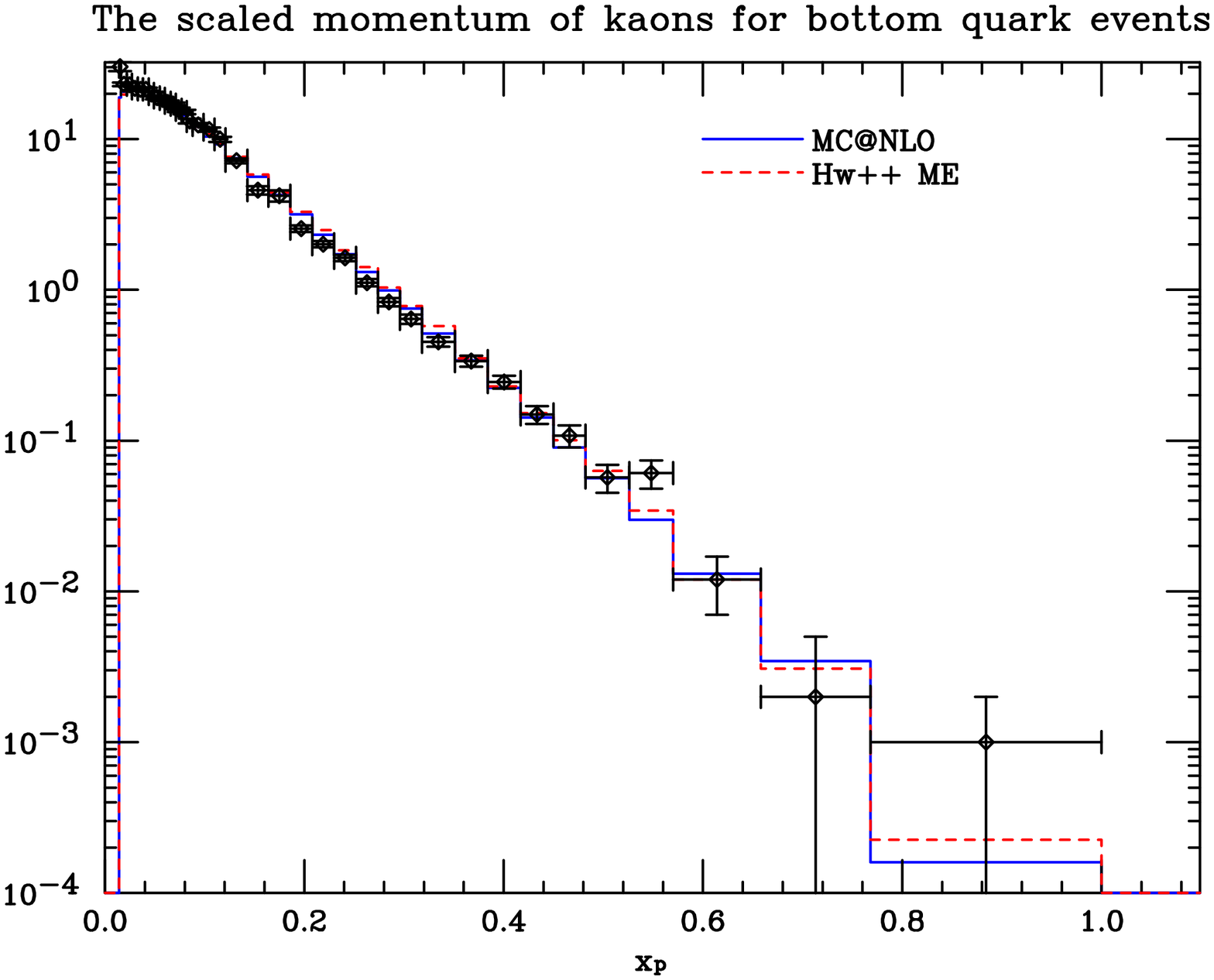,%
width=3.03in,height=2.93in,angle=90}
\caption{The scaled momentum distributions of kaons.}
\label{fig:cdlh3}    
\end{figure}

\begin{figure}[!ht]
\vspace{2cm}
\hspace{0.5cm}
\psfig{figure=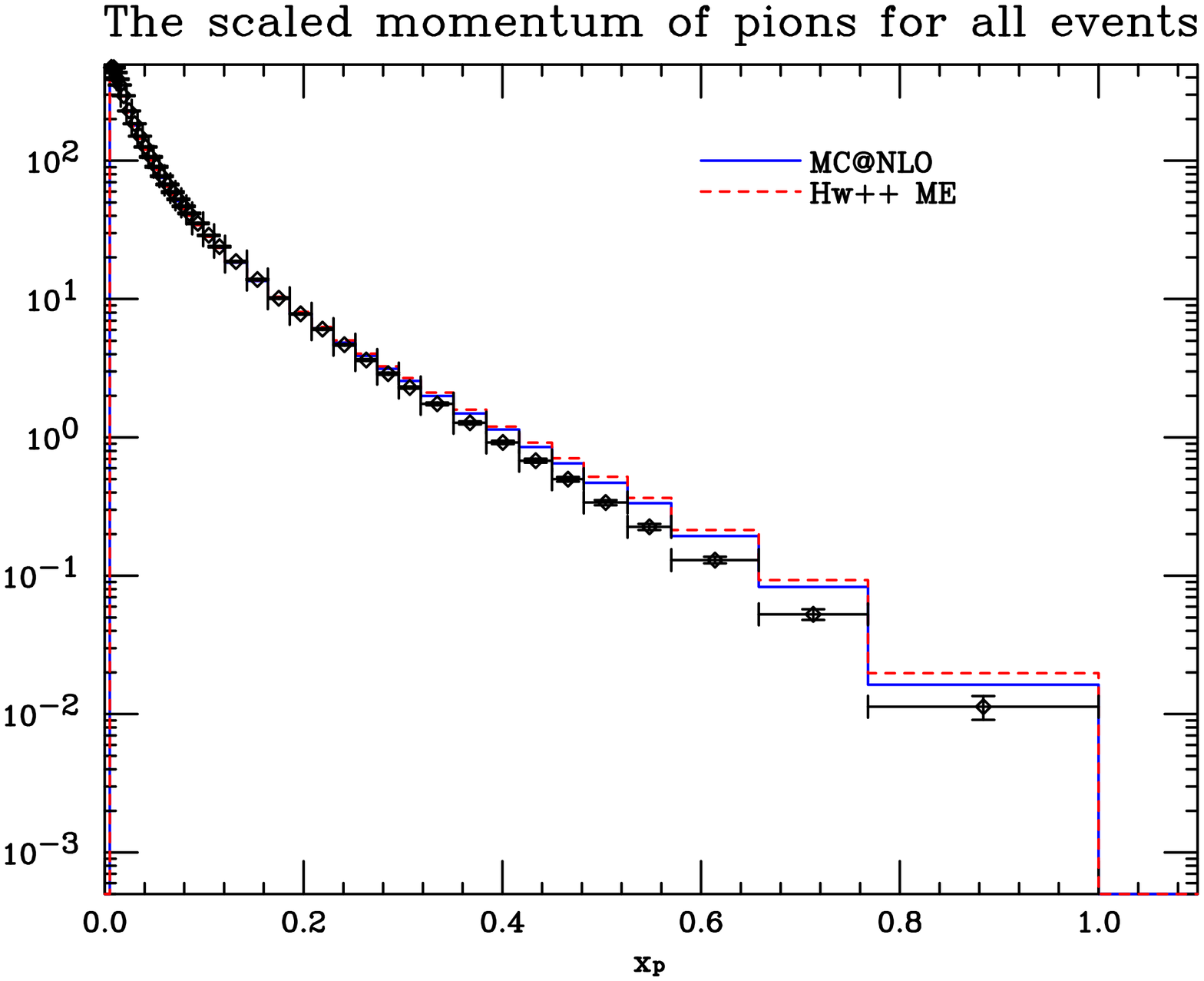,%
width=3.03in,height=2.93in,angle=90}
\psfig{figure=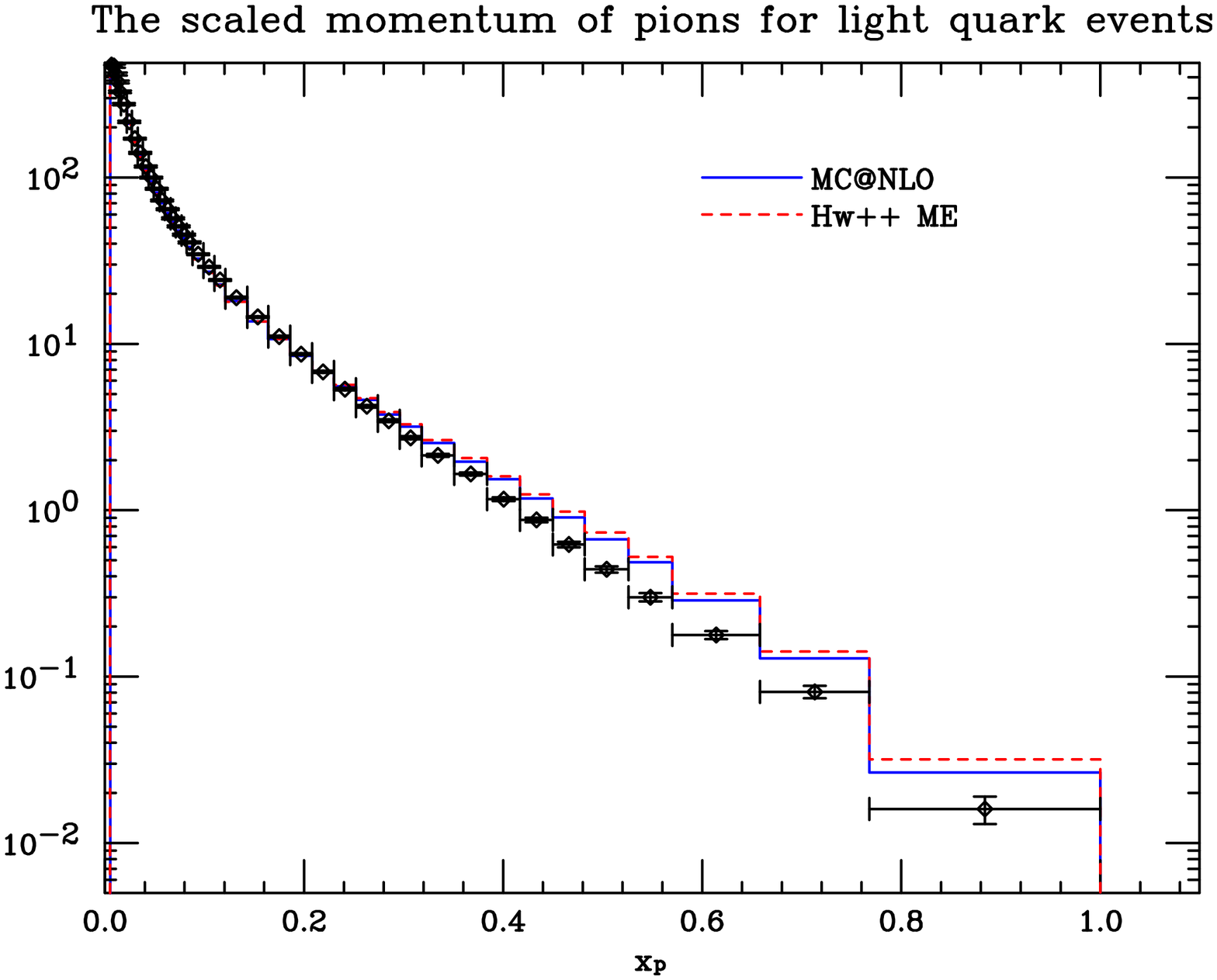,%
width=3.03in,height=2.93in,angle=90} 

\hspace{0.5cm}
\psfig{figure=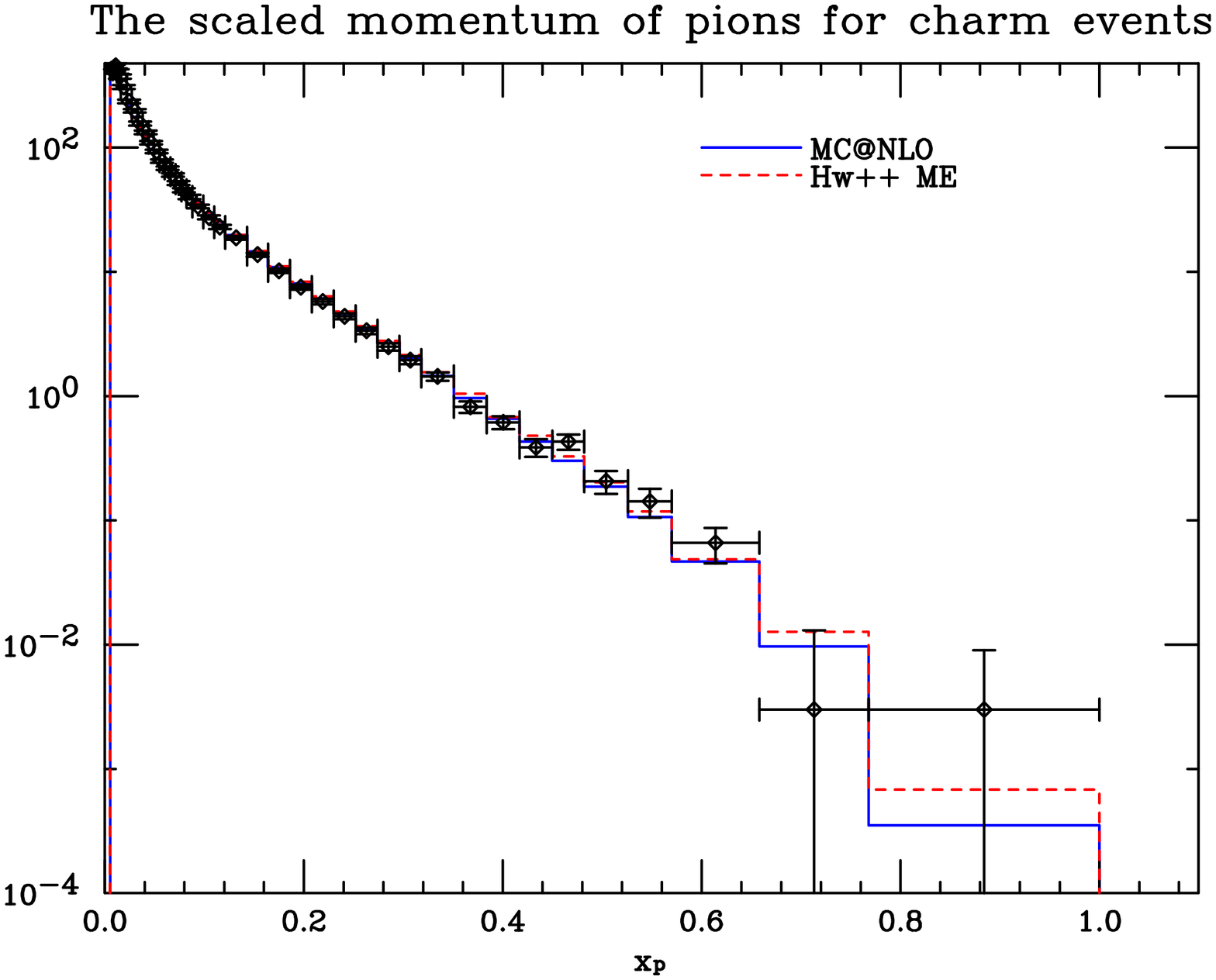,%
width=3.03in,height=2.93in,angle=90}
\psfig{figure=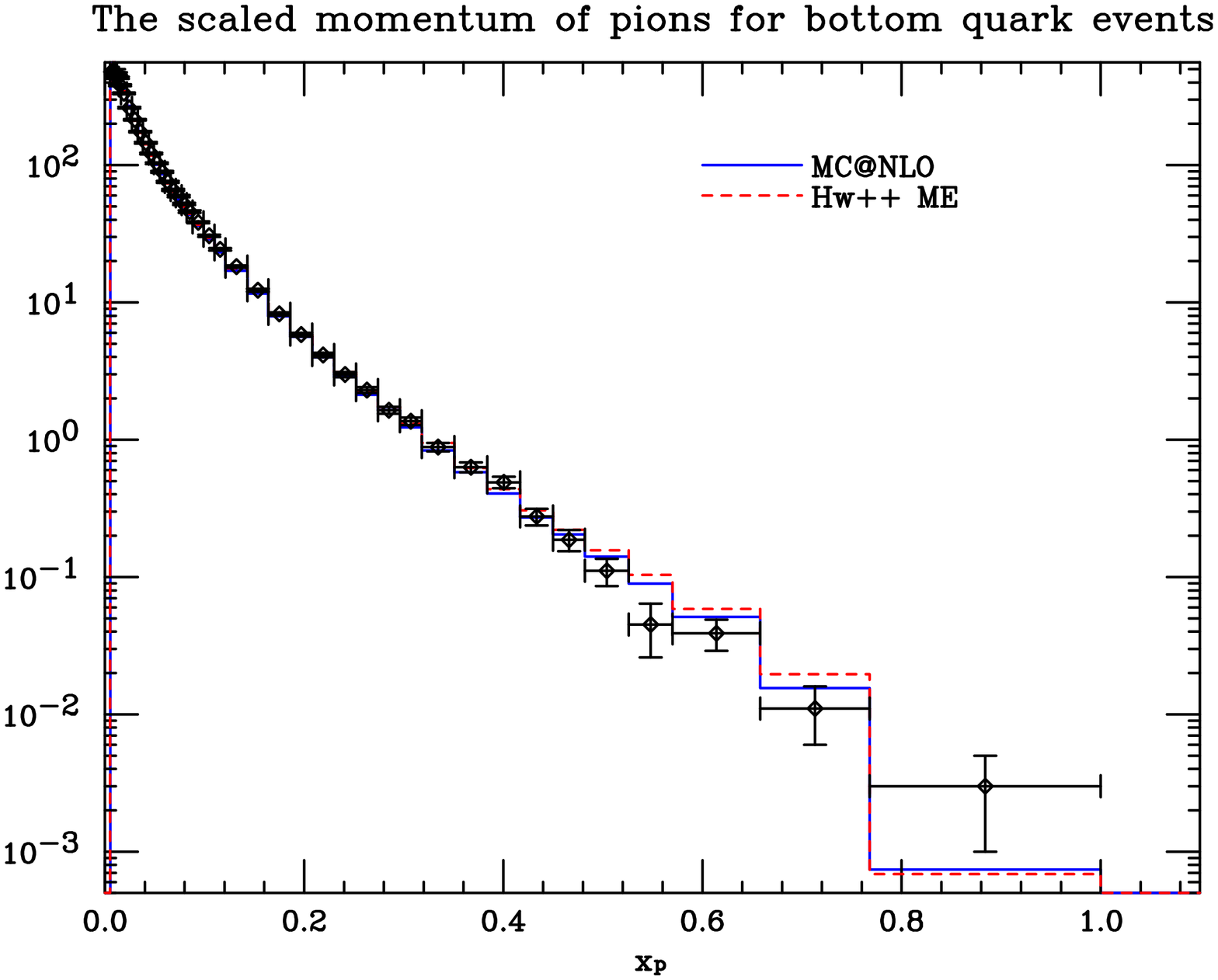,%
width=3.03in,height=2.93in,angle=90}
\caption{The scaled momentum distributions of pions.}
\label{fig:cdlh4}    
\end{figure}

Figures \ref{fig:cdlh2}-\ref{fig:cdlh4} show comparisons of identified particle spectra from events of different
flavour with SLD data \cite{Abe:2003iy}. 
In general we are able to give a good description of the data with the MC@NLO method. The MC@NLO results for the LEP event shape distributions do not differ greatly from the matrix element
correction results. For example, the thrust distribution appears to suffer from the same problem of generating too much
transverse structure, leading to less two-jet like event shapes. 

However, despite the
similarity in results, we can be confident that the MC@NLO results are normalised to
the full NLO cross-section including  virtual corrections unlike the matrix element
correction. 

The
identified particle spectra includes hadron momenta distributions from heavy quark
production. Although the results are similar, in some cases the MC@NLO results are slightly better than the matrix
element correction results. This can be attributed to the better treatment of the heavy quark
production cross-sections in the relevant plots.

\begin{figure}[!ht]
\vspace{2cm}
\hspace{0.5cm}
\psfig{figure=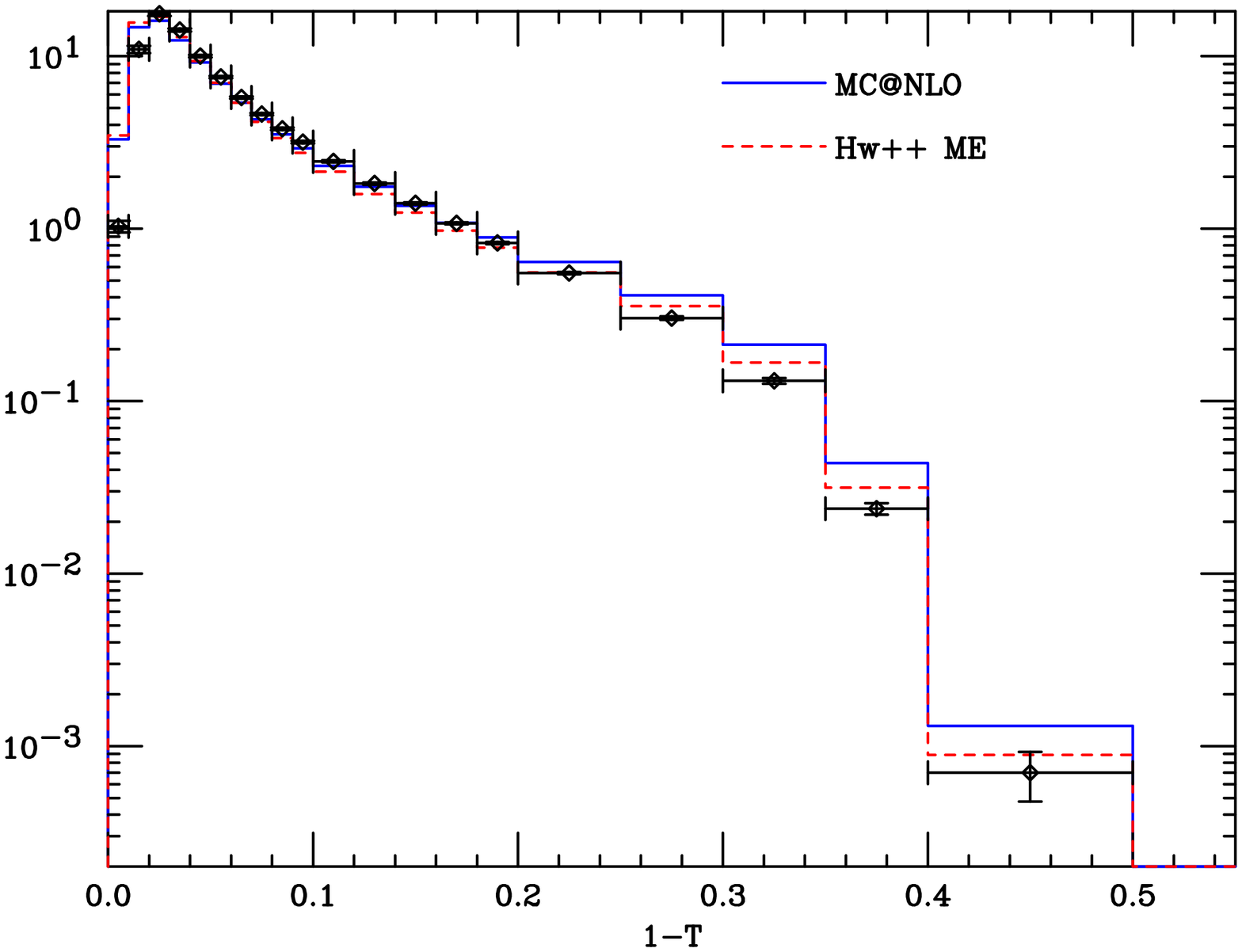,%
width=3.03in,height=2.93in,angle=90}
\psfig{figure=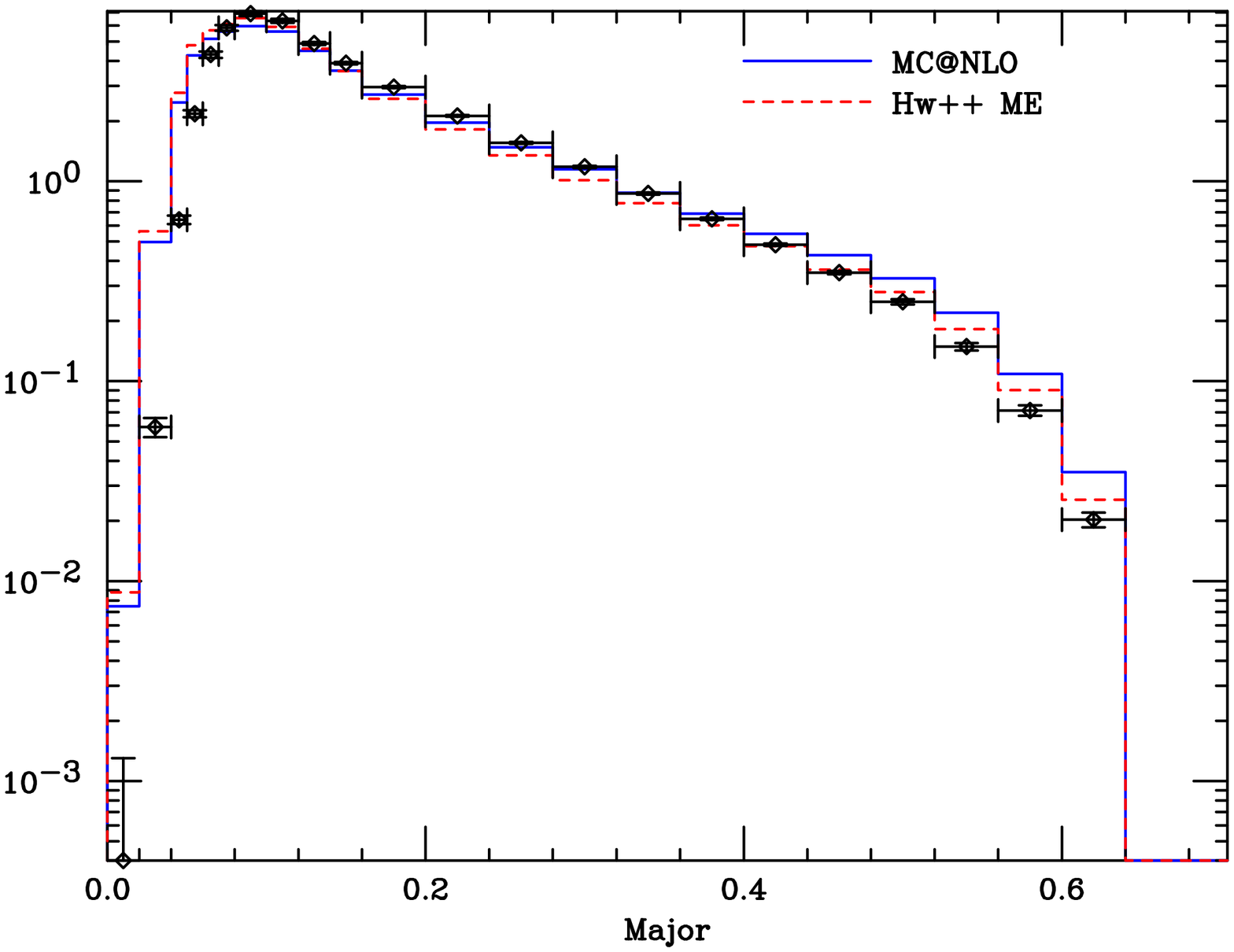,%
width=3.03in,height=2.93in,angle=90} 

\hspace{4cm}
\psfig{figure=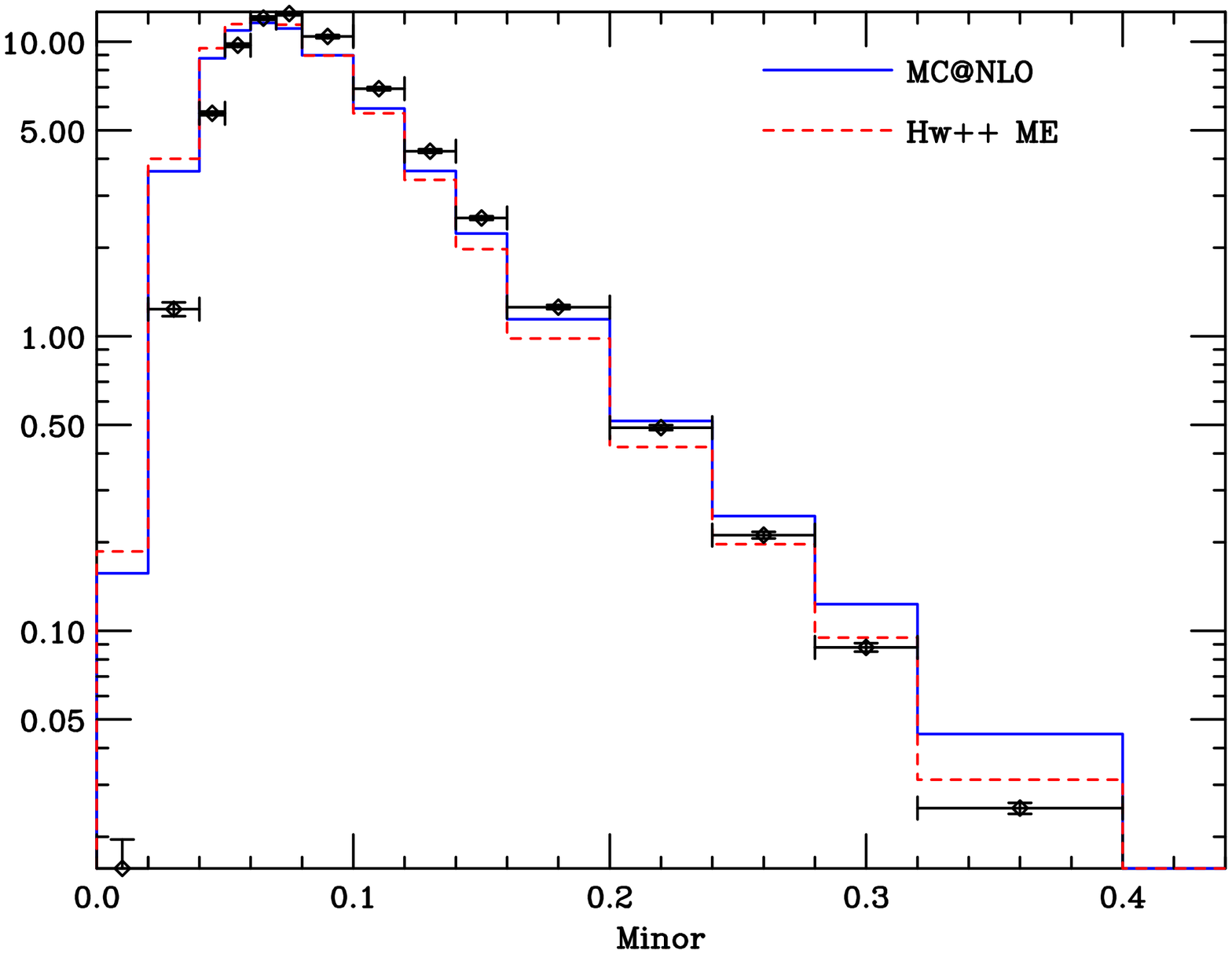,%
width=3.03in,height=2.93in,angle=90}
\caption{Thrust, Thrust major and Thrust minor. Data from \cite{Abreu:1996na}.}
\label{fig:tt}    
\end{figure}

\begin{figure}[!ht]
\vspace{2cm}
\hspace{0.5cm}
\psfig{figure=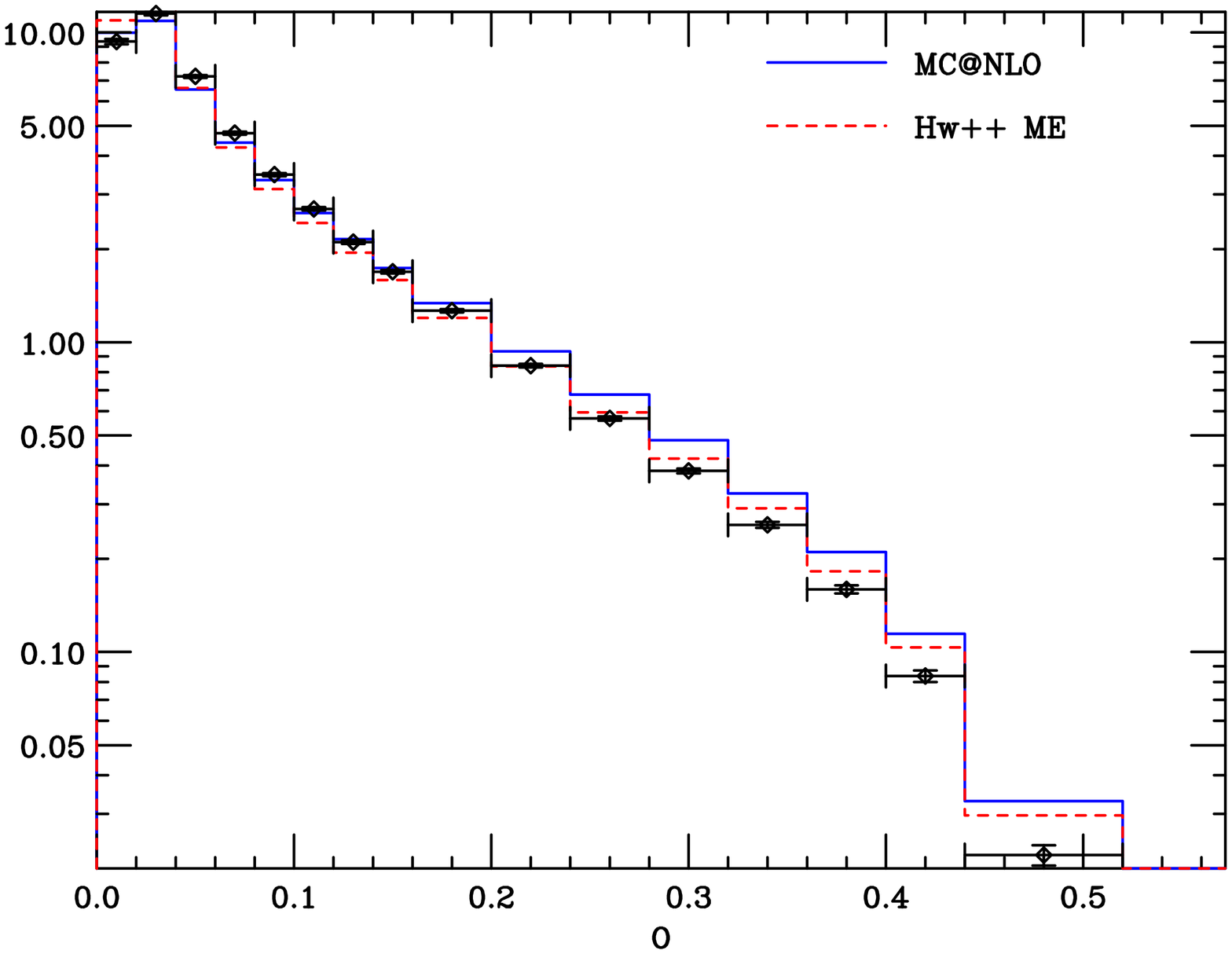,%
width=3.03in,height=2.93in,angle=90}
\psfig{figure=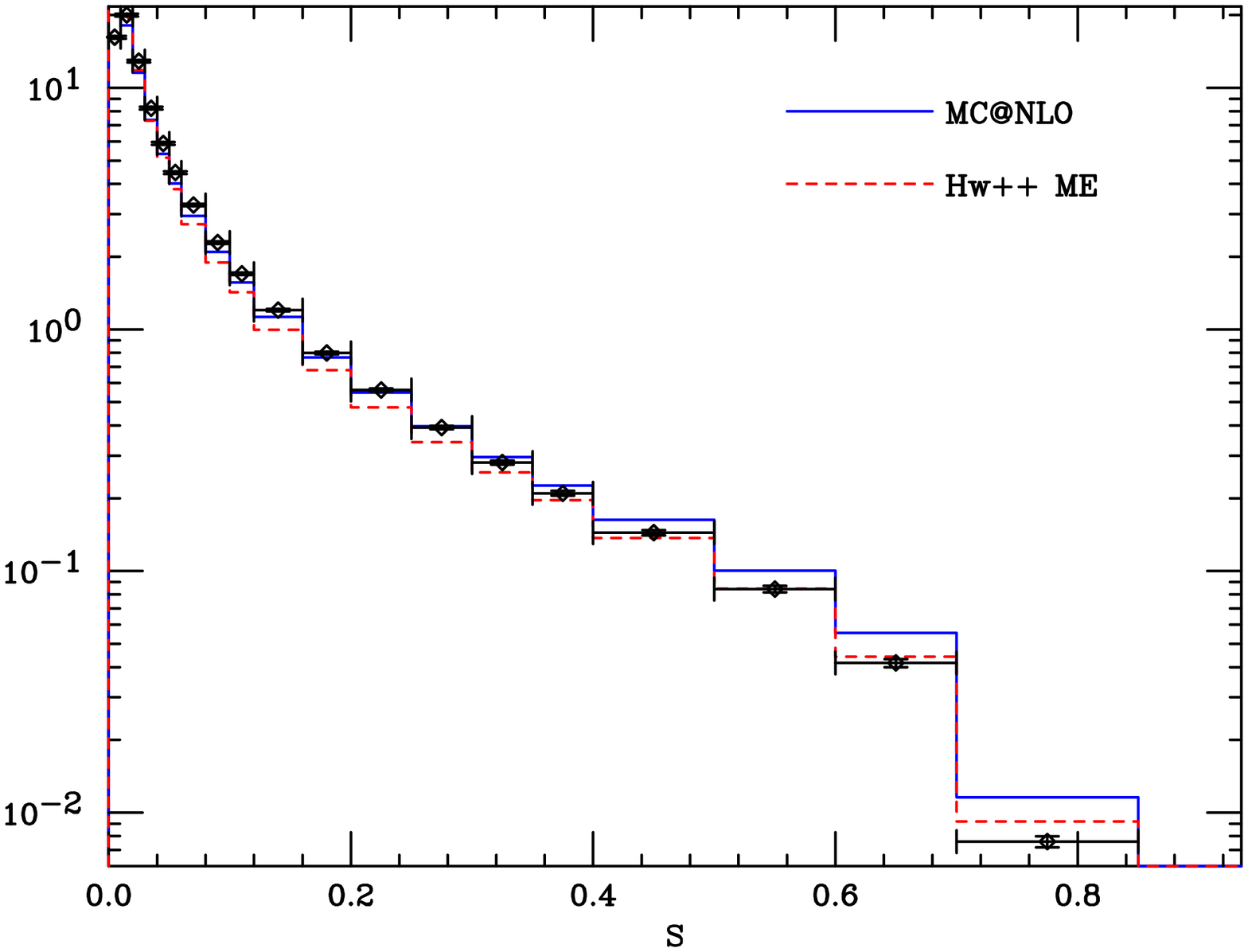,%
width=3.03in,height=2.93in,angle=90} 

\hspace{0.5cm}
\psfig{figure=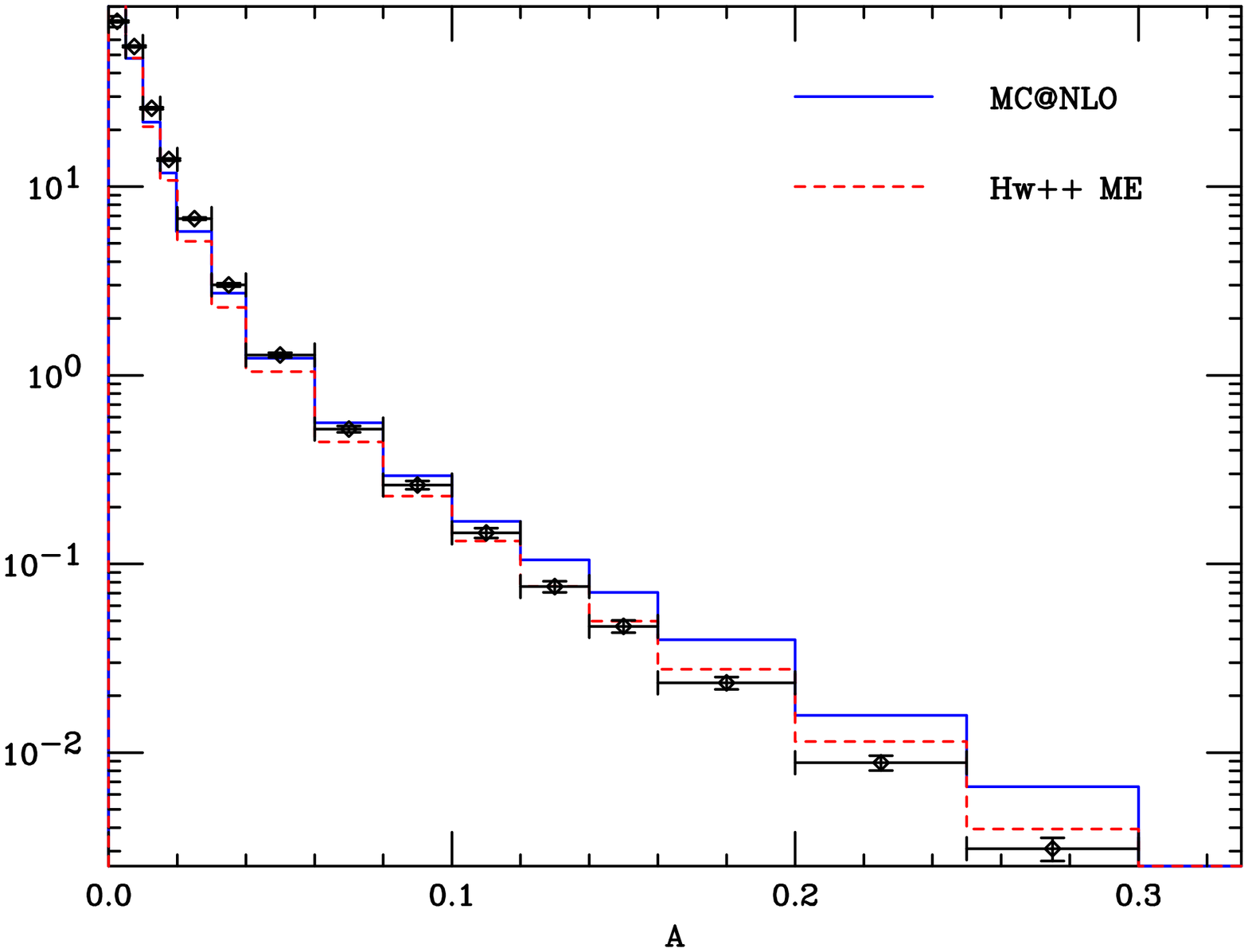,%
width=3.03in,height=2.93in,angle=90}
\psfig{figure=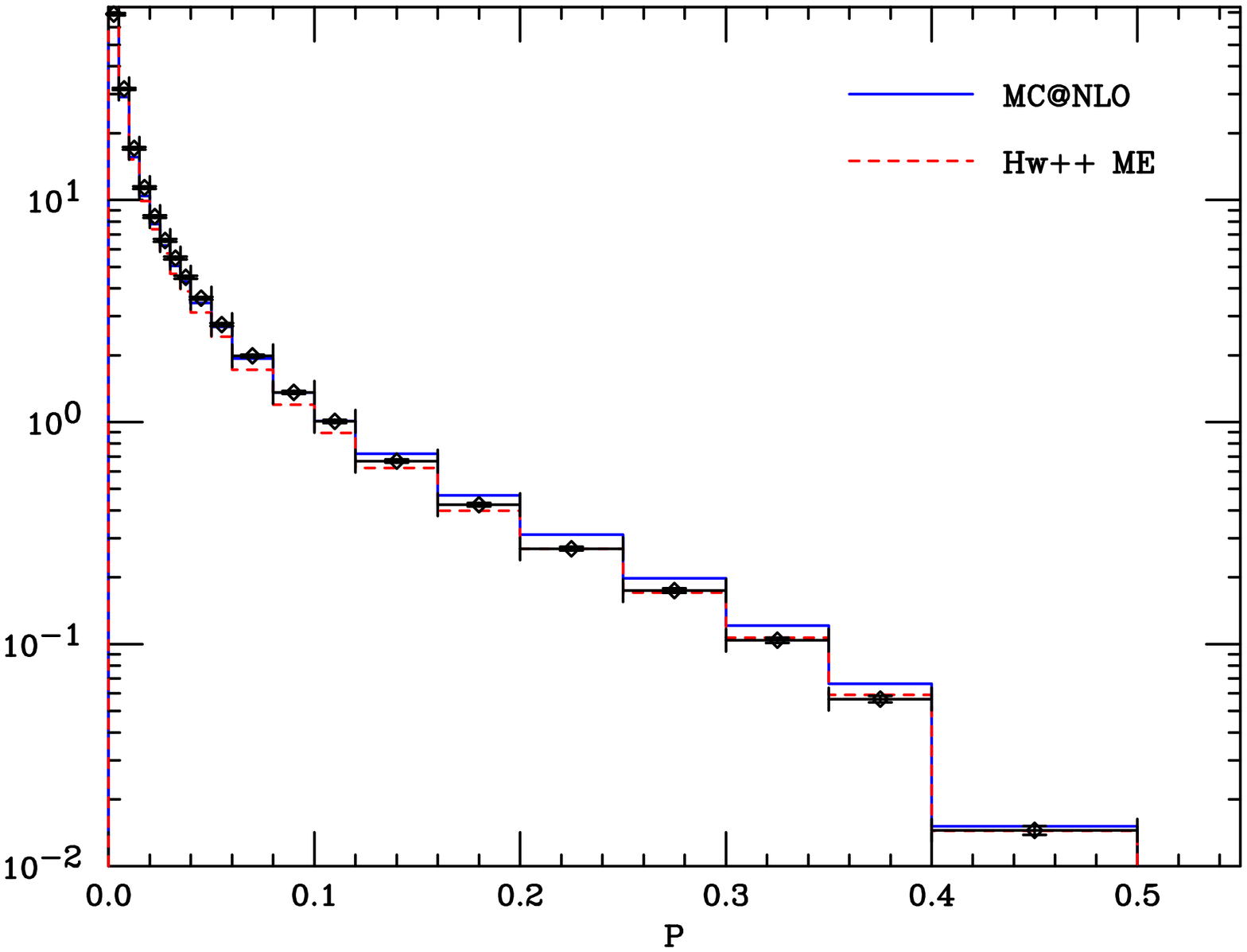,%
width=3.03in,height=2.93in,angle=90}
\caption{Oblateness, Sphericity, Aplanarity and Planarity distributions. Data from
  \cite{Abreu:1996na}.}
\label{fig:osap}    
\end{figure}

\begin{figure}[!ht]
\vspace{2cm}
\hspace{0.5cm}
\psfig{figure=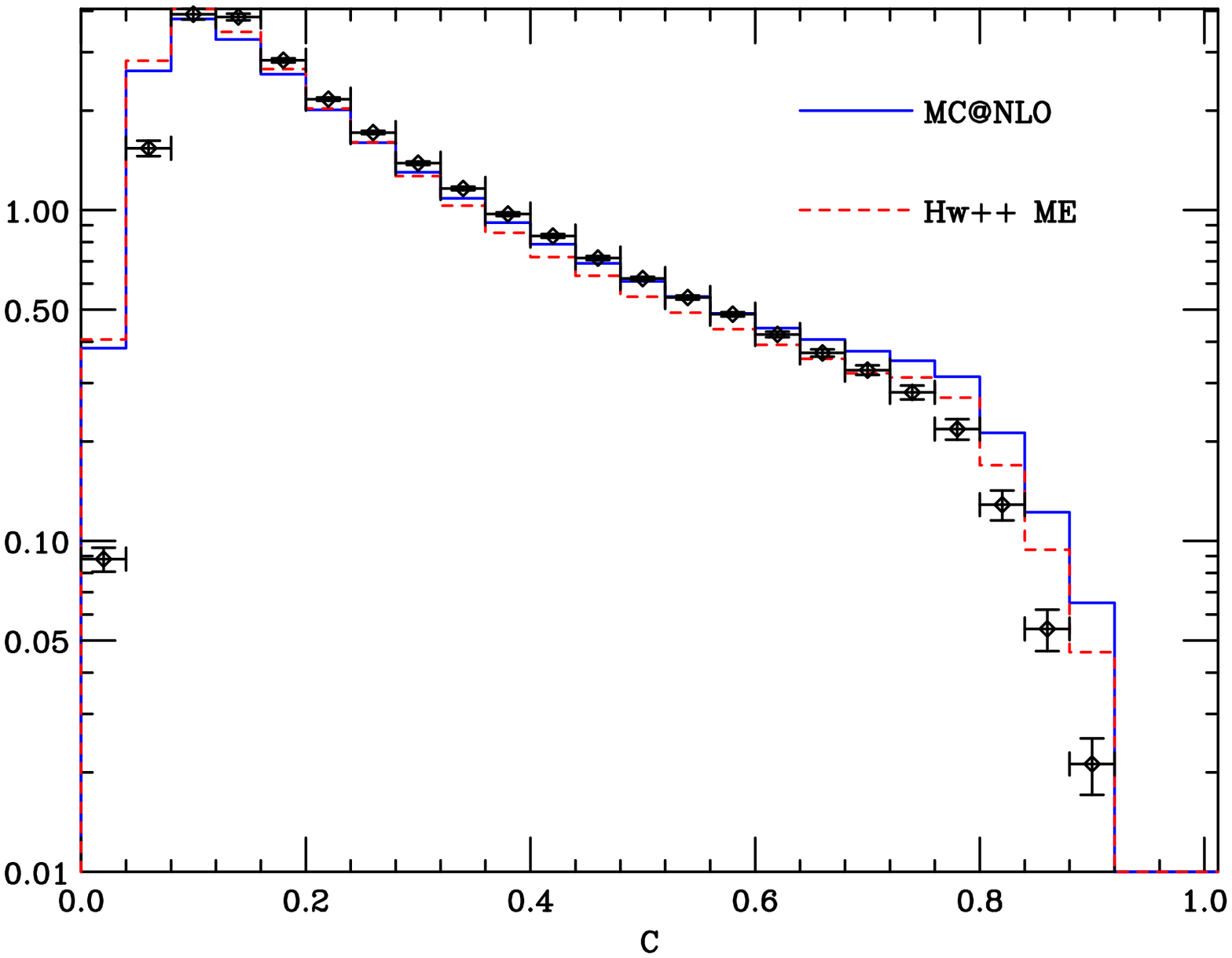,%
width=3.03in,height=2.93in,angle=90}
\psfig{figure=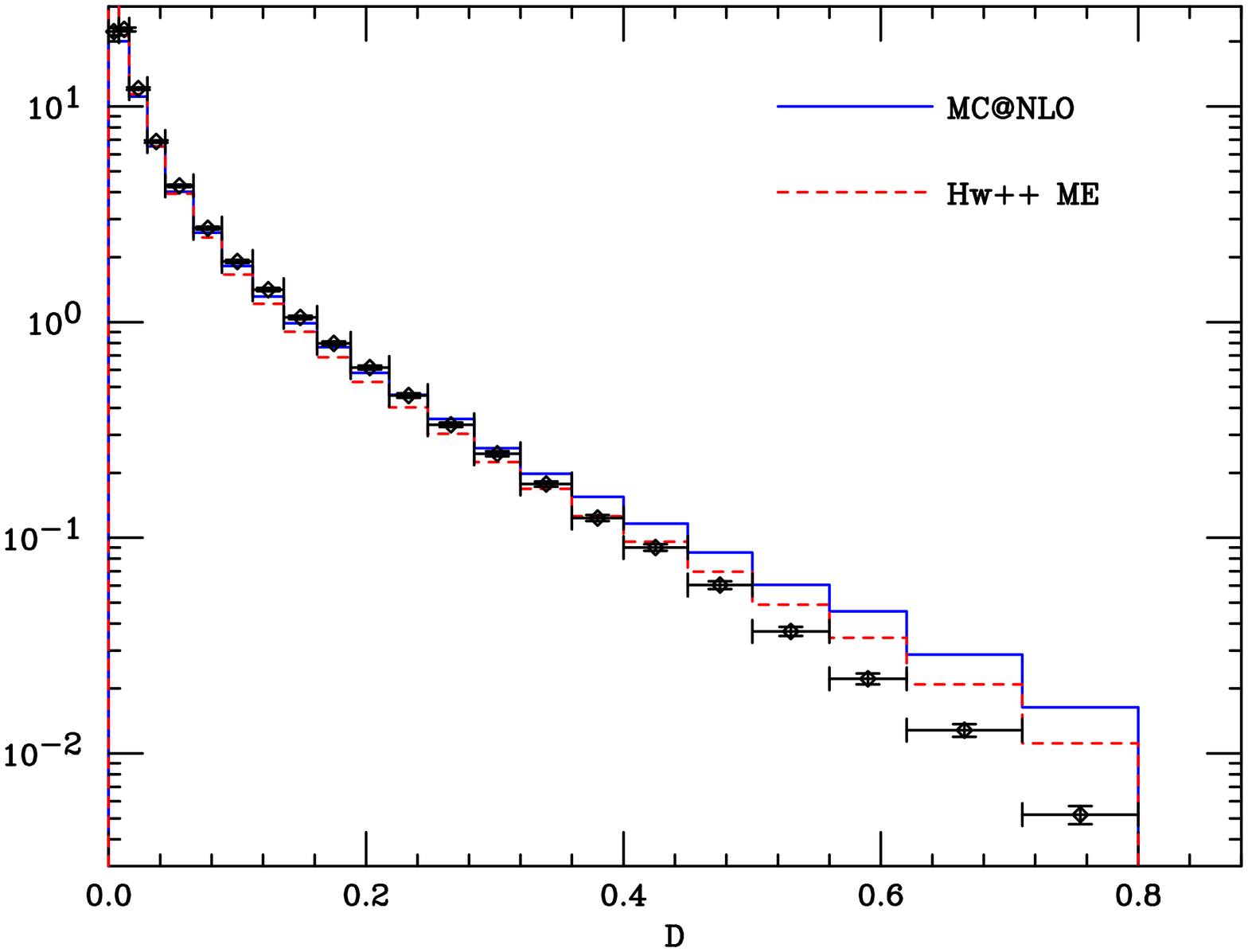,%
width=3.03in,height=2.93in,angle=90} 

\hspace{0.5cm}
\psfig{figure=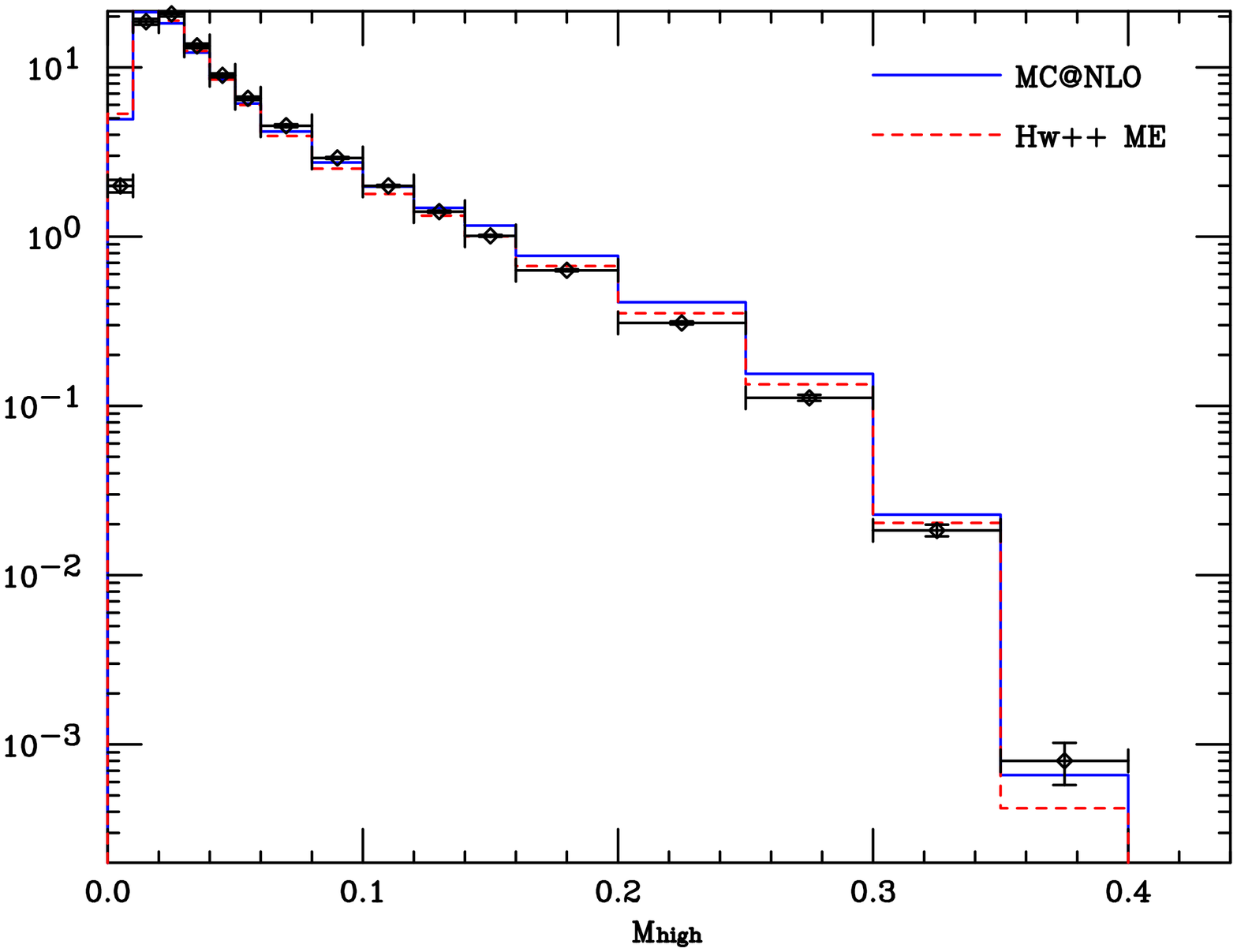,%
width=3.03in,height=2.93in,angle=90}
\psfig{figure=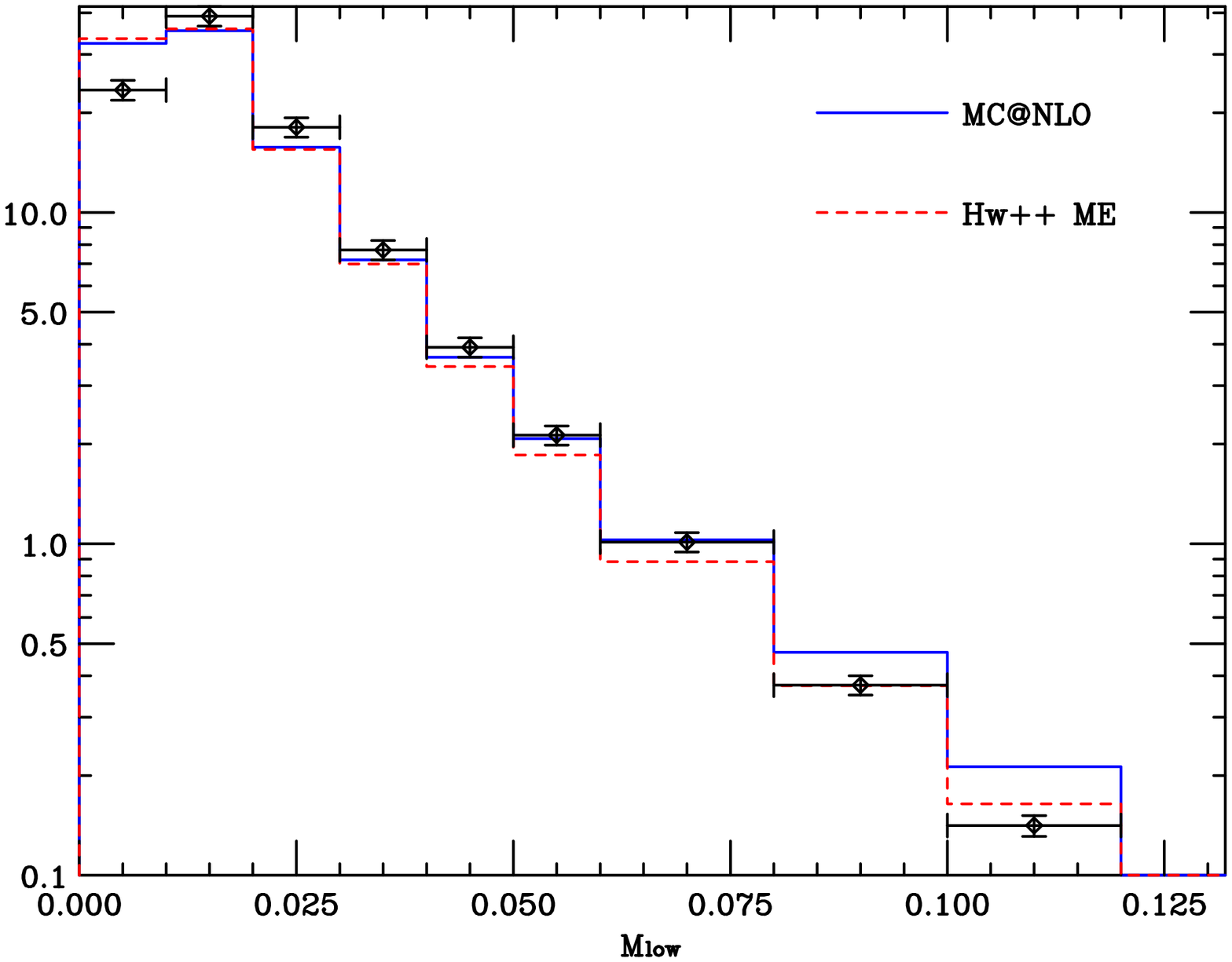,%
width=3.03in,height=2.93in,angle=90}
\caption{$C$ Parameter and $D$ Parameter distributions and the high, $M_{\rm high}$ and
  low, $M_{\rm low}$  hemisphere masses. Data from \cite{Abreu:1996na}.}
\label{fig:cdlh1}    
\end{figure}

\begin{figure}[!ht]
\vspace{2cm}
\hspace{0.5cm}
\psfig{figure=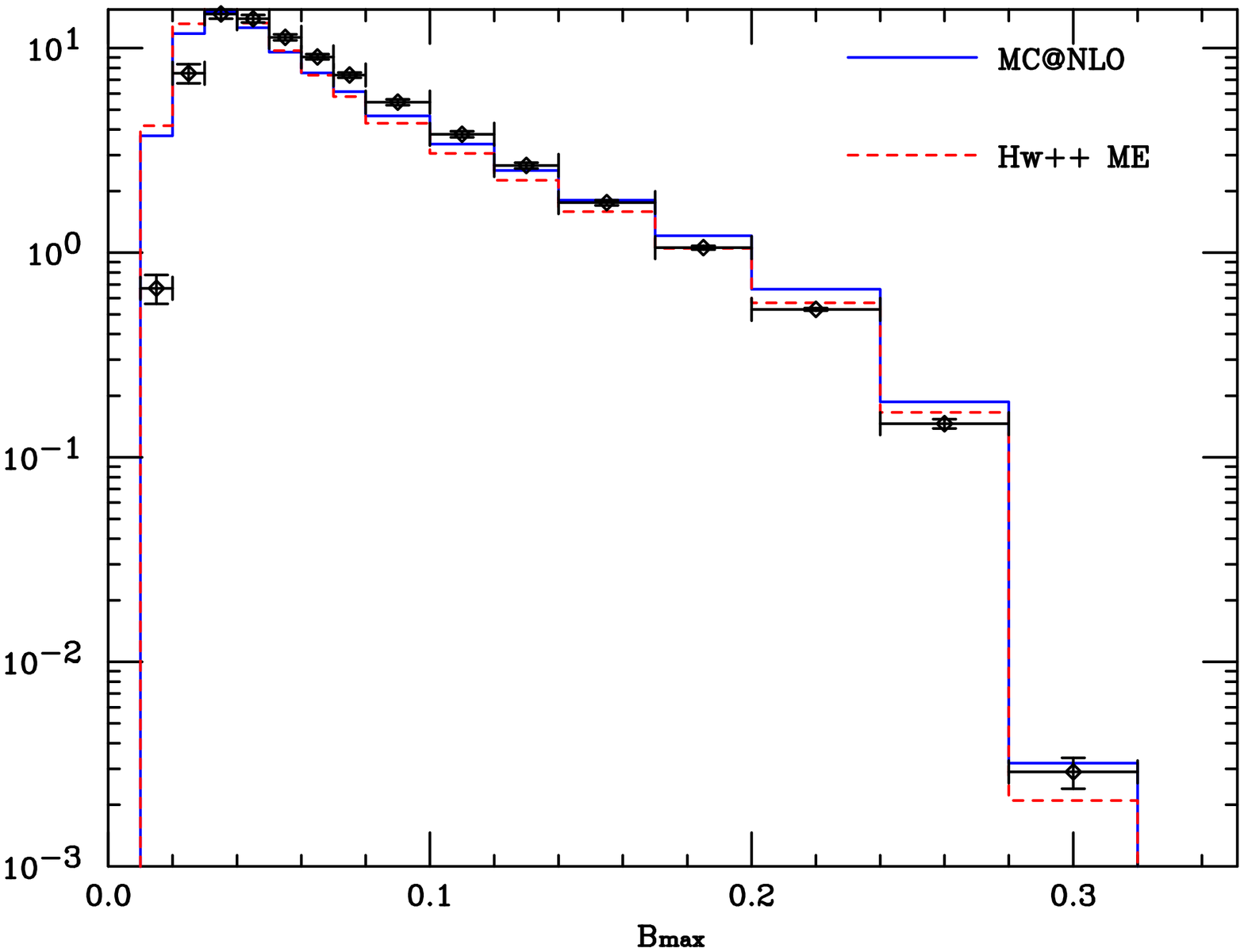,%
width=3.03in,height=2.93in,angle=90}
\psfig{figure=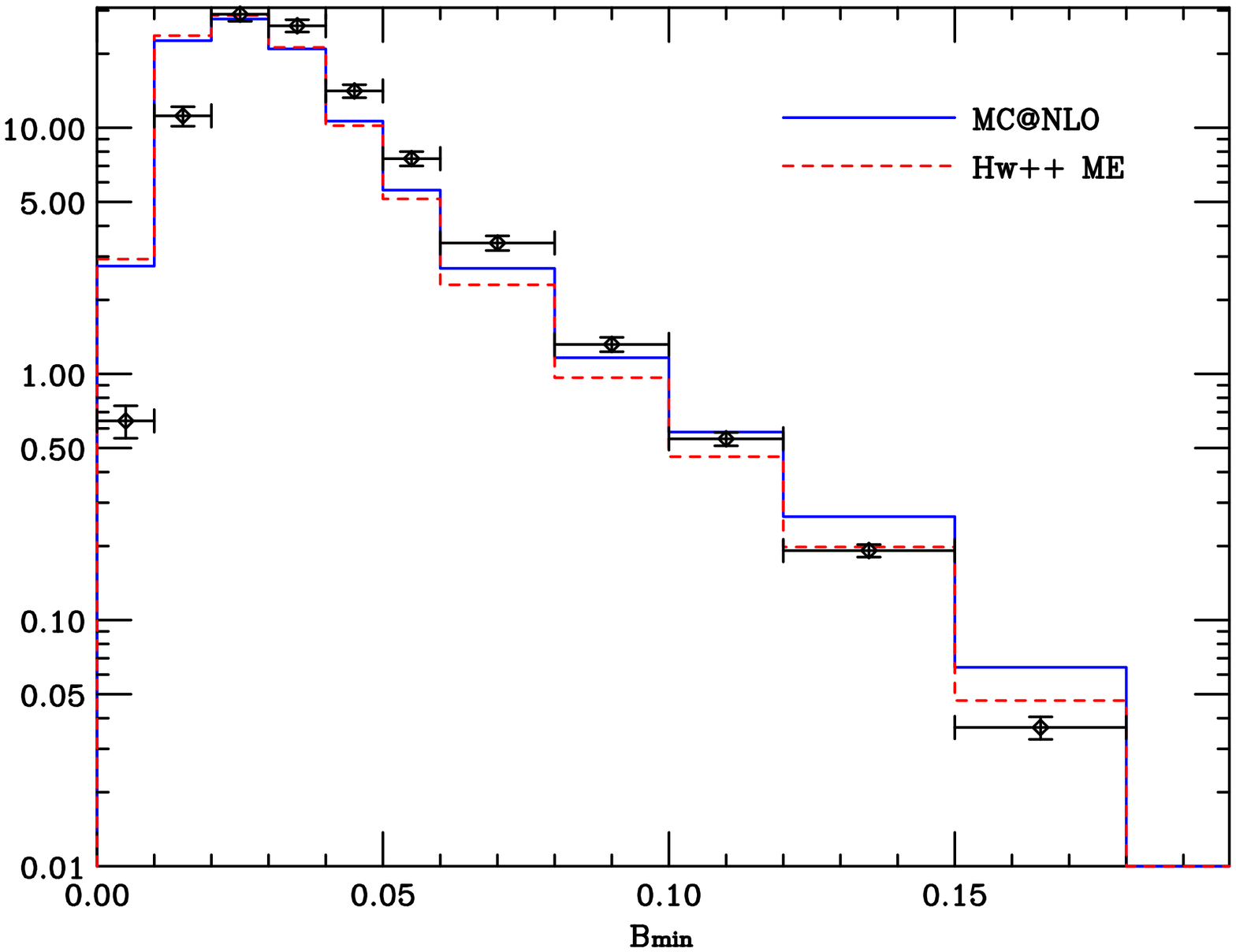,%
width=3.03in,height=2.93in,angle=90} 

\hspace{4cm}
\psfig{figure=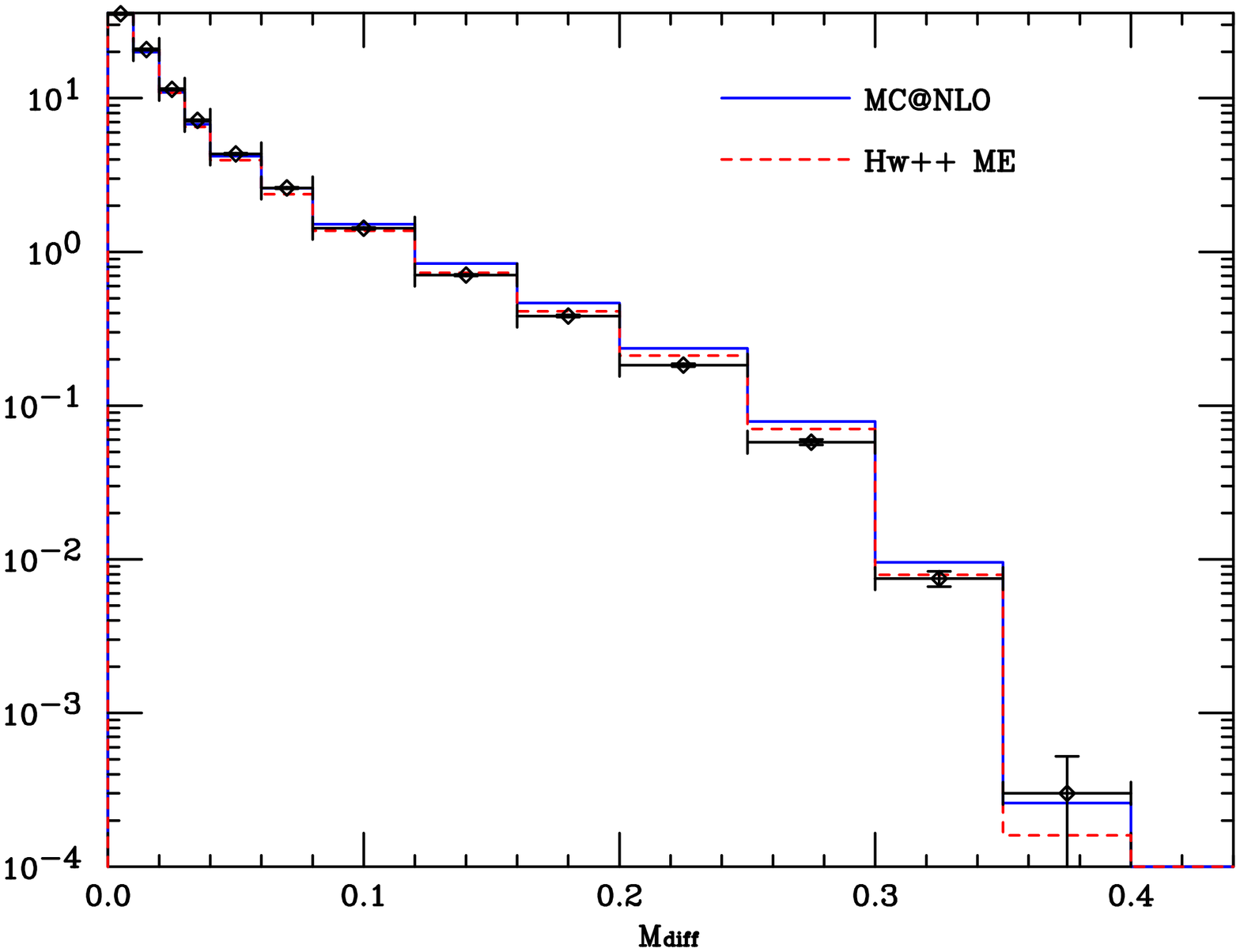,%
width=3.03in,height=2.93in,angle=90}
\caption{The wide and narrow jet broadening measures $B_{\rm max}$ and $B_{\rm min}$ and
  the difference in hemisphere masses, $M_{\rm diff}$. Data from \cite{Abreu:1996na}.}
\label{fig:xnmd} 
 
\end{figure}
\begin{figure}[!ht]
\hspace{0.5cm}
\psfig{figure=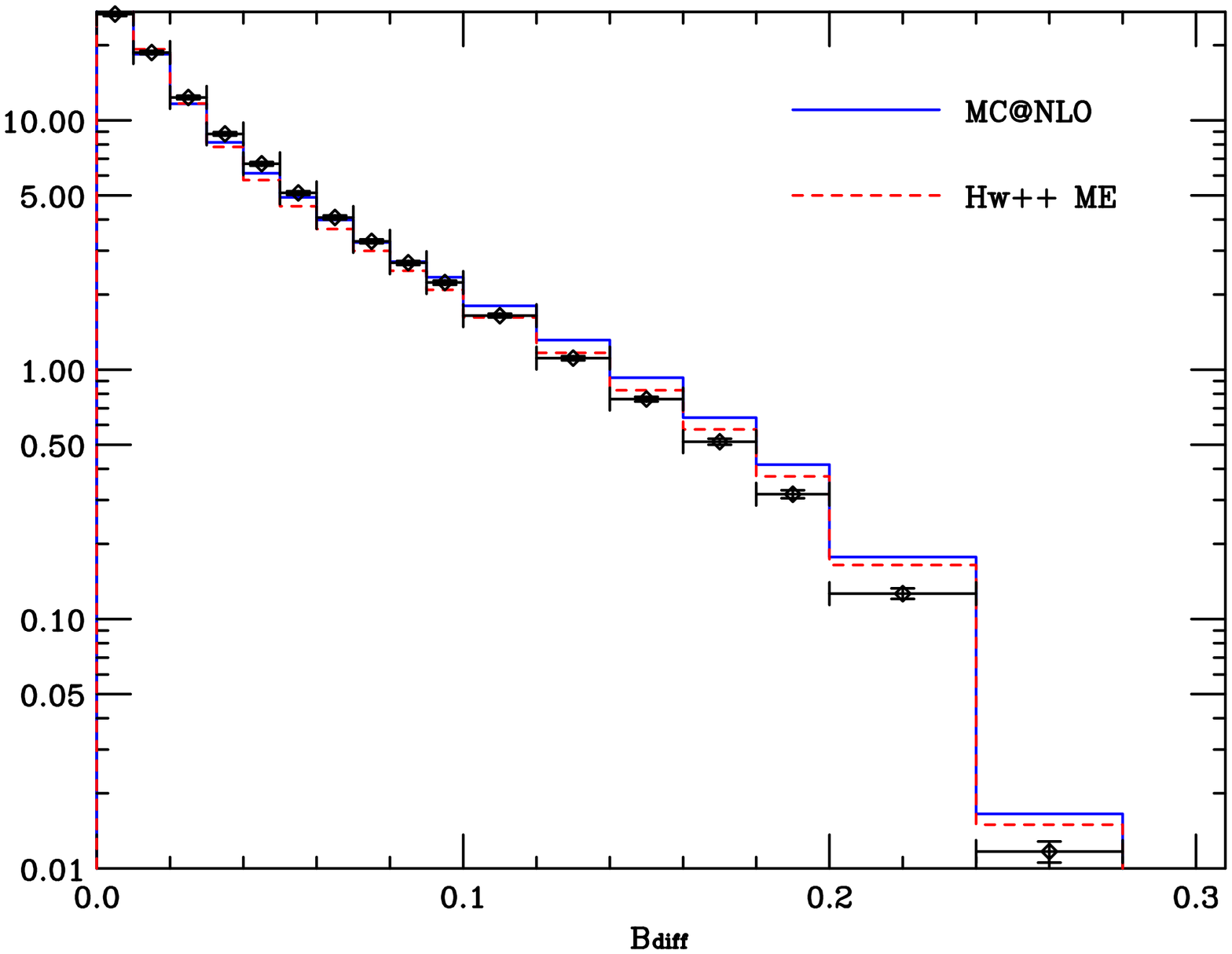,%
width=3.03in,height=2.93in,angle=90}
\psfig{figure=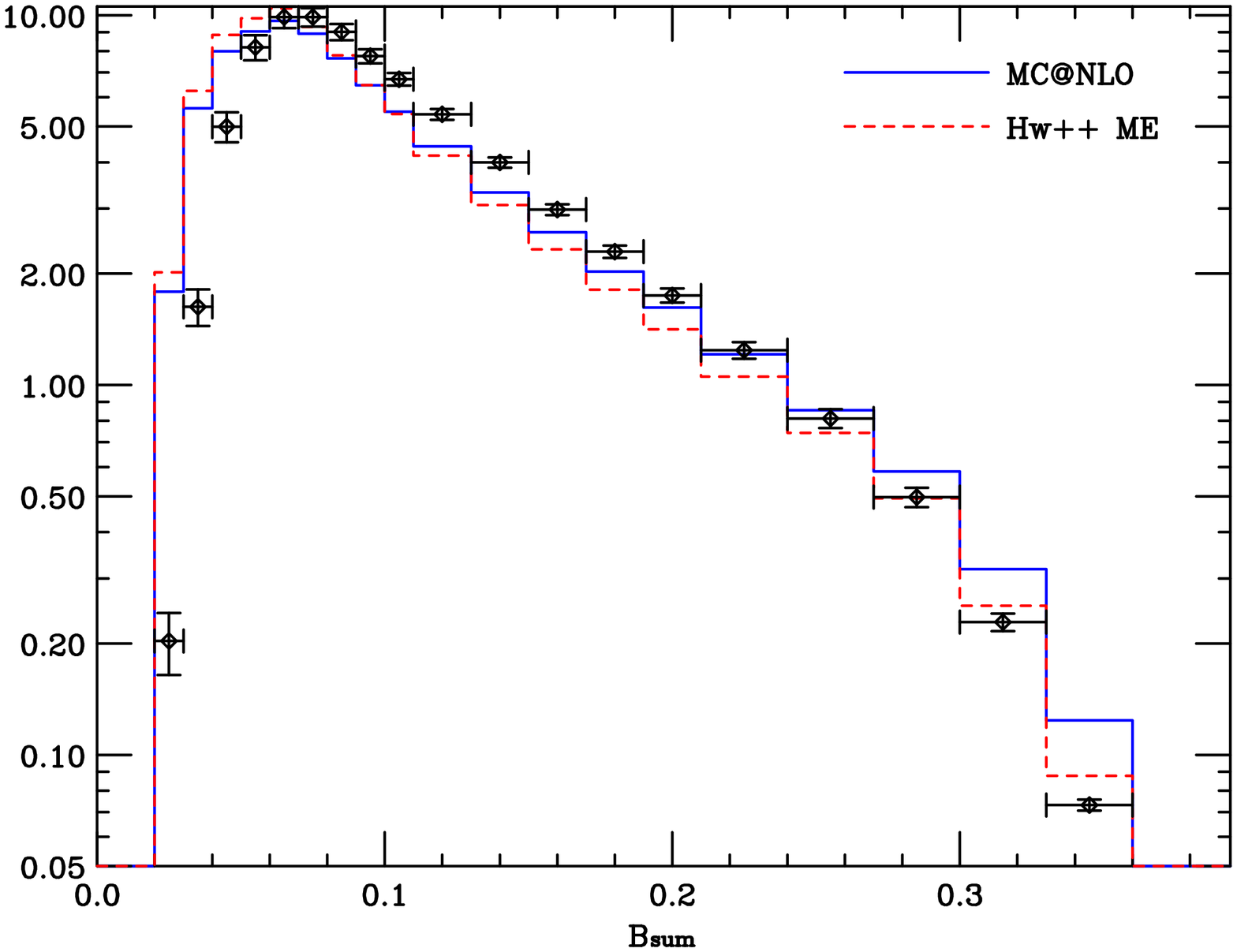,%
width=3.03in,height=2.93in,angle=90} 
\caption{The difference and sum of jet broadenings $B_{\rm diff}$ and $B_{\rm sum}$. Data
  from \cite{Abreu:1996na}.}
\label{fig:bsbd}    
\end{figure}
\newpage
\clearpage
\pagebreak
\pagebreak
\clearpage
\section{Drell-Yan lepton pair production}
\label{DY}
\subsection{Kinematics}
The method described above was then applied to Drell-Yan electron pair production at the
Tevatron.
At leading order, the relevant subprocess is illustrated in Figure \ref{fig:LOO} is
$q+\bar{q} \rightarrow V$.
The invariant mass $Q$ and the rapidity $Y$ 
of the boson V can be written in terms of the momentum
fractions of the incoming partons, $x_q$ and $x_{\bar{q}}$ as
\begin{eqnarray}
Q^2&=&x_qx_{\bar{q}}S \,, \nonumber \\
Y&=&\frac{1}{2}\ln\frac{x_q}{x_{\bar{q}}}
\end{eqnarray}
where $S$ is the proton-antiproton center-of-mass energy.
Inverting this, we have 
\begin{eqnarray}
\label{xq}
x_q&=&\sqrt{\frac{Q^2}{S}}e^{Y}\,, \nonumber \\
x_{\bar{q}}&=&\sqrt{\frac{Q^2}{S}}e^{-Y}\;.
\end{eqnarray}
Next we consider the real emission subprocesses illustrated in Figure \ref{fig:NLOO}. It is convenient to express the kinematics of the NLO diagrams in terms of
Mandelstam invariants. For gluon emission, we define
\begin{eqnarray}
s&=&(p_q+p_{\bar{q}})^2, \nonumber \\
t&=&(p_q-p_g)^2, \nonumber \\
u&=&(p_{\bar{q}}-p_g)^2 
\end{eqnarray}
with $Q^2=s+t+u$. For the QCD Compton process, these are given by
\begin{eqnarray}
s&=&(p_q+p_g)^2, \nonumber \\
t&=&(p_{q}^{'}-p_g)^2, \nonumber \\
u&=&(p_q-p_{q}^{'})^2 \;.
\end{eqnarray}
We can also express the kinematics in terms of the variables $x$ and
$y$ which are given by
\begin {eqnarray}
x&=&\frac{Q^2}{s}; 0 \leq x \leq 1 \,,\nonumber \\
y&=&\cos \theta; -1 \leq y \leq 1 \,,
\end{eqnarray}
where $\theta$ is the scattering angle of the emitted parton in the partonic
center-of-mass system. Using these
definitions, we can show that
\begin{eqnarray}
Q^2&=&xx_1x_2S \,, \nonumber \\
Y&=&\frac{1}{2}\ln\frac{x_1}{x_2}\frac{2-(1-x)(1+y)}{2-(1-x)(1-y)} 
\end{eqnarray}
where $x_1,x_2$ are now the momentum fractions of the incoming partons in the NLO
subprocess and are given in terms of $x_q,x_{\bar{q}}$ by
\begin{eqnarray}
x_1&=&\frac{x_q}{\sqrt{x}}\sqrt{\frac{2-(1-x)(1-y)}{2-(1-x)(1+y)}} \,, \nonumber \\
x_2&=&\frac{x_{\bar{q}}}{\sqrt{x}}\sqrt{\frac{2-(1-x)(1+y)}{2-(1-x)(1-y)}} \;.
\end{eqnarray}
At the parton level the Born cross-section for the production of a virtual photon is given by:
\begin{equation}
\sigma^{i}_0=\frac{4 \pi \alpha_{em}^2 e_{q_{i}}^2}{9Q^2}
\end{equation}
where $Q$ is the electron pair invariant mass.
Extending this to the hadronic level, the Born cross-section becomes
\begin{equation}
\label{sigB}
\frac{d\sigma(S,Q^2)^{H}_0}{dQ^2}=\int dx_{q}dx_{\bar{q}}\sum_i\sigma^{i}_{0}[f_{q_{i}/A}(x_{q_{i}})f_{q_{i}/B}(x_{\bar{q}_{i}})+q\leftrightarrow
\bar{q}]\delta(Q^2-x_{q_{i}}x_{\bar{q}_{i}}S)
\end{equation}
where $f_{q_{i}/A}(x_{q_{i}},Q^2)$ is the distribution function for parton $i$ in the
hadron $A$ evaluated at the Born scale, $Q$.

\begin{figure}[h!]
\[
\psfig{figure=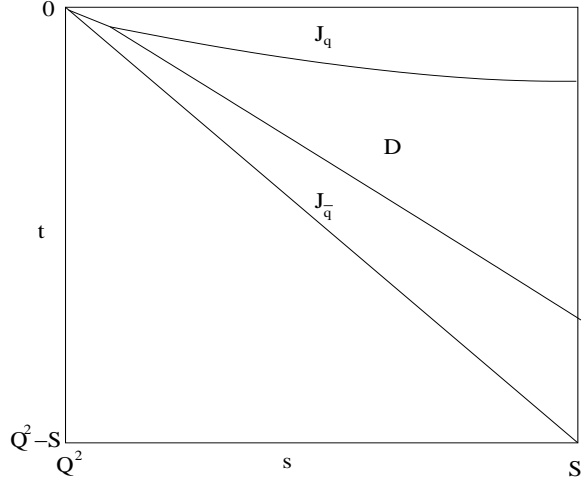,%
width=3in,height=2.5in,angle=0}
\]
\caption{Jet and dead regions in s-t phase space for $q +
\bar{q} \rightarrow V + g$. Not to scale.}
\label{stps}
\end{figure}

\begin{figure}[h!]
\[
\psfig{figure=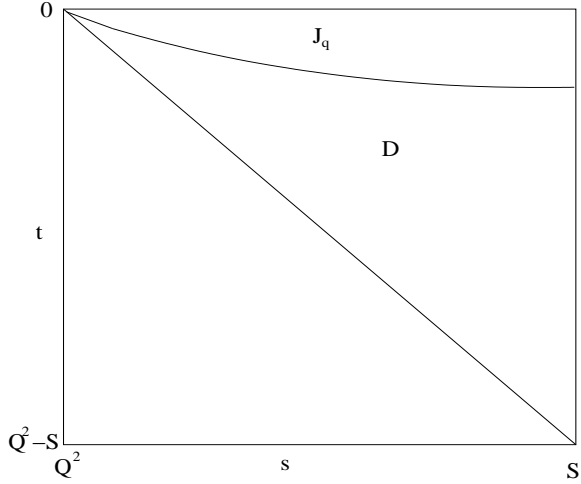,%
width=3in,height=2.5in,angle=0}
\]
\caption{Jet and dead regions in s-t phase space for $q +
g \rightarrow V + q$. Not to scale.}
\label{stg}
\end{figure}

\begin{figure}[h!]
\[
\psfig{figure=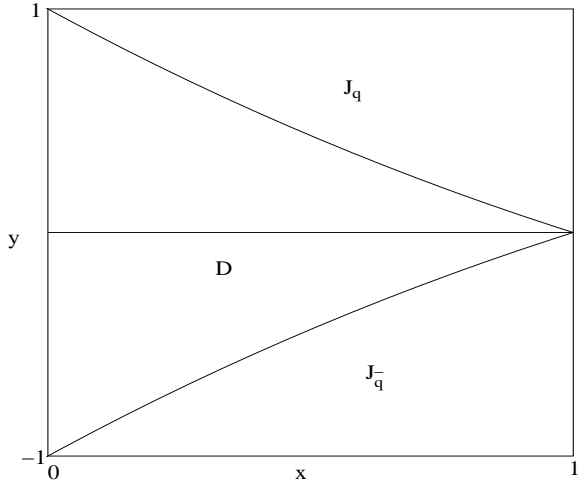,%
width=3in,height=2.5in,angle=0}
\]
\caption{Jet and dead regions in x-y phase space for $q +
\bar{q} \rightarrow V + g$. Not to scale.}
\label{xyps}
\end{figure}

The differential cross-section for real gluon emission is given by:
\begin{equation}
\label{eqn:me}
\frac{d^2 \sigma}{\sigma_{0}ds
  dt}=M_{q\bar{q}}=\frac{D_q(x_1)D_{\bar{q}}(x_2)}{D_q(x_q)D_{\bar{q}}(x_{\bar{q}})}\frac{\alpha_S}{2
  \pi}C_F\frac{Q^2}{s^3tu}[(s+t)^2+(s+u)^2] 
\end{equation}
where  $C_F=4/3$ and $D_q(x_1)=x_1f_q(x_1)$ etc. Note that in this and subsequent
equations, $\sigma_0$ is actually $\frac{d^2\sigma_0}{dQ^2dY}$. The PDF ratio takes account of the
change of kinematics from the Born momentum fractions $x_{q},x_{\bar{q}}$ to $x_1,x_2$.
The corresponding differential cross-section for the QCD Compton subprocess is given by
\begin{equation}
\frac{d^2 \sigma}{\sigma_{0}ds
  dt}=M_{qg}=\frac{D_q(x_1)D_{g}(x_2)}{D_q(x_q)D_{\bar{q}}(x_{\bar{q}})}\frac{\alpha_S}{2
  \pi}T_F\frac{Q^2}{s^4t}[s^2+t^2+2Q^2u]
\end{equation}
where $T_F=1/2$.
The shower variables, $z$ and $\tilde \kappa$ for the Drell-Yan processes are discussed in detail in \cite{Gieseke:2003rz}. The invariant
mass and rapidity of the boson are chosen to be preserved in the definition of the shower
variables.  Also discussed is the choice of the jet regions (where gluon emission is soft
and/or collinear with the parent parton) for the quark $q$, and
antiquark, $\bar{q}$. In terms of the shower variables for the quark jet, the Mandelstam
variables become
\begin{eqnarray}
s&=&\frac{Q^2[1+(1-z)\tilde{\kappa}]}{z}, \nonumber \\
t&=&-Q^2(1-z)\tilde{\kappa}, \nonumber \\
u&=&-(1-z)s
\end{eqnarray}
The jet region is then defined as the area of the $s-t$ phase space where 
\begin{equation}
\tilde {\kappa}=\frac{st}{uQ^2} < \tilde {\kappa}_{q} \;.
\label{kp}
\end{equation}
 For the antiquark jet, we have $t \leftrightarrow u$ and thus 
\begin{equation}
\tilde {\kappa}=\frac{su}{tQ^2}< \tilde {\kappa}_{\bar{q}} \;.
\end{equation}
In order to ensure that the jet regions touch without overlapping
we require $\tilde {\kappa}_{q}=1/\tilde {\kappa}_{\bar{q}}$. In the discussion that
follows, we choose the symmetrical choice $\tilde {\kappa}_{q}=\tilde
{\kappa}_{\bar{q}}=1$. The jet and dead regions corresponding to this choice are labeled
$J_q,J_{\bar{q}}$ and $D$ respectively in Figure
\ref{stps}.
Now the gluon emission probability off the quark in the parton shower approximation of {\tt Herwig++}
is
\begin{equation}
\label{eqn:ps}
\frac{d^2 P}{dz d \tilde{\kappa}}=\frac{\alpha_S}{2
  \pi}C_F\frac{1+z^2}{\tilde {\kappa}(1-z)}
\end{equation}
which gives a differential cross-section
\begin{equation}
\label{eqn:ps1}
\frac{d^2 \sigma}{\sigma_{0}ds
  dt}=M_{{C}_{q\bar{q}}}=\frac{D_q(x_1)D_{\bar{q}}(x_2)}{D_q(x_q)D_{\bar{q}}(x_{\bar{q}})}\frac{\alpha_S}{2
  \pi}C_F\frac{(s+u)[s^2+(s+u)^2]}{s^3tu} \;.
\end{equation}
Interchange $t \leftrightarrow u$ for the corresponding emission cross-section off the
antiquark. Note that the parton shower approximation in (\ref{eqn:ps1}) overestimates the
matrix element
expression (\ref{eqn:me}) and becomes exact in the collinear and soft limit $t \rightarrow 0$. For the
Compton subprocess $q+g \rightarrow V+q$, the parton shower approximation is given by
\begin{equation}
\frac{d^2 \sigma}{\sigma_{0}ds
  dt}=M_{{C}_{qg}}=\frac{D_q(x_1)D_g(x_2)}{D_q(x_q)D_{\bar{q}}(x_{\bar{q}})}\frac{\alpha_S}{2
  \pi}T_F\frac{(s+u)[u^2+(s+u)^2]}{s^4t} \;.
\end{equation}
Interchange $t \leftrightarrow u$ for the subprocess $g+q \rightarrow V+q$. 
In this case there is only one jet region which corresponds to the emitted quark being
collinear with the gluon. The same jet definition in (\ref{kp}) is used and the corresponding region is shown in
Figure \ref{stg}. 
Now, we can further re-write the above expressions for the exact differential
cross-sections and the parton shower approximation in terms of variables $x$ and $y$.
Using these definitions,
\begin {eqnarray}
\label{tu}
t&=&-\frac{Q^2(1-x)(1-y)}{2x}\,, \nonumber \\
u&=&-\frac{Q^2(1-x)(1+y)}{2x}\;.
\end{eqnarray} 
In these variables the differential cross-sections become
\begin{eqnarray}
M_{q\bar{q}}&=&\frac{D_q(x_1)D_{\bar{q}}(x_2)}{D_q(x_q)D_{\bar{q}}(x_{\bar{q}})}\frac{\alpha_S}{2
  \pi}C_F\frac{y^2(1-x)^2+(1+x)^2}{(1-x)(1-y^2)} \,, \nonumber \\
M_{{C}_{q\bar{q}}}&=&\frac{D_q(x_1)D_{\bar{q}}(x_2)}{D_q(x_q)D_{\bar{q}}(x_{\bar{q}})}\frac{\alpha_S}{2\pi}C_F\frac{(4+(1-y)^2+2x(1-y^2)+x^2(1+y)^2)(1-y+x(1+y))}{4(1-x)(1-y^2)}\,, \nonumber \\
M_{qg}&=&\frac{D_q(x_1)D_{g}(x_2)}{D_q(x_q)D_{\bar{q}}(x_{\bar{q}})}\frac{\alpha_S}{2
  \pi}T_F\frac{(3+y^2)(1-x)^2-2y(1-x^2)+2(1+x^2)}{4(1-y)}\,, \nonumber \\
M_{{C}_{qg}}&=&\frac{D_q(x_1)D_{g}(x_2)}{D_q(x_q)D_{\bar{q}}(x_{\bar{q}})}\frac{\alpha_S}{2
  \pi}T_F\frac{1+y^2-2xy(1+y)+x^2(1+y)^2}{x(1-y)}\;.
\end{eqnarray} 
Interchange $y \leftrightarrow -y$ in $M_{{C}_{q\bar{q}}}$ for the antiquark jet and in
$M_{qg}$ and $M_{C_{qg}}$ for the process $g+q \rightarrow V +q$. The corresponding jet
regions of $x-y$ phase space are shown in Figure \ref{xyps}. 
As we shall see in
Section \ref{MCNLO}, expressing the cross-sections in these variables makes it
easier to carry out the MC@NLO subtractions and divergence mappings.

The scale at which the parton distribution functions $D_{i}(x_1),D_{i}(x_2)$
are evaluated was set to 
\begin{equation}
M=\sqrt {\frac{ut}{s}}
\end{equation}
 which in terms of the {\tt Herwig++} variables is given by
$\sqrt {(1-z)^2
\tilde {\kappa}Q^2}$. This is equal to $\mid{\bf k_{\perp} }\mid$, the transverse momentum
of the emitted parton in the partonic center of mass frame. This is the same scale used in
the parton shower and is also the scale at which $\alpha_S$ was determined. The Drell-Yan
cross-section can be computed in 2 different factorization schemes: the DIS scheme and the
$\overline{\rm MS}$
scheme. For comparison, both cross-sections were used for the event generation in this
paper. The NLO parton distribution functions were obtained from the CTEQ5d (DIS) and CTEQ5m ($\overline{\rm MS}$)
PDF sets which are frozen at a
scale of $1$ GeV. For $M < 1$ GeV, $f(x,M)$ was set equal to $f(x,1)$. 

\subsection{Next-to-leading order cross-section}
\begin{figure}[h!]
\[
\psfig{figure=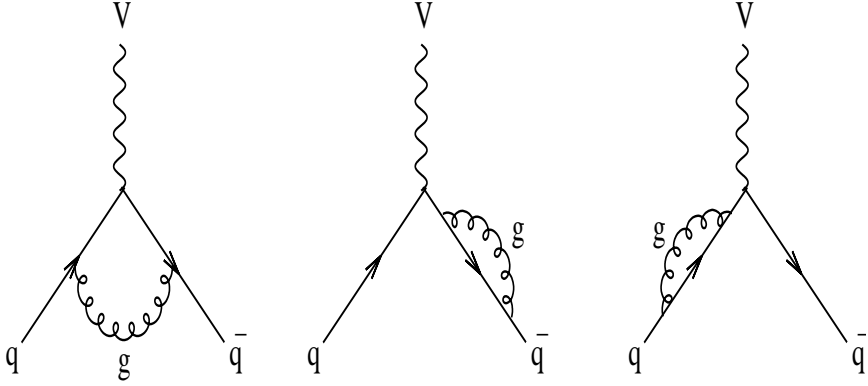,%
width=4.5in,height=2in,angle=0}
\]
\caption{Virtual gluon corrections to the quark-antiquark annihilation Born term
  $q+\bar{q} \rightarrow V$}
\label{virt}
\end{figure}

In the massless limit, we can integrate the differential cross-section (\ref{eqn:me}) for
the real emission process $q+\bar{q} \rightarrow V+g$ over $y$ using the
dimensional regularization scheme to regulate the divergences. The result is \cite{Altarelli:1978id}  
\begin{eqnarray}
\label{eqn:R}
\frac{\sigma^{NLO}_R}{\sigma_0}&=&\frac{\alpha_s}{2\pi}C_F\left(\frac{4\pi}{\mu^2}\right)^\epsilon\frac{\Gamma(1-\epsilon)}{\Gamma(1-2\epsilon)}\left[\frac{2}{\epsilon^2}\delta(1-x)-\frac{2}{\epsilon}\frac{1+x^2}{(1-x)_+}
  \nonumber \right.\\
&+& \left. 4(1+x^2)\left(\frac{\ln(1-x)}{1-x}\right)_+-2\frac{1+x^2}{1-x}
  \ln x
\right]
\end{eqnarray}
where $\epsilon$ is as defined in (\ref{eqe}) and the plus-prescription is defined by 
\begin{equation}
\int_0^1dx(a(x))_+b(x)=\int_0^1a(x)[b(x)-b(1)] \;.
\end{equation}
Now in addition to the real emission diagrams in Figure \ref{fig:NLOO}, we have virtual gluon
corrections arising from self-energy and vertex corrections. These are illustrated in
Figure \ref{virt}.
Integrating these diagrams using dimensional regularization we obtain
\begin{equation}
\label{eqn:V}
\frac{\sigma^{NLO}_V}{\sigma_0}=\frac{\alpha_S}{2\pi}C_F\left(\frac{4\pi}{\mu^2}\right)^\epsilon\delta(1-x)\frac{\Gamma(1-\epsilon)}{\Gamma(1-2\epsilon)}\left[-\frac{2}{\epsilon^2}-\frac{3}{\epsilon}-8+\frac{2}{3}\pi^2\right]\;.
\end{equation}
The product of an infrared and collinear divergence  is contained in the $1/ \epsilon^2$ term and is cancelled out between
the real and virtual diagrams to leave a pure collinear divergence, proportional to $1/
\epsilon$. This is cancelled out by QCD corrections to the quark distribution functions
due to real and virtual gluon emission. In the DIS factorization scheme, the NLO distribution function evaluated at a
scale $\mu_F$
is given in terms of the bare distribution function $f^{0}_{q}(x)$ as
\begin{eqnarray}
f_q(x,\mu_F^2)&=&f^{0}_{q}(x) \nonumber \\
&+&\frac{\alpha_S}{2\pi}C_F\frac{1}{1-\epsilon}\int_x^1
\frac{dw}{w}f^{0}_{q}(w)\left[\left(\frac{1+z^2}{(1-z)_+}+\frac{3}{2}\delta(1-z) \right)\left(-\frac{1}{\epsilon}\frac{\Gamma(1-\epsilon)}{\Gamma(1-2\epsilon)}+\ln
    \frac{\mu_F^2}{4\pi} \right) \right.\nonumber \\
&+&(1+z^2)\left(\frac{\ln (1-z)}{1-z} \right )_+-\frac{3}{2}\frac{1}{(1-z)_+}-\frac{1+z^2}{1-z}
  \ln z \nonumber \\
&+&\left. 3+2z-\left(\frac{9}{2}+\frac{\pi^2}{3}\right)\delta(1-z) \right]
\end{eqnarray}
where $z=x/w$.
Combining these corrections gives the full NLO cross-section ratio \cite{Altarelli:1978id}
\begin{eqnarray}
\frac{\sigma^{NLO}}{\sigma_0}&=&\sum_{q}\int dx_1dx_2\frac{x[D_q(x_1,\mu_F^2)D_{\bar{q}}(x_2,\mu_F^2)+q\leftrightarrow
\bar{q}]}{D_q(x_q)D_{\bar{q}}(x_{\bar{q}})}\left[\delta(1-x)+\frac{\alpha_S}{2\pi}C_F\left\{\frac{3}{(1-x)_+} \right.\right.
  \nonumber \\
&-&6-4x+2(1+x^2)\left(\frac{\ln(1-x)}{1-x}\right)_++\left(1+\frac{4}{3}\pi^2\right)\delta(1-x)\nonumber \\
&+&\left.\left.\left(\frac{1+x^2}{(1-x)_+}+\frac{3}{2}\delta(1-x)\right)2\ln\frac{\mu^2}{\mu_F^2}\right\}\right]
\label{cs}
\end{eqnarray}
where $\sum_q$ signifies a sum over parton flavours $q$.
The residual collinear divergence has been cancelled as expected. The last term in the
$O(\alpha_S)$ term proportional to $\ln \frac{\mu^2}{\mu_F^2}$ can be eliminated via the
DGLAP equation \cite{Lipatov:1974qm,Gribov:1972rt,Altarelli:1977zs,Dokshitzer:1977sg} which describes how the distribution functions evolve with the
scale $\mu$.
\begin{equation}
\frac{d}{d\ln\mu^2}f_q(x,\mu^2)=\frac{\alpha_S}{4\pi}\int_x^1\frac{dw}{w}f^{0}_{q}(w)P_{qq}\left(\frac{x}{w}\right)
\end{equation}
where the splitting function
\begin{equation}
P_{qq}(z)=C_F\left[\frac{1+z^2}{(1-z)_+}+\frac{3}{2}\delta(1-z)\right]
\end{equation}
describes the probability of a quark coming from the splitting, $q \rightarrow qg$.
We can use the above expressions to replace $D_q(x,\mu_F^2)$ with $D_q(x,\mu^2)$ in
(\ref{cs}). The logarithmic term is then cancelled to give
\begin{eqnarray}
\label{css}
\frac{\sigma^{NLO}}{\sigma_0}&=&\sum_{q}\int
dx_1dx_2 \frac{x[D_q(x_1,\mu^2)D_{\bar{q}}(x_2,\mu^2)+q\leftrightarrow
\bar{q}]}{D_q(x_q)D_{\bar{q}}(x_{\bar{q}})}\left[\delta(1-x)+\frac{\alpha_S}{2\pi}C_F\left\{\frac{3}{(1-x)_+}
  \right. \right.\nonumber \\
&-&\left.\left.6-4x+2(1+x^2)\left(\frac{\ln(1-x)}{1-x}\right)_++\left(1+\frac{4}{3}\pi^2\right)\delta(1-x)\right\}\right] \;.
\end{eqnarray}
\section{MC@NLO method}
\label{MCNLO}
Now by writing the virtual and PDF corrections in terms of the hard matrix element (\ref{eqn:me}), we
can rewrite (\ref{css}) in integral from as
\begin{eqnarray}
\frac{\sigma^{NLO}}{\sigma_0}&=&\sum_{q}\int
dxdy\left[\left\{\frac{x[D_q(x_1,\mu^2)D_{\bar{q}}(x_2,\mu^2)+q\leftrightarrow
\bar{q}]}{D_q(x_q)D_{\bar{q}}(x_{\bar{q}})}\frac{1}{2}\left (\delta(1-x)+\frac{\alpha_S}{2\pi}C_F\left(\frac{3}{(1-x)_+}
  \right. \right. \right.\right.\nonumber \\
&-&\left.\left.\left.6-4x+2(1+x^2)\left(\frac{\ln(1-x)}{1-x}\right)_++\left(1+\frac{4}{3}\pi^2\right)\delta(1-x)\right)\right)-M_{q\bar{q}}\right\}\nonumber \\
&+&\left.M_{q\bar{q}}\right] \;.
\end{eqnarray}
The first term in the curly brackets is the sum of the Born term, virtual and QCD PDF
corrections expressed as an integral over the variables $x$ and $y$. Since the area of the
$x-y$ phase space is $2$, there is a factor of $1/2$ in the integrand. The remaining term is
the real emission contribution to the cross-section. 
Now we can define a functional $F$ as in (\ref{eqF}) which represents final states
generated from the 2 different starting configurations; $q+\bar{q} \rightarrow V$ and
$q+\bar{q} \rightarrow V+g$  as
\begin {eqnarray}
\frac{\sigma^{NLO}}{\sigma_0}&=&\sum_{q}\int
dxdy\left[F_V\left\{\frac{x[D_q(x_1,\mu^2)D_{\bar{q}}(x_2,\mu^2)+q\leftrightarrow
\bar{q}]}{D_q(x_q)D_{\bar{q}}(x_{\bar{q}})}\frac{1}{2}\left (\delta(1-x)+\frac{\alpha_S}{2\pi}C_F\left(\frac{3}{(1-x)_+}
  \right. \right. \right.\right.\nonumber \\
&-&\left.\left.\left.6-4x+2(1+x^2)\left(\frac{\ln(1-x)}{1-x}\right)_++\left(1+\frac{4}{3}\pi^2\right)\delta(1-x)\right)\right)-M_{q\bar{q}}\right\}\nonumber \\
&+&\left.F_{Vg}M_{q\bar{q}}\right] \;.
\end{eqnarray}
where $F_V$ and $F_{Vg}$ are functionals which represent final states generated from  $q+\bar{q} \rightarrow V$ and
$q+\bar{q} \rightarrow V+g$ starting configurations respectively. As discussed in Section
\ref{section3}, this is not entirely correct because of double counting in the final states
represented by $F_{V}$ arising from the parton shower. To resolve this issue,  we
subtract the contribution from the parton shower contributions, $M_{{C}_{q\bar{q}}}$ from the
integrals in
jet regions $J_q$ and $J_{\bar{q}}$ in Figure \ref{xyps} and integrate the full matrix element
$M_{q\bar{q}}$ over the hard emission region, $D$. The modified generating functional then becomes
\begin {eqnarray}
F^{q\bar{q}}&=&\sum_{q}\int_J\left[F_V\left\{\frac{x[D_q(x_1,\mu^2)D_{\bar{q}}(x_2,\mu^2)+q\leftrightarrow
\bar{q}]}{D_q(x_q)D_{\bar{q}}(x_{\bar{q}})}\frac{1}{2}\left (\delta(1-x)+\frac{\alpha_S}{2\pi}C_F\left(\frac{3}{(1-x)_+}
  \right. \right. \right.\right.\nonumber \\
&-&\left.\left.\left.6-4x+2(1+x^2)\left(\frac{\ln(1-x)}{1-x}\right)_++\left(1+\frac{4}{3}\pi^2\right)\delta(1-x)\right)\right)-M_{q\bar{q}}+M_{{C}_{q\bar{q}}}\right\}\nonumber \\
&+&\left.F_{Vg}\left \{M_{q\bar{q}}-M_{{C}_{q\bar{q}}}\right\}\right] \nonumber \\
&+&\sum_{q}\int_D \left[F_V\left\{\frac{x[D_q(x_1,\mu^2)D_{\bar{q}}(x_2,\mu^2)+q\leftrightarrow
\bar{q}]}{D_q(x_q)D_{\bar{q}}(x_{\bar{q}})}\frac{1}{2}\left (\delta(1-x)+\frac{\alpha_S}{2\pi}C_F\left(\frac{3}{(1-x)_+}
  \right. \right. \right.\right.\nonumber \\
&-&\left.\left.\left.6-4x+2(1+x^2)\left(\frac{\ln(1-x)}{1-x}\right)_++\left(1+\frac{4}{3}\pi^2\right)\delta(1-x)\right)\right)-M_{q\bar{q}}\right\}\nonumber \\
&+&\left.F_{Vg}M_{q\bar{q}}\right] \;.
\end{eqnarray}
where $J=J_q \cup J_{\bar{q}}$. 
A similar procedure can be adopted for the Compton subprocess which as discussed has one
jet region. In this case the QCD PDF corrections cancel out the collinear divergence in
the matrix element. The final result is
\begin {eqnarray}
F^{qg}&=&\sum_{q}\int_J\left[F_V\left\{\frac{x[D_q(x_1,\mu^2)D_g(x_2,\mu^2)+ q\leftrightarrow
g]}{D_q(x_q)D_{\bar{q}}(x_{\bar{q}})}\frac{\alpha_S}{2\pi}T_F\frac{1}{2}\left (\frac{3}{2}-5x+\frac{9}{2}x^2 \right. \right. \right.
 \nonumber \\
&+& \left.\left.\left.(x^2+(1+x^2))\ln(1-x)\right)-
M_{qg}+M_{C_{qg}}\right\}+F_{Vq}\{M_{qg}-M_{C_{qg}}\}\right]
\nonumber \\
&+& \sum_{q}\int_D\left[F_V\left\{\frac{x[D_q(x_1,\mu^2)D_g(x_2,\mu^2)+ q\leftrightarrow
g]}{D_q(x_q)D_{\bar{q}}(x_{\bar{q}})}\frac{\alpha_S}{2\pi}T_F\frac{1}{2}\left (\frac{3}{2}-5x+\frac{9}{2}x^2 \right. \right. \right.
 \nonumber \\
&+& \left.\left.\left.(x^2+(1+x^2))\ln(1-x)\right)-
M_{qg}\right\}+F_{Vq}M_{qg}\right] 
\end{eqnarray}
where $T_F=1/2$ and $F_{Vq}$ is the functional which represents final states generated from a
$q+g \rightarrow V + q$ starting configuration. Details of the algorithm used for event
generation can be found in Appendix \ref{HMM}.

\section{Intrinsic \boldmath{$p_T$}}
\label{ipt}
In QCD, the transverse momentum of partons arises in two ways. The first which has been
discussed above is due to the real emission of gluons and involves large momentum
transfers. This is often termed the perturbative component of the transverse momentum and
at large $p_T$ behaves as $1/{p_T}^2$. At low $p_T$ values, with the resummation of the double logarithms
from soft gluon emission (as is done in parton shower generators like {\tt Herwig++}),
this component of the $p_T$ vanishes as $p_T \rightarrow 0$.


The
second way in which partons acquire transverse momentum is non-perturbative. It involves small momentum transfers and cannot
be calculated by perturbation theory. Therefore this has to be modelled to fit the
observed data at low $p_T$ values. A small part of
this contribution can be attributed to quarks being
confined in the transverse direction to within the radius of the proton and therefore
gaining some intrinsic transverse momentum due to the uncertainty principle. 

Data on lepton pair transverse momentum from the CFS collaboration \cite{Ito:1980ev} suggest that a Gaussian distribution (\ref{eqn:intrinsic}) best describes
the intrinsic $p_T$ distribution.
\begin{equation}
\label{eqn:intrinsic}
h(p_{T})=\frac{2}{{p_{{T}_{\rm rms}}}^2}e^{\left(-\frac{p_{T}}{p_{{T}_{\rm rms}}}\right)^2}
\end{equation}
where $p_{{T}_{\rm rms}}$ is the root-mean-square $p_T$ of the Gaussian distribution. 


The intrinsic $p_T$ (\ref{eqn:intrinsic}) was implemented in {\tt Herwig++}. The intrinsic
$p_T$ component is generated according to the distribution using
a random number generator and added to the parent partons obtained at the end of the
space-like shower originating from the hard QCD process. 

\begin{figure}[!ht]
\vspace{2cm}
\hspace{3.5cm}
\psfig{figure=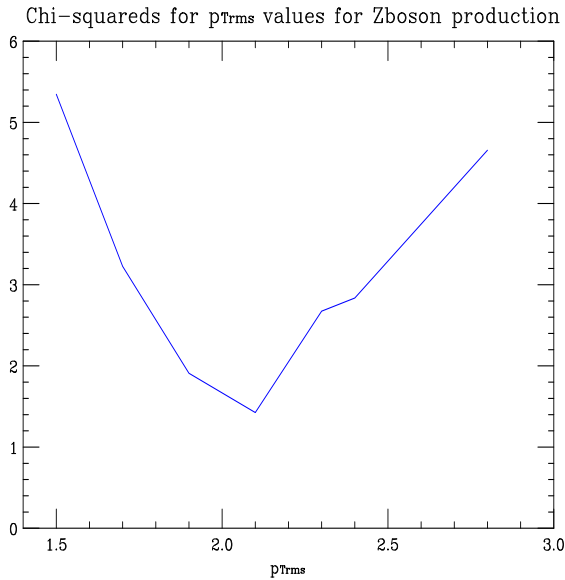,%
width=3.03in,height=2.93in,angle=90}
\caption{$\chi^2$ per degree of freedom.}
\label{fig:chi}    
\end{figure}

Figure \ref{fig:chi} shows the distribution of the $\chi^2$ per degree of freedom obtained for different
$p_{T_{\rm rms}}$ fits to run I
($1800$ GeV)
CDF data  \cite{Affolder:1999jh} for
Drell-Yan $Z$ boson production and the best fit value can be seen to be $p_{{T}_{\rm rms}}=2.1$ GeV. 
\section{Results on Drell-Yan production}
\begin{figure}[h!]
\[
\psfig{figure=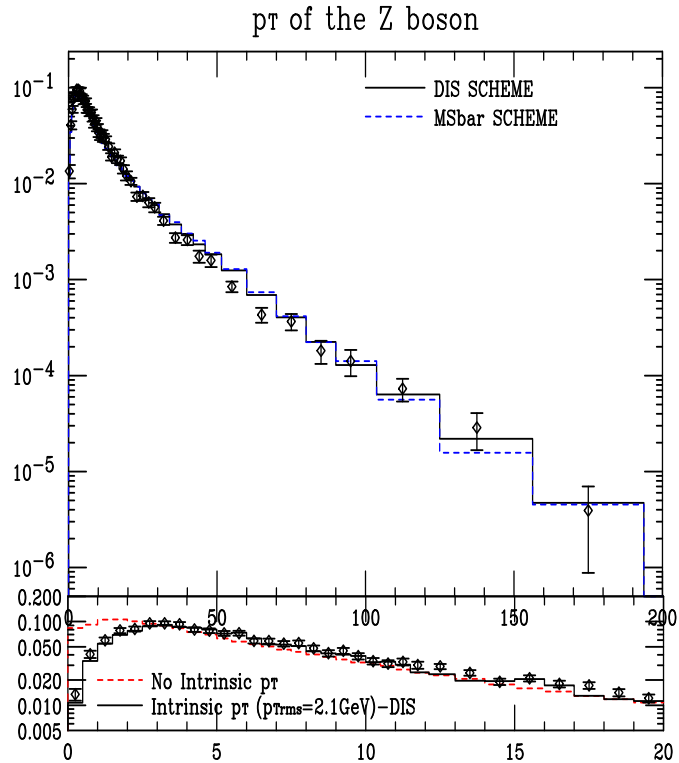,%
width=4in,height=3.5in,angle=-270}
\]
\caption{Transverse momentum of the Z boson. Data from \cite{Affolder:1999jh}.}
\label{ptz}
\end{figure}

Details of event generation and partonic final state properties are described in
Appendices \ref{Born}, \ref{HMM} and \ref{divmap}. Once generated the events were showered
using {\tt Herwig++} version 2.0.1 and the
distribution of the transverse momentum of the $Z$ boson was obtained. The hadronization scale was set
to the default value of $0.631$ GeV and the 2-loop $\alpha_S$ value was used. Figure \ref{ptz} shows a comparison of the
distributions obtained for both factorization schemes with CDF Run I data \cite{Affolder:1999jh}.

\begin{figure}[!ht]
\vspace{2cm}
\hspace{0.5cm}
\psfig{figure=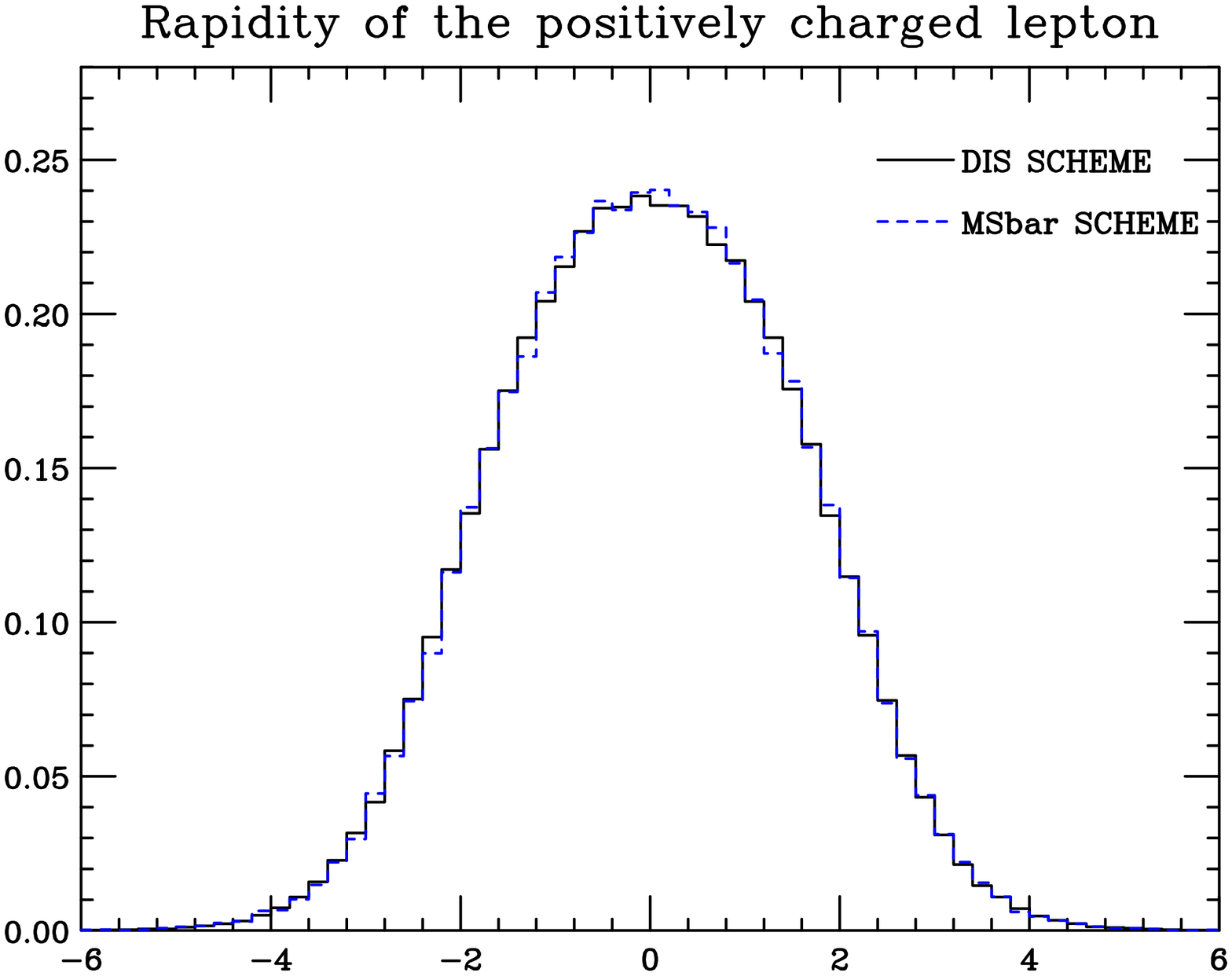,%
width=3.03in,height=2.93in,angle=90}
\psfig{figure=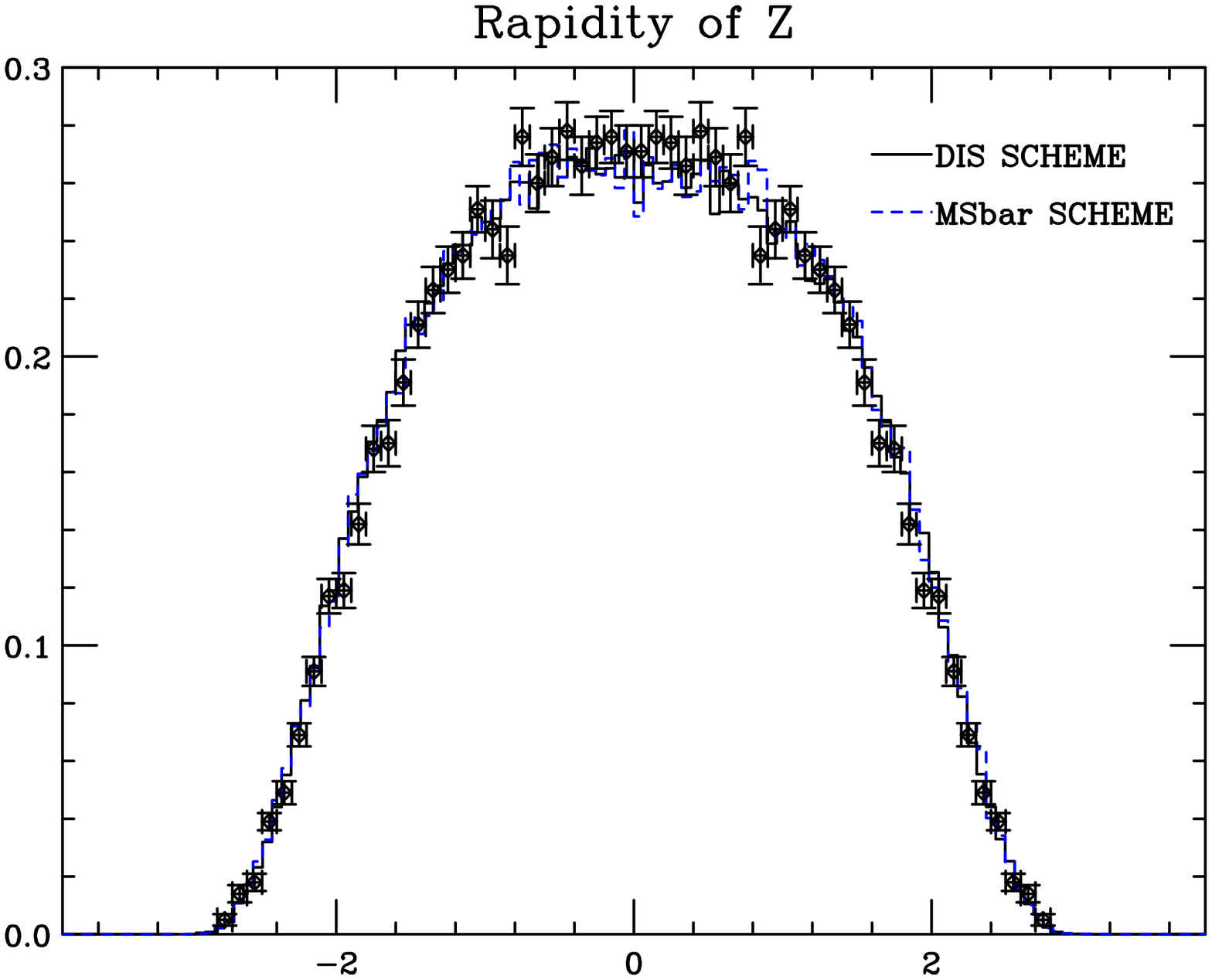,%
width=3.03in,height=2.93in,angle=90}
\caption{Rapidity distributions of the positively charged lepton and the Z boson.}
\label{fig:AR}    
\end{figure}

Also shown in Figure \ref{fig:AR} are the rapidity
distributions of the Z boson and the positively charged lepton arising from its decay. for comparison
distributions are compared against Run II ($\sqrt{s}=1.96$ TeV) data from the D0 collaboration \cite{Abazov:2007jy}. 

As can be seen in Figure \ref{ptz} the MC@NLO method provides a good description of the
CDF data for the transverse momentum of the Z boson. It also proves to be stable with
respect to change of scheme. Figure \ref{ptz} also shows the
effect of the $p_T$ distribution in the low $p_T$ region. The red dashed line corresponds
to setting
$p_{T_{\rm rms}}=0$ GeV whilst the black line corresponds to setting  $p_{T_{\rm rms}}=2.1$ GeV. Comparing the
two, one can see the
effect of adding the non-perturbative intrinsic $p_T$ to the parton shower which gives a
better description of the data. 

In addition, Figure \ref{fig:AR} shows that the predicted rapidity distributions of the Z boson and the positive
lepton produced are stable with respect to the change of scheme.

\section{Summary and conclusions}
We have successfully applied the MC@NLO method to $e^+e^-$ annihilation and Drell-Yan
processes modelled by the {\tt Herwig++} event generator. In general, we conclude that the MC@NLO method provides an improved description of event
shape distributions when compared to pure leading order Monte Carlo results. As we have
seen, the MC@NLO results for $e^+e^-$ annihilation do not differ greatly from the matrix element correction results
although we are now confident that the results are normalised to the full NLO
cross-section. Better results were obtained using the positive weight Nason@NLO method where the hardest
gluon emission was generated first \cite{LatundeDada:2006gx}. The MC@NLO method applied to the Drell-Yan
process gives a good description of the transverse momentum and stable predictions for Z
boson and lepton. The differences between the performance of the method for $e^+e^-$ annihilation
and Drell-Yan production may be attributed to the different shower variables for initial
state and final state radiation.

In addition, we have also successfully implemented the intrinsic $p_T$ component for hadron-hadron
collisions into {\tt Herwig++}.

\section{Acknowledgements}
I am grateful to the other members of the {\tt Herwig++} collaboration for
developing the program that underlies the present work and for helpful comments. I am particularly grateful to
Bryan Webber for constructive comments and discussions throughout.
\appendix
\section{Monte Carlo algorithm for \boldmath{$e^+e^-$} annihilation}
\label{HM}

The integrals in (\ref{eq12}) can be evaluated using a variety of Monte Carlo methods. In
this report, the `Hit or Miss' Monte Carlo method is used. This is the simplest and
oldest form of Monte Carlo integration and essentially involves finding the area of a
region in phase space by integrating over a larger region, a binary function which is 1 in
the region and 0 elsewhere.  The sampling method used for the points $x_q,x_{\bar{q}}$ is
the importance sampling method whereby more samples are taken from regions where the
integrand is large and less from regions where it is small. This ensures that the sampled
points have the same distribution as the integrand. 

First, let us investigate what each of the terms in (\ref{eq12}) signify. The program {\tt
  Herwig++} generates n-jets from an inputed set of n momentum space points $(x_{i}, \ldots
,x_{n})$. So for 2 and 3-jet formation, we require sets of 2 and 3 momentum space points
with each set corresponding to an event in the phase space. The relative numbers of these
events as well as the $x_{i}$ values can be obtained from (\ref{eq12}) as outlined. (2-jet events have $x_{q},x_{\bar{q}}=1$). In the discussion that follows,
$\sqrt{s}$ was set to $M_Z=91.2$ GeV and $\alpha_S(M_z)=0.118$.

\begin{enumerate}
\item
Randomly sample points $x_q,x_{\bar{q}}$, in each of regions
$J_{q},J_{\bar{q}}$ and $D$ of the phase space
and using the ``Hit Or Miss'' Monte Carlo method, evaluate the 4 integrals,
$I_{J}^{(2)},I_{J}^{(3)},I_{D}^{(2)}$ and $I_{D}^{(3)}$ as well as their absolute sum, $I$.
\begin{eqnarray}
\label{eq13}
I_{J}^{(2)}&=&\int_{J}dx_q
dx_{\bar{q}}\left[2-\frac{\alpha_S}{2\pi}C_F\left\{M-M_{C}-3\right\}\right] \,,\nonumber\\
I_{J}^{(3)}&=&\int_{J}dx_q dx_{\bar{q}}\frac{\alpha_S}{2\pi}C_F[M-M_{C}]\,, \nonumber\\
I_{D}^{(2)}&=&\int_{D}dx_q dx_{\bar{q}}\left[2-\frac{\alpha_S}{2\pi}C_F\left\{M-3\right\}\right]\,,\nonumber\\
I_{D}^{(3)}&=&\int_{D}dx_q dx_{\bar{q}}\frac{\alpha_S}{2\pi}C_FM \,,\nonumber\\
I&=&\mid{I_{J}^{(2)}}\mid+\mid{I_{J}^{(3)}}\mid+\mid{I_{D}^{(2)}}\mid+\mid{I_{D}^{(3)}}\mid
\;.
\end{eqnarray}
Note also the maximum values of the integrands in $I_{J}^{(3)}$ and $I_{D}^{(3)}$.
\item
The eventual proportion of 2-jet Monte Carlo events will be determined by the ratio $\frac
{\mid{I_{J}^{(2)}}\mid+\mid{I_{D}^{(2)}}\mid}{I}$. Likewise, the proportions of 3-jet events in the soft regions
$J_{q},J_{\bar{q}}$ and the hard region $D$ are determined by the ratios $\frac{\mid{I_{J}^{(3)}}\mid}{I}$ and $\frac{\mid{I_{D}^{(3)}}\mid}{I}$ respectively. The algorithm
below is then used to importance-sample the 3-jet events so that the corresponding
($x_{q},x_{\bar{q}}$) values of the Monte Carlo events have the same distribution as the integrands in $I_{J}^{(3)}$ and
$I_{D}^{(3)}$:  
\begin{enumerate}
\item
For event generation in region $R$ ($R=D,J_q$ or $J_{\bar{q}}$), randomly select a point
$x_q,x_{\bar{q}}$ in that region.\\
\item
Evaluate the absolute value of the  integrand in $I_{R}^{(3)}$ for this point, $\mid{w(x_q,x_{\bar{q}})}\mid$.\\
 Is $\mid {w(x_q,x_{\bar{q}})}\mid$ $> R$ $\mid{w_{\rm max}}\mid$ ? ($R$ is a random number between 0
 and 1 and $\mid {w_{\rm max}}\mid$ is the maximum value of
 $\mid{w(x_q,x_{\bar{q}})}\mid$ determined in Step 1).\\
\item 
If NO, return to (a). If YES, accept the event and set
$w^{\rm unw}$= sgn $w(x_q,x_{\bar{q}})$ i.e. $w^{\rm unw} = 1$ if $w(x_q,x_{\bar{q}})$ is positive and $-1$ if
negative. (In regions $J_{q}$ and $J_{\bar{q}}$, $M < M_{C}$, hence the integrands and the
integral, $I_{J}^{(3)}$ in these regions are negative). This process is called {\tt unweighting}.\\
\item
Repeat the process until the correct proportion of 2-jet and 3-jet events have been generated.\\ 
\item
Using the importance-sampled points, obtain an estimate for the integral,
$I_R^{(2,3)}=\frac{\sum{w^{\rm unw}}}{N}\times I$, where $N$ is the total number of Monte Carlo
events generated. We typically use $N=10^6$.\\
\end{enumerate}
\end{enumerate}
This method of generating Monte Carlo events is termed `Monte Carlo at Next to Leading
Order' or MC@NLO. In this way, for a  total of $N$ events, the correct proportion of 2-jet and 3-jet
events with $\pm$ unit weight is generated  with the same distribution as the integrands in
(\ref{eq13}). All of these integrals are finite, but the integrands are divergent at
isolated points within the integration regions. Before the sampling could
be carried out, the divergences  in the integrands (which cause problems in the sampling
process) had to be taken care of. This is the described in section \ref{section4}. 

\section{Divergences and mappings for \boldmath{$e^+e^-$} annihilation}
\label{section4}
\subsection{Divergences in dead region \boldmath{$D$}}
In region $D$, the hard matrix element:
\begin{equation}
M(x_q,x_{\bar{q}})=\frac{x_q^2+x_{\bar{q}}^2}{(1-x_q)(1-x_{\bar{q}})}
\label{eqM}
\end{equation}
diverges as $(x_q,x_{\bar{q}}) \rightarrow (1,0),(0,1)$ and $(1,1)$. To avoid these
divergences, one can map the divergent regions into another region in such a way that the
divergence is regularized. This is ensured by the fact that the region of integration
vanishes as the singularity is approached. The mappings used are presented. 
\subsubsection{Region \boldmath{$D: (1,1)$}}
\begin{figure}[h!]
\begin{center}
\[
\psfig{figure=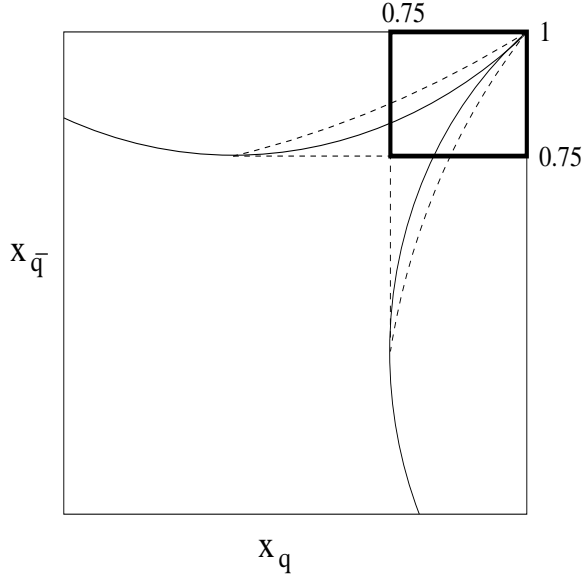,%
width=3in,height=3in,angle=0}
\]
\caption{The mapped region (solid) and the region mapped onto (dashed)}
\label{mapC1}
\end{center}
\end{figure}

There is a double pole in $M$ at $(x_q,x_{\bar{q}})=(1,1)$. To avoid this pole, the region
$x_q,x_{\bar{q}}>\frac{3}{4}$ is mapped into a region which includes $D$ but whose width vanishes quadratically as $x_q,x_{\bar{q}}\rightarrow1$ \cite{Gieseke:2003rz}. The mapping used is: 
\begin{eqnarray}
x_q^{'}&=&1-\left[\frac{1}{4}-(1-x_q)\right]=\frac{7}{4}-x_q \,,\nonumber \\
x_{\bar{q}}^{'}&=&1-2(1-x_q^{'})\left[\frac{3}{4}-(1-x_{\bar{q}})\right]=\frac{5}{8}+\frac{1}{2}x_q+\frac{3}{2}x_{\bar{q}}-2x_qx_{\bar{q}}
\label{eqC1}
\end{eqnarray}
when $x_q>x_{\bar{q}}>\frac{3}{4}$. This mapping also introduces an extra weight factor of
$2(1-x_q^{'})$ in the integrand. (Interchange $x_q$ and $x_{\bar{q}}$ in both the mapping
and weight factor when $x_{\bar{q}}>x_q>\frac{3}{4}$). Figure \ref{mapC1} shows the region
mapped (solid) and the region mapped onto (dashed).  
\subsubsection{Region \boldmath{$D: (1,0), (0,1)$}}
\label{example}
\begin{figure}[h!]
\begin{center}
\[
\psfig{figure=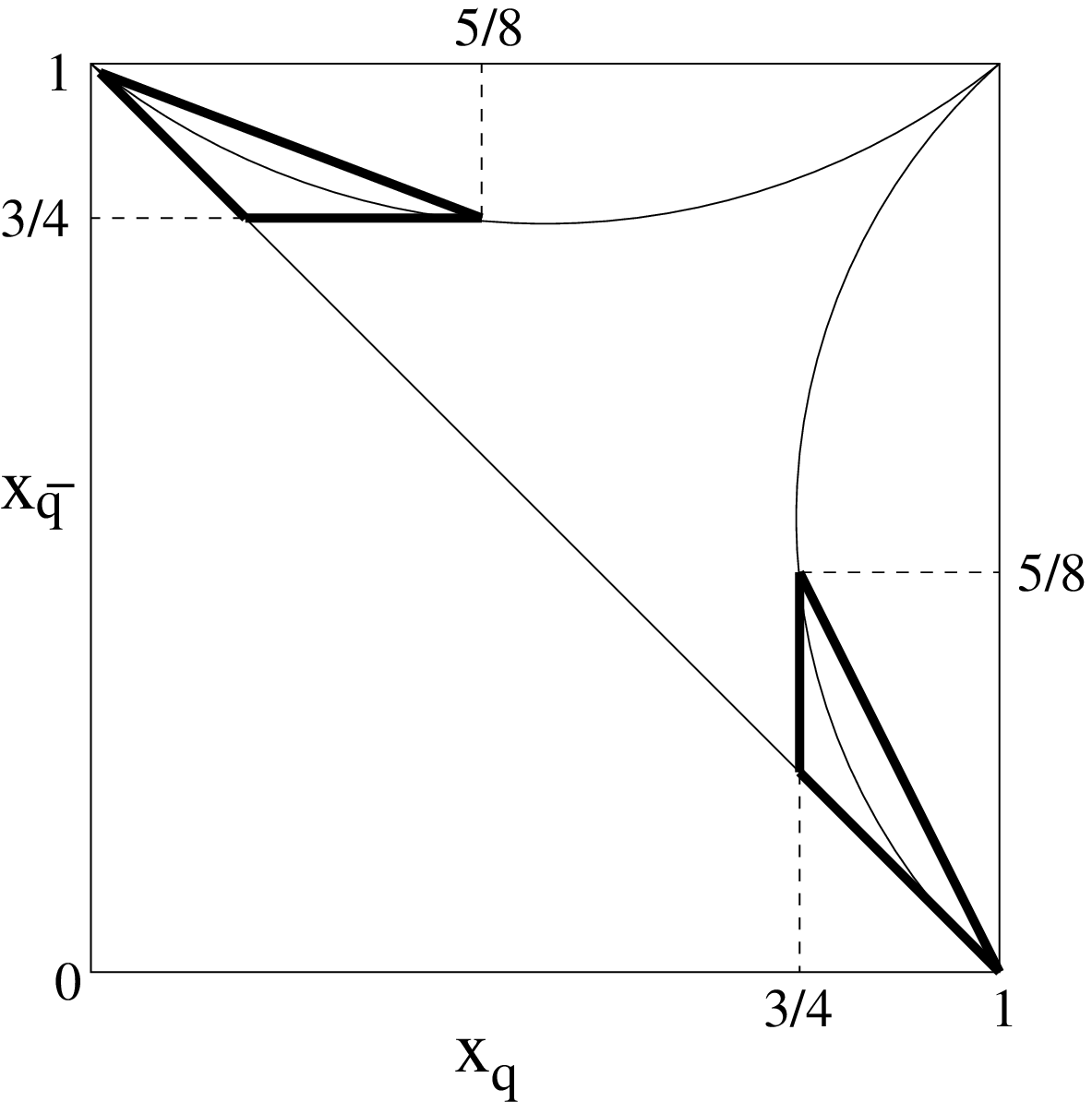,%
width=3in,height=3in,angle=0}
\]
\caption{Mapped regions}
\label{mapC2}
\end{center}
\end{figure}

A simple pole is approached in $M$ as $(x_q,x_{\bar{q}})\rightarrow(1,0)$ and $(0,1)$. In
the region $x_q<\frac{5}{8}$, $x_{\bar{q}}>\frac{3}{4}$, a new set of random points is
generated for points which fall between the lines $x_q=2.5 (1-x_{\bar{q}})$ and
$x_q=1-x_{\bar{q}}$. These points have an extra weight factor to cancel the divergence as
the pole is approached. The mapping used is: 
\begin{eqnarray}
x_{\bar{q}}^{'}&=&1-0.25r_2 \,,\nonumber \\
x_q^{'}&=&(1+1.5r_1)(1-x_{\bar{q}}^{'})
\label{eqC2}
\end{eqnarray}
where $r_1$ and $r_2$ are random numbers in the range $[0,1]$.
This mapping introduces a weight factor of $2r_2$ in the integrand. Interchange $x_q$ and
$x_{\bar{q}}$ in  the mapping for the region where
$x_{\bar{q}}<\frac{5}{8},x_q>\frac{3}{4}$. The mapped regions are shown with solid
boundaries in  Figure \ref{mapC2}.  
\subsection{Divergences in jet regions \boldmath{$J_{q}$} and \boldmath{$J_{\bar{q}}$}}
\label{AB}
\begin{figure}[h!]
\begin{center}
\[
\psfig{figure=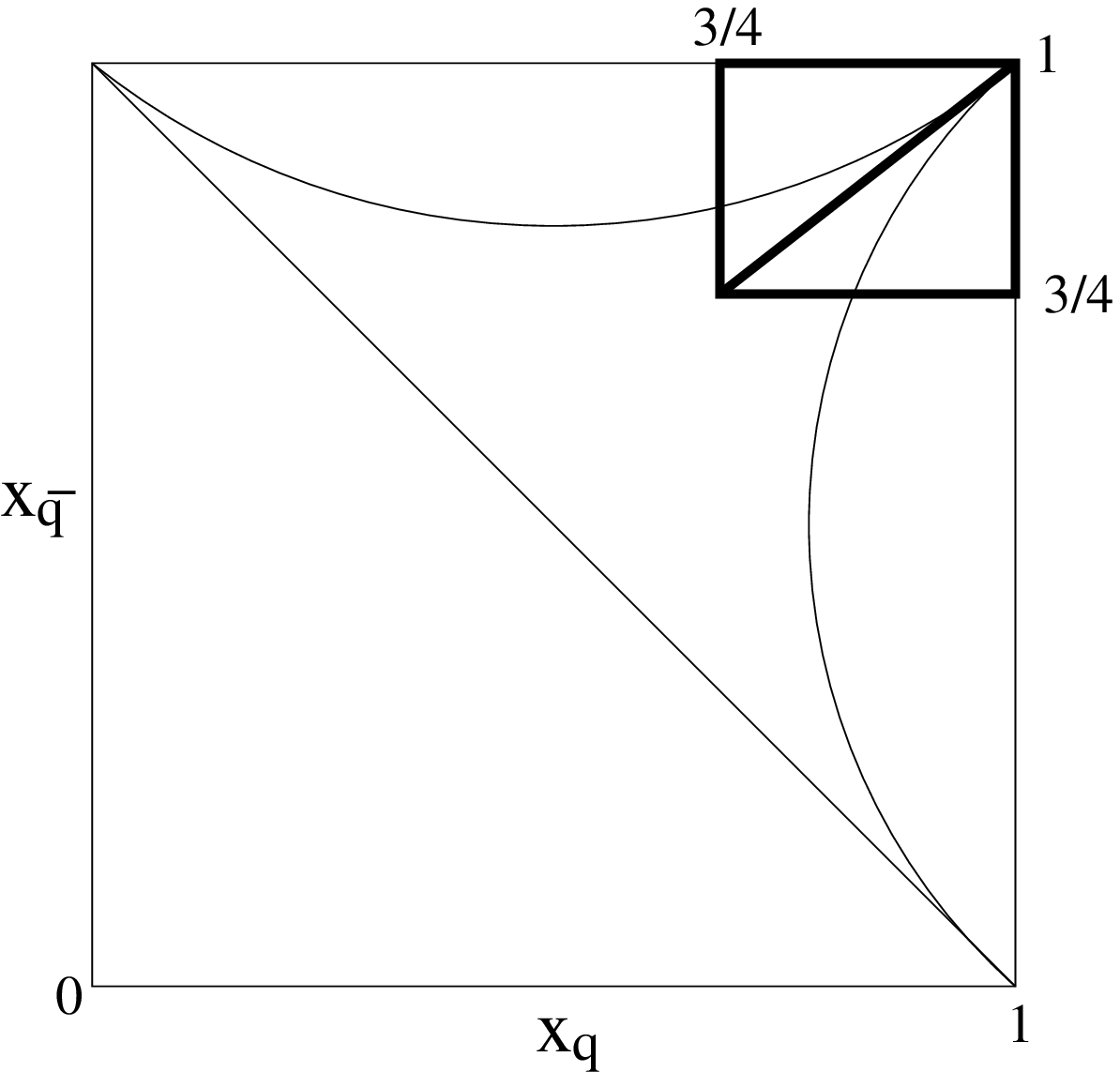,%
width=3in,height=3in,angle=0}
\]
\caption{Mapped regions}
\label{mapAB}
\end{center}
\end{figure}

In both regions $J_{q}$ and $J_{\bar{q}}$, there is a simple pole in the term $(M-M_{C})$ at
$(x_q,x_{\bar{q}})=(1,1)$. In the region $x_q,x_{\bar{q}}>\frac{3}{4}$, a new set of
random points are generated which have a weight factor to cancel the divergence. The
mapping used in region $J_{q}$ where $x_{\bar{q}}>x_q$  is: 
\begin{eqnarray}
x_q^{'}&=&1-0.25r_1 \,,\nonumber \\
x_{\bar{q}}^{'}&=&1-(1-x_q^{'})r_2
\label{eqB}
\end{eqnarray}
where $r_1$ and $r_2$ are random numbers in the range $[0,1]$. The weight factor for this
mapping is $2r_1$. For region $J_{\bar{q}}$, where $x_q>x_{\bar{q}}$, interchange $x_q$ and
$x_{\bar{q}}$ in the mapping. The mapped regions are shown with solid boundaries in Figure
\ref{mapAB}. 
There are no poles at (1,0) and (0,1) because at these points the singularities
in $M$ and $M_{C}$ are cancelled out in the subtraction. 

\subsection{Mapping method}
To illustrate the method behind determining the mappings, the mapping in section
\ref{example} will be explicitly calculated here. Figure \ref{mapC2} shows the mapped
region which is between the lines $x_{\bar{q}}=2.5(1-x_{q})$ and
$x_{\bar{q}}=1-x_{q}$. The integrand in this region goes as
$\frac{1}{1-x_{\bar{q}}}$ and hence is divergent as $x_{\bar{q}} \rightarrow 1$. By performing the
change of variables shown in (\ref{eqmap}), this divergence can be regularized. 
\begin{eqnarray}
&&\int_{0.75}^{1}dx_{\bar{q}}\int_{1-x_{\bar{q}}}^{2.5(1-x_{\bar{q}})}dx_{q}f(x_{q},x_{\bar{q}})\nonumber \\
&=&\int_{0}^{1}dr_{2}\int_{0}^{1}dr_{1}(0.25^{2} \times
1.5r_{2})f(x_{\bar{q}}=1-0.25r_{2}, x_{q}=(1+1.5r_{1})(1-x_{\bar{q}})) \;.
\label{eqmap}
\end{eqnarray}
$r_{1}$ and $r_{2}$ are random numbers between $[0,1]$ as discussed in section
\ref{example} and the factor $0.25^{2} \times 1.5r_{2}$ is the Jacobian factor arising
from the change of variables. It can be seen that this Jacobian factor explicitly removes
the $\frac{1}{1-x_{\bar{q}}}$ divergence in the integrand. Since the `Hit Or Miss' Monte
Carlo method is used to perform the integration, the weights of points in the mapped
region must be multiplied by an area factor equal to inverse of the area of the mapped
region i.e. $\frac{64}{3}$ in this case. Hence the weight of a sampled point in
this region is
\begin{equation}
 0.25^{2} \times 1.5r_{2} \times \frac{64}{3} \times f= 2r_{2}f \;.
\end{equation}
A similar
treatment is followed for all the other divergences. 
Now that we have taken care of the divergences, we need to check that the mappings give
the true value of the integral. 

\subsection{Testing the mappings}
The integration package {\tt VEGAS} \cite{Lepage:1980dq} was used to obtain estimates for the integrals in
(\ref{eq12}). {\tt VEGAS} is an iterative and adaptive Monte Carlo integration algorithm which is based on importance
sampling and is good for the evaluation of multidimensional integrals. 

The algorithm was used to compute integrals $I_{J}^{(3)}$ and $I_{D}^{(3)}$ in section
\ref{HM}. Forty iterations were carried out with  $10^7$ program calls per iteration. The results are
compared with the values obtained with the mappings and unweighting procedure described in sections
\ref{section3} and \ref{section4}. Also worth comparing are the integrals obtained before the unweighting
in step 2 of section \ref{HM}. This gives a measure of how efficient
the unweighting process is. These are presented in the Table 1. The errors in the
{\tt VEGAS} results are not to be trusted as they appear to be underestimated.
\begin{table}
\small
\label{tab:t1}
\begin{center}
\begin{tabular}{|c|c|c|}  \hline 
\ INTEGRAL    & \ $I_{J}^{(3)}$          &  \ $I_{D}^{(3)}$  \\ \hline
\ With mappings      &\ $-0.0393 \pm 1 \times 10^{-4}$   &\ $0.0339 \pm 1 \times 10^{-4}$ \\
\ \tt VEGAS        &\ $-0.03923063 \pm 6 \times 10^{-8}$ &\ $0.0339753 \pm 9 \times 10^{-7}$  \\ 
\ Unweighted events     &\ $-0.0393 \pm 1 \times 10^{-4}$   &\ $0.0339 \pm 1 \times 10^{-4}$\\ 
\hline
\end{tabular}
\end{center}
\caption{Comparison of {\tt VEGAS} with weighted and mapping integrals}
\end{table}

The estimated value for the ratio, $\frac{\sigma_{\rm total}}{\sigma_0}$ after the mappings
can also be compared with the true value,  $[1+\frac{\alpha_S}{\pi}]$ to
$O(\alpha_S)$. This ratio is the sum of the four integrals, $I_{J}^{(2)},I_{J}^{(3)},
I_{D}^{(2)}$ and $I_{D}^{(3)}$ outlined in section \ref{HM}. 
\begin{table}
\small
\begin{center}
\begin{tabular}{|c|c|}  \hline 
{\ True cross-section ratio}   &{\ Estimated cross-section ratio} \\ \hline
{\ 1.0375}                 &{\ $1.0367 \pm 1.3 \times 10^{-3}$}            \\ \hline
\end{tabular}
\end{center}  
\caption{Comparison of the cross-section ratio with the true cross-section ratio to $O(\alpha_S)$}
\end{table}    
Table 3 shows the relative number of 2-jet and 3-jet events generated from a total
of $10^6$ events. 
\begin{table}
\small
\begin{center}
\begin{tabular}{|c|c|c|}  \hline 
{\sc Number of 2-jet events}    & {\sc Number of 3-jet events}  \\  \hline
{\ 934,567}                      &{\ 65,433}   \\  \hline
\end{tabular}
\end{center}
\caption{Relative number of 2-jet and 3-jet events per 1,000,000 events}
\end{table}


\section {Heavy quark integrals} 
\label{HQ}
Event generation for heavy quark production follows the same lines as discussed in Section \ref{HM} for the massless case and the soft divergence mappings used are also implemented
in the heavy parton case.
$10^{7}$ events were generated in this way
to obtain a better estimate of the cross-section. The 3-jet integrals obtained are presented in Tables 4 and 5 for both charm and bottom quark production from vector boson exchange. The {\tt VEGAS} results are also
presented for comparison. The maximum absolute weights in
the two regions $J$ and $D$ are also presented.\\
\begin{table}[h!]
\small
\begin{center}
\begin{tabular} {|c|c|c|}  \hline 
{\ Integral}     &{\ $I_{J}^{(3)}$ }                  &{\ $I_{D}^{(3)}$}  \\ \hline
{\ With mappings}        &{\ $-0.0403 \pm 3 \times 10^{-4}$}   &{\ $0.03215 \pm 4 \times 10^{-5}$} \\
{\ \tt VEGAS}      &{\ $-0.03989509 \pm 7 \times 10^{-8}$} &{\ $0.0321728 \pm 7 \times 10^{-7}$}  \\
{\ Maximum $\mid$weight$\mid$}  &{\ 2360}       &{\ 15} \\ \hline
\end{tabular}
\end{center}
\caption{Comparison of {\tt VEGAS} and mapping integrals for charm quarks} 
\end{table}
\begin{table}[h!!]
\small
\begin{center}
\begin{tabular} {|c|c|c|}  \hline 
{\ Integral}     &{\ $I_{J}^{(3)}$}                  &{\ $I_{D}^{(3)}$} \\ \hline
{\ With mappings}        &{\ $-0.0483 \pm2 \times 10^{-3}$}   &{\ $0.02809 \pm 2 \times 10^{-3}$} \\
{\ \tt VEGAS}      &{\ $-0.0459681 \pm 6 \times 10^{-7}$} &{\ $0.0281315 \pm 1 \times 10^{-7}$}  \\
{\ Maximum $\mid$weight$\mid$}  &{\ 13338}       &{\ 9.3} \\ \hline
\end{tabular}
\end{center}
\caption{Comparison of {\tt VEGAS} and mapping integrals for bottom quarks} 
\end{table}
\\
$I_{J}^{(3)}$ and $I_{D}^{(3)}$ here are the corresponding heavy parton integrals to the integrals
$I_{J}^{(3)}$ and $I_{D}^{(3)}$ discussed in section \ref{HM} in the massless limit.
The estimated value for the ratio, $\frac{\sigma_{total}}{\sigma_V}$ after
the mappings
can also be compared with the true value at $O(\alpha_S)$, $1+c_1\frac{\alpha_S}{\pi}$.\\
\begin{table}[h!]
\small
\begin{center}
\begin{tabular} {|c|c|c|}  \hline 
{\ Quark flavour} &{\ True ratio}   &{\ Estimated ratio} \\ \hline 
{\ Charm}           &{\ 1.0373}                 &{\ $1.0374 \pm 6 \times 10^{-4}$}     \\
{\ Bottom}          &{\ 1.0385}                 &{\ $1.0387 \pm 3 \times 10^{-3}$}     \\ 
\hline
\end{tabular}
\end{center}   
\caption{Comparison of the cross-section ratio with the true cross-section ratio to $O(\alpha_S)$}
\end{table}
\\
The problem with this method arises during the unweighting process described in Section \ref{HM}. The efficiency of the unweighting process for a particular
can be defined as the ratio of the integral over the region to the maximum
value of its integrand $\frac{I}{w_{\rm max}}$. This is a measure of the rejection rate in
the unweighting process. So the smaller the value of $w_{\rm max}$, the greater the
unweighting efficiency. Now, the divergence mappings described in section \ref{AB}
are not as effective in smoothing out the
distribution of event weights in regions $J_{Q}$ and $J_{\bar{Q}}$ (where there is a soft gluon singularity)
as was the case in the massless limit. This results in a
relatively peaked distribution in the soft gluon region and hence a relatively large
absolute value for the maximum
weight (see Tables 4 and 5). Due to this, the unweighting process is comparatively inefficient.

To resolve this, we can impose a limit on the gluon `softness' allowed in 3-jet
events from the soft and collinear regions, $J_{Q}$ and $J_{\bar{Q}}$. 3-jet events with
gluon energy fractions, $x_{g}$ below this limit are then considered
to be 2-jet events with $x_{Q}$ and $x_{\bar{Q}}$ equal to 1 and $x_{g}=0$. For charm and
bottom quarks, a
limiting value of $1 \times 10^{-4}$ was chosen such that the maximum absolute weight is sufficiently lowered
(giving a smoother distribution) whilst the estimated cross-section is not too far off from the true cross-section (Table 7).
\begin{table}
\small
\begin{center}
\begin{tabular} {|c|c|c|}  \hline 
{\ Quark flavour} &{\ Maximum 'soft' $\mid$weight$\mid$}   &{\ Estimated ratio} \\ \hline
{\ Charm}           &{\ 4.8}                 &{\ $1.0381 \pm 5 \times 10^{-4}$}            \\ 
{\ Bottom}          &{\ 43.1}                 &{\ $1.0439 \pm 2 \times 10^{-3}$}     \\ 
\hline
\end{tabular}
\end{center}   
\caption{Cross-section ratio and maximum weights after phase-space cut-offs}
\end{table}
The same method was applied for the axial vector boson coupling.
\section{Assigning  parton properties}
\label{dress}
Having chosen a phase space point, to generate a full event we have to assign full
4-momenta, as well as flavour, spin and colour information to the partons. In this section
we address these issues in turn.
\subsection{Momentum 4-vectors}
\begin{figure}[h]
\[
\psfig{figure=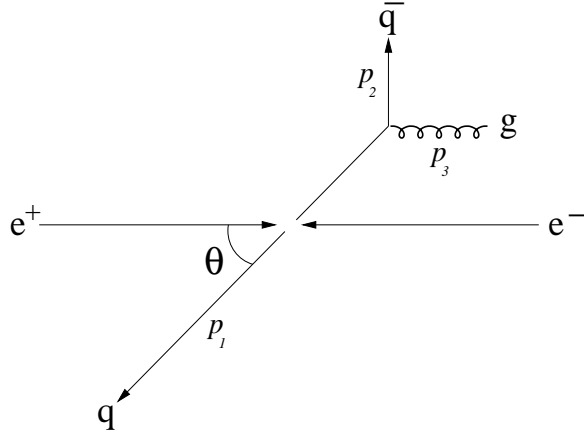,%
width=3in,height=3in,angle=0}
\]
\caption{Annihilation in centre-of-mass frame}
\label{figreaction}
\end{figure}
Figure \ref{figreaction} shows the production of a quark of 4-momentum $p_{1}$, an
antiquark of 4-momentum $p_{2}$ and a gluon of 4-momentum $p_{3}$ from an $e^{+}e^{-}$
annihilation reaction in the centre-of-mass frame. 
It can be shown that the angles $\theta_{ij}$ between partons of momenta $p_{i}$ and $p_{j}$ satisfy
the relations,   
\begin{eqnarray}
 \cos\theta_{12}&=&\frac{x_qx_{\bar{q}}-2(1-x_g)+4\rho}{\sqrt{(x_q^2-4\rho)(x_{\bar{q}}^2-4\rho)}} \,,\nonumber\\
 \cos\theta_{13}&=&\frac{x_qx_g-2(1-x_{\bar{q}})}{x_g\sqrt{x_q^2-4\rho}}\,,\nonumber\\
 \cos\theta_{23}&=&\frac{x_{\bar{q}}x_g-2(1-x_q)}{x_g\sqrt{x_{\bar{q}}^2-4\rho}}\;.
\end{eqnarray}
The angular differential cross-section for the process
$e^{+}e^{-} \rightarrow q\bar{q}$ is defined in (\ref{eqntheta}) \cite{Djouadi:1989uk}. This is the distribution of the angle $\Theta$ between the initial $q\bar{q}$ axis (before gluon
emission) and the $e^+e^-$ axis.
\begin{equation}\label{eqntheta}
\frac{d\sigma}{d\cos\Theta}=(1+\cos^2\Theta)\sigma_U+2\sin^2\Theta\sigma_L+2\cos\Theta\sigma_F\;.
\end{equation}
In the Born approximation,
\begin{eqnarray}
\sigma_U&=&\beta\sigma_{VV}+\beta^3\sigma_{AA}\,,\nonumber\\
\sigma_L&=&\frac{1}{2}(1-\beta^2)\beta\sigma_{VV}\,,\nonumber\\
\sigma_F&=&\beta^2\sigma_{VA}
\end{eqnarray}
with
\begin{eqnarray}
\sigma_{VV}&=&\frac{\pi\alpha^2}{2s}\left[Q_f^2-2Q_fV_eV_f\chi_1(s)+(A_e^2+V_e^2)V_f^2\chi_2(s)\right]\,,\nonumber\\
\sigma_{AA}&=&\frac{\pi\alpha^2}{2s}\left[(A_e^2+V_e^2)A_f^2\chi_2(s)\right]\,,\nonumber\\
\sigma_{VA}&=&\frac{\pi\alpha^2}{2s}\left[-2Q_fA_eA_f\chi_1(s)+4A_eV_eA_fV_f\chi_2(s)\right]\,,\nonumber\\
\beta&=&\sqrt{1-4\rho}\,,\nonumber\\
\chi_1(s)&=&\delta\frac{s(s-M_Z^2)}{(s-M_Z^2)^2+\Gamma^2_ZM_Z^2}\,,\nonumber\\
\chi_2(s)&=&\delta^2\frac{s^2}{(s-M_Z^2)^2+\Gamma^2_ZM_Z^2}\,,\nonumber\\
\delta&=&\frac{\sqrt{2}G_FM_Z^2}{4\pi\alpha}\;,
\end{eqnarray}
$G_F$ is the Fermi constant, $\alpha$ is the electromagnetic coupling, $M_Z$ and
$\Gamma_Z$ are the mass and total decay width of the Z boson respectively and $V_f$ and
$A_f$ are the vector and axial couplings of fermion, $f$ to the Z boson. 

By applying the unweighting procedure described in step 2 of section \ref{HM} to
(\ref{eqntheta}), the angles $\Theta$ for each event can be distributed according to angular
differential cross-section. Since the azimuthal angles for $q\bar{q}$ production and
the $q \rightarrow qg$ process are isotropic, 4-momentum vectors for the quark, antiquark
and gluon can therefore straightforwardly be constructed for each event.
  
Initially all parton 4-momenta were generated in the massless limit. Though this is
not essential for the parton shower, the parton 3-momenta were rescaled by a common factor
using the Newton-Raphson iteration method to give the right
parton masses (for heavy partons) and a gluon virtuality of $0.75$ GeV.
\subsection{Flavour}
To assign flavour to the quarks, we need to investigate the flavour dependence of the
total cross-section for $e^+e^- \rightarrow q\bar{q}$. Integrating (\ref{eqntheta}) over
all angles $\Theta$ gives; 
\begin{equation}
\sigma_f=(\beta+\frac{1}{2}\beta(1-\beta^2))\sigma_{VV}+\beta^3\sigma_{AA}\;.
\label{eqncs}
\end{equation}
$\sigma_f$ is the contribution of a quark of flavour {\em f} to the total
cross-section. Hence the quarks are assigned flavours according to the relative values of $\sigma_f$.
\subsection{Spins}
\label{Spins}

To assign spins to the partons, the matrix element for  $e^+e^- \rightarrow q\bar{q}$ is
calculated for all  possible helicity configurations and its  modulus squared is used as
a weight in allocating spins to the quarks and antiquarks. In other words, the helicity configurations are
allocated according to their contributions to the total cross-section. It is assumed that
the electron and positron beams are unpolarised so that the helicities of the electrons and
positrons are assigned randomly.  

In the massless limit, the chiral components are the helicity eigenstates and so when a Z
boson (which couples to chiral components) is exchanged, the matrix elements for different
spin configurations are straightforward to calculate. In the heavy parton case, the chiral
components are a mixture of helicity eigenstates so that the matrix element calculations
become more complicated. Also, there are 4 possible helicity configurations in the massless
limit whilst there are 8 helicity configurations in the heavy parton case (electrons and positrons are assumed to be massless). 

As an illustration, the matrix element for the helicity configuration, $e^{-}_\uparrow$
$e^{+}_\downarrow$ $q_\uparrow$ $\bar{q}_\downarrow$ is given in (\ref{M}).
\begin{eqnarray}\label{M}
M_{\uparrow\downarrow\uparrow\downarrow}&=&(1+\cos\Theta)^{2}\left[{c_{\rm R}^{e}}^{2}\chi_{2}\lambda^{2}\left[c_{\rm
      L}^{q}\left(\frac{\sqrt{s}}{2}-p_{\rm cm}\right)+c_{\rm
      R}^{q}\left(\frac{\sqrt{s}}{2}+p_{\rm cm}\right)\right]^{2} \right.\nonumber\\
&-& \left. 2eQ_{q}c_{\rm R}^{e}\chi_{1}\lambda\left[c_{\rm L}^{q}\left(\frac{\sqrt{s}}{2}-p_{\rm
      cm}\right)+c_{\rm R}^{q}\left(\frac{\sqrt{s}}{2}+p_{\rm cm}\right)\right]+e^{2}Q_{q}^{2}\right]
\end{eqnarray}
where 
\begin{equation}
\lambda^{2}=\frac{2\sqrt{32}\pi\alpha}{s} \,,
\end{equation}
$c_{R}^{q},c_{L}^{q},c_{R}^{e}$ and
$c_{L}^{e}$ are the right and left-handed couplings of the quarks and leptons to the Z
boson respectively and $p_{\rm cm}$ is the centre-of-mass
momentum of the quark.

So far we have assigned the particles into helicity eigenstates. This is because the
interface of the event record to the event generator {\tt Herwig++}, requires that the electrons, positrons and partons are in
definite helicity eigenstates. In practice however, for an unpolarised beam of electrons and positrons, the spins of the partons
resulting from the annihilation reaction are superpositions of the helicity
eigenstates. The spin density matrix, $\rho$ for spin-$\frac{1}{2}$ particles is given by;
\begin{equation}
\rho=\frac{1}{2}[I+\langle{\sigma}\rangle\cdot\sigma]
\end{equation}
which can be written as
 \begin{equation}
        \left[
          \begin{array}{ccc}
          \frac{1}{2}(1+a_1) & \frac{1}{2}(a_2-ia_3) \\
          \frac{1}{2}(a_2+ia_3) & \frac{1}{2}(1-a_1)
          \end{array}
          \right]
\end{equation}
where $\langle{\sigma}\rangle$ is the matrix of spin expectation values, $\sigma=(\sigma_x,\sigma_y,\sigma_z)$ which are the Pauli matrices and
$\frac{1}{2}a_1, \frac{1}{2}a_2$ and $\frac{1}{2}a_3$ are the expectation values of the spin
matrices $S_z, S_x, S_y$ respectively. Since the helicity eigenstates are also eigenstates
of $S_z$ (taking the z axis to be along the quark momentum direction) and since $\frac{1}{2}(1+a_1)$ and $\frac{1}{2}(1-a_1)$ are the probabilities
of a parton being in either state, we can assign spin states to the partons by distributing values of $a_1$
between 0 and 1 according to their corresponding matrix element contributions.
\subsection {Gluon emission}
For the 3-jet events, the gluon can be radiated either from the quark or
anti-quark. Intuitively, we expect the fastest or more energetic of the two partons to be
the least likely to radiate the gluon i.e. most likely to keep its original direction. In
the massless limit, it has been shown that the relative probabilities of a parton
retaining its direction after production (not emitting the gluon) are in the ratio of their
respective energies squared. i.e $\propto x_{i}^2$ \cite{Kleiss:1986re}. This notion is used to
assign mother partons to the gluons in the 3-jet events. (For heavy quarks, this procedure is approximate). In addition it can be shown that
the  helicity of the gluon equals the non-emitting parton's helicity which is opposite to the
spin of the mother parton \cite{Kleiss:1986re}. 

\subsection{Colour}
\begin{figure}[h!!]
\[
\psfig{figure=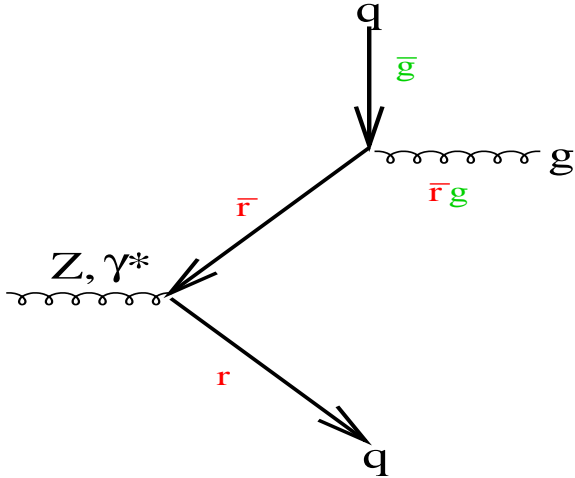,%
width=3in,height=2.5in,angle=0}
\]
\caption{Colour Flow}
\label{figcolour1}
\end{figure}
The colour of a parton flows out of its production vertex  whilst its anticolour
flows into its
production vertex. The gluon carries the anticolour of the
quark colour and the colour corresponding to the antiquark anticolour. This is illustrated
in Figure \ref{figcolour1} for the specific case where a $\bar{r}$ antiquark
radiates a $g\bar{r}$ gluon and becoming a $\bar{g}$ antiquark in the process. (The colour
flow lines are essentially the same for a radiating quark). 

Colours were assigned to the partons taking into account the restrictions on the colour
flow discussed above. 
Note that this colour treatment is the planar approximation, which is correct only to
$O(\frac{1}{{N_{C}}^2})$ where $N_C =3$ is the number of
colours. In this approximation, it is always possible to draw any Feynman diagram such
that no colour lines need cross as in the figures above.

\section{Born variables and parton flavours for Drell-Yan production}
\label{Born}
The Born variables
$Q^2$ and  $Y$  for each event were distributed according to the
Born cross-section in (\ref{sigB}). Hence the momentum fractions $x_q$ and $x_{\bar{q}}$ given
in (\ref{xq}) are generated. The flavour of the quarks involved in each event is also
determined at the Born level according to the product,
$e_{q_i}^{2}f_q(x_{q_i},Q^2)f_{\bar{q}}(x_{\bar{q_{i}}},Q^2)$. Colours are assigned in the planar
approximation which is correct to $O(\frac{1}{{N_C}^2})$ where $N_C=3$.
\section{Monte Carlo algorithm for Drell-Yan production}
\label{HMM}
The eight integrals listed were evaluated by the `Hit Or Miss' method and used to
generate a set of unweighted events using the importance sampling method described in
Section \ref{HM}.
\begin{eqnarray}
I^{(2)}_{Jq\bar{q}}&=&\sum_{q}\int_J dx dy\left[\frac{x[D_q(x_1,\mu^2)D_{\bar{q}}(x_2,\mu^2)+q\leftrightarrow
\bar{q}]}{D_q(x_q)D_{\bar{q}}(x_{\bar{q}})}\frac{1}{2}\left \{\delta(1-x)+\frac{\alpha_S}{2\pi}C_F\left(\frac{3}{(1-x)_+}
  \right. \right. \right.\nonumber \\
&-&\left.\left.\left.6-4x+2(1+x^2)\left(\frac{\ln(1-x)}{1-x}\right)_++\left(1+\frac{4}{3}\pi^2\right)\delta(1-x)\right)\right\}-M_{q\bar{q}}+M_{{C}_{q\bar{q}}}\right]\nonumber\,, \\
I^{(3)}_{Jq\bar{q}}&=&\sum_{q}\int_J dx dy [M_{q\bar{q}}-M_{C_{q\bar{q}}}] \,,\nonumber \\
I^{(2)}_{Dq\bar{q}}&=& \sum_{q}\int_J dx dy\left[\frac{x[D_q(x_1,\mu^2)D_{\bar{q}}(x_2,\mu^2)+q\leftrightarrow
\bar{q}]}{D_q(x_q)D_{\bar{q}}(x_{\bar{q}})}\frac{1}{2}\left \{\delta(1-x)+\frac{\alpha_S}{2\pi}C_F\left(\frac{3}{(1-x)_+}
  \right. \right. \right.\nonumber \\
&-&\left.\left.\left.6-4x+2(1+x^2)\left(\frac{\ln(1-x)}{1-x}\right)_++\left(1+\frac{4}{3}\pi^2\right)\delta(1-x)\right)\right\}-M_{q\bar{q}}\right]\nonumber\,, \\
I^{(3)}_{Dq\bar{q}}&=&\sum_{q}\int_J dx dy M_{q\bar{q}} \,,\nonumber \\
I^{(2)}_{Jqg}&=&\sum_{q}\int_J dx dy\left[\frac{x[D_q(x_1,\mu^2)D_g(x_2,\mu^2)+ q\leftrightarrow
g]}{D_q(x_q)D_{\bar{q}}(x_{\bar{q}})}\frac{\alpha_S}{2\pi}T_F\frac{1}{2}\left \{\frac{3}{2}-5x+\frac{9}{2}x^2 \right. \right. 
 \nonumber \\
&+& \left.\left.(x^2+(1+x^2))\ln(1-x)\right\}-
M_{qg}+M_{C_{qg}}\right]\,,\nonumber \\
I^{(3)}_{Jqg}&=&\sum_{q}\int_J dx dy [M_{qg}-M_{C_{qg}}]\,,
 \nonumber \\
I^{(2)}_{Dqg}&=&\sum_{q}\int_J dx dy\left[\frac{x[D_q(x_1,\mu^2)D_g(x_2,\mu^2)+ q\leftrightarrow
g]}{D_q(x_q)D_{\bar{q}}(x_{\bar{q}})}\frac{\alpha_S}{2\pi}T_F\frac{1}{2}\left \{\frac{3}{2}-5x+\frac{9}{2}x^2 \right. \right.
 \nonumber \\
&+& \left.\left.(x^2+(1+x^2))\ln(1-x)\right\}-
M_{qg}\right]\,,\nonumber \\
I^{(3)}_{Dqg}&=&\sum_{q}\int_D dx dy M_{qg} \;.
\end{eqnarray}
As in the $e^+e^-$ case, the $I^{(3)}_{J}$ integrals are negative because
$M<M_{C}$. The integrals are all finite but there are divergences in the integrands which need to
be regularized to make
the sampling process efficient. This is
the subject of Section \ref{divmap}. So far we have discussed event generation in the DIS
scheme. In the $\overline{\rm MS}$ factorization scheme, the full cross-section ratio is
\begin{eqnarray}
\frac{\sigma^{NLO}}{\sigma_0}&=&\sum_{q}\int
dx_1dx_2 \left[\frac{x[D_q(x_1,\mu^2)D_{\bar{q}}(x_2,\mu^2)+q\leftrightarrow
\bar{q}]}{D_q(x_q)D_{\bar{q}}(x_{\bar{q}})}\left\{\delta(1-x)+\frac{\alpha_S}{2\pi}C_F\left(-2\frac{1+x^2}{1-x} \ln x
  \right. \right.\right.\nonumber \\
&+&\left.\left.4(1+x^2)\left(\frac{\ln(1-x)}{1-x}\right)_++\left(-8+\frac{2}{3}\pi^2\right)\delta(1-x)\right)\right\}
\nonumber \\
&+&\frac{x[D_q(x_1,\mu^2)D_g(x_2,\mu^2)+ q\leftrightarrow
g]}{D_q(x_q)D_{\bar{q}}(x_{\bar{q}})}\frac{\alpha_S}{2\pi}T_F\left \{\frac{1}{2}+3x-\frac{7}{2}x^2 \right. 
 \nonumber \\
&+& \left.\left.(x^2+(1+x^2))\ln\frac{(1-x)^2}{x}\right\}\right] \;.
\label{csm}
\end{eqnarray}
\clearpage
Hence the corresponding integrals
for $\overline{\rm MS}$  scheme event generation are 
\begin{eqnarray}
I^{(2)}_{Jq\bar{q}}&=&\sum_{q}\int_J dx dy\left[\frac{x[D_q(x_1,\mu^2)D_{\bar{q}}(x_2,\mu^2)+q\leftrightarrow
\bar{q}]}{D_q(x_q)D_{\bar{q}}(x_{\bar{q}})}\frac{1}{2}\left
\{\delta(1-x)+\frac{\alpha_S}{2\pi}C_F\left(-2\frac{1+x^2}{1-x} \ln x
  \right. \right. \right.\nonumber \\
&+&\left.\left.\left.4(1+x^2)\left(\frac{\ln(1-x)}{1-x}\right)_++\left(-8+\frac{2}{3}\pi^2\right)\delta(1-x)\right)\right\}-M_{q\bar{q}}+M_{{C}_{q\bar{q}}}\right]\nonumber\,, \\
I^{(3)}_{Jq\bar{q}}&=&\sum_{q}\int_J dx dy [M_{q\bar{q}}-M_{C_{q\bar{q}}}] \,,\nonumber \\
I^{(2)}_{Dq\bar{q}}&=& \sum_{q}\int_J dx dy\left[\frac{x[D_q(x_1,\mu^2)D_{\bar{q}}(x_2,\mu^2)+q\leftrightarrow
\bar{q}]}{D_q(x_q)D_{\bar{q}}(x_{\bar{q}})}\frac{1}{2}\left \{\delta(1-x)+\frac{\alpha_S}{2\pi}C_F\left(-2\frac{1+x^2}{1-x} \ln z
  \right. \right. \right.\nonumber \\
&+&\left.\left.\left.4(1+x^2)\left(\frac{\ln(1-x)}{1-x}\right)_++\left(-8+\frac{2}{3}\pi^2\right)\delta(1-x)\right)\right\}-M_{q\bar{q}}\right]\nonumber\,, \\
I^{(3)}_{Dq\bar{q}}&=&\sum_{q}\int_J dx dy M_{q\bar{q}} \,,\nonumber \\
I^{(2)}_{Jqg}&=&\sum_{q}\int_J dx dy\left[\frac{x[D_q(x_1,\mu^2)D_g(x_2,\mu^2)+ q\leftrightarrow
g]}{D_q(x_q)D_{\bar{q}}(x_{\bar{q}})}\frac{\alpha_S}{2\pi}T_F\frac{1}{2}\left \{\frac{1}{2}+3x-\frac{7}{2}x^2 \right. \right. 
 \nonumber \\
&+& \left.\left.(x^2+(1+x^2))\ln \frac{(1-x)^2}{x}\right\}-
M_{qg}+M_{C_{qg}}\right]\,,\nonumber \\
I^{(3)}_{Jqg}&=&\sum_{q}\int_J dx dy [M_{qg}-M_{C_{qg}}]\,,
 \nonumber \\
I^{(2)}_{Dqg}&=&\sum_{q}\int_J dx dy\left[\frac{x[D_q(x_1,\mu^2)D_g(x_2,\mu^2)+ q\leftrightarrow
g]}{D_q(x_q)D_{\bar{q}}(x_{\bar{q}})}\frac{\alpha_S}{2\pi}T_F\frac{1}{2}\left \{\frac{1}{2}+3x-\frac{7}{2}x^2 \right. \right.
 \nonumber \\
&+& \left.\left.(x^2+(1+x^2))\ln\frac{(1-x)^2}{x}\right\}-
M_{qg}\right]\,,\nonumber \\
I^{(3)}_{Dqg}&=&\sum_{q}\int_D dx dy M_{qg} \;.
\end{eqnarray}
The divergences and mappings used are discussed in Section \ref{divmap}.
\section{Divergences and mappings for Drell-Yan production}
\label{divmap}
\subsection{Divergences in \boldmath{$M_{q\bar{q}}$} in dead region \boldmath{$D$}}
In region $D$, the hard matrix element:;
\begin{equation}
M_{q\bar{q}}(x,y)=\frac{D_{q}(x_1)D_{\bar{q}}(x_2)}{D_{q}(x_q)D_{\bar{q}}(x_{\bar{q}})}\frac{\alpha_S}{2
  \pi}C_F\frac{y^2(1-x)^2+(1+x)^2}{(1-x)(1-y^2)}
\end{equation}
diverges as $(x,y) \rightarrow(0,1), (0,-1)$ and $(1,0)$. The mappings used to regularize
these divergences are presented.
\subsubsection{Region \boldmath{$D: (0,1), (0,-1)$}}
\begin{figure}[h!]
\[
\psfig{figure=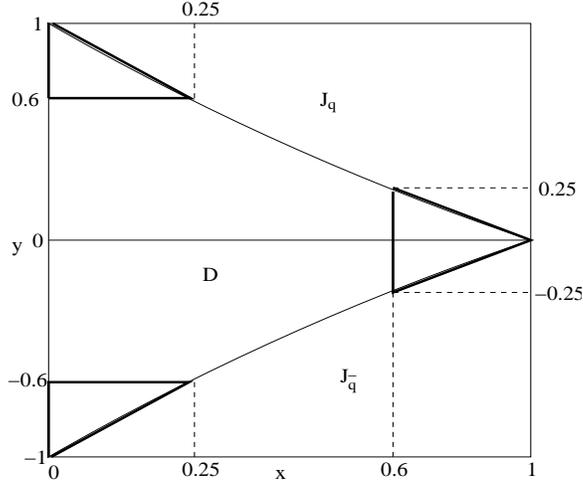,%
width=3in,height=2.5in,angle=0}
\]
\caption{Mapped regions}
\label{map1}
\end{figure}

A simple pole is approached in $M_{q\bar{q}}$ as $(x,y)\rightarrow(0,1),(0,-1)$. In the
region $x<\frac{1}{4}, y>\frac{3}{5}$, a new set of random points are generated for events
which fall between the lines $x=\frac{5}{8}(1-y)$ and $x=0$. The mapping used and weight
factor, $w$ for these points is:
\begin{eqnarray}
y^{'}&=&1-\frac{2}{5}r_1 \,,\nonumber \\
x^{'}&=&\frac{5}{8}r_2(1-y^{'})\,,\nonumber \\
w&=&2r_1
\end{eqnarray}
where $r_1,$ and $r_2$ are random numbers in the range $[0:1]$. For the region $x<\frac{1}{4}, y<-\frac{3}{5}$, interchange $y \leftrightarrow -y$. The mapped
regions are shown with solid boundaries in Figure \ref{map1}.
\subsubsection{Region \boldmath{$D: (1,0)$}}
As $x\rightarrow1$, there is a pole in $M_{q\bar{q}}$. This divergence is regularized by
applying the following mapping and weight factor to points which fall between the lines
$y=\frac{5}{8}(1-x)$ and $y=0$ within the region
$x>\frac{3}{5},0<y<\frac{1}{4}$.
\begin{eqnarray}
x^{'}&=&1-\frac{2}{5}r_1 \,,\nonumber \\
y^{'}&=&\frac{5}{8}r_2(1-x^{'})\,,\nonumber \\
w&=&2r_1 \;.
\end{eqnarray}
In the region $x>\frac{3}{5},0>y>-\frac{1}{4}$, a similar mapping is used;
\begin{eqnarray}
x^{'}&=&1-\frac{2}{5}r_1 \,,\nonumber \\
y^{'}&=&-\frac{5}{8}r_2(1-x^{'})\,,\nonumber \\
w&=&2r_1 \;.
\end{eqnarray}
 The mapped
regions are shown with solid boundaries in Figure \ref{map1}.
\subsection{Divergences in \boldmath{$M_{qg}$} in dead region \boldmath{$D$}}
\begin{figure}[h!]
\[
\psfig{figure=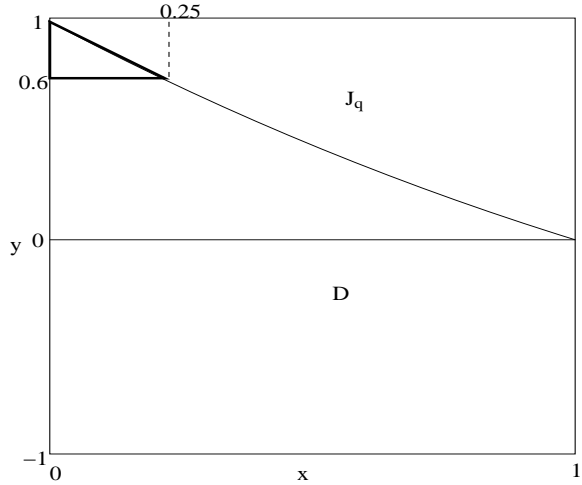,%
width=3in,height=2.5in,angle=0}
\]
\caption{Mapped regions}
\label{map2}
\end{figure}

In region $D$, the hard matrix element:
\begin{equation}
M_{qg}(x,y)=\frac{D_{q}(x_1)D_g(x_2)}{D_{q}(x_q)D_{\bar{q}}(x_{\bar{q}})}\frac{\alpha_S}{2
  \pi}T_F\frac{(3+y^2)(1-x)^2-2y(1-x^2)+2(1+x^2)}{4(1-y)}
\end{equation}
diverges as $(x,y) \rightarrow(0,1)$. for points within the region, $x<\frac{1}{4}, y>\frac{3}{5}$, a new set of random points are generated for events
which fall between the lines $x=\frac{5}{8}(1-y)$ and $x=0$. The mapping used to
regularize this divergence is:
\begin{eqnarray}
y^{'}&=&1-\frac{2}{5}r_1\,, \nonumber \\
x^{'}&=&\frac{5}{8}r_2(1-x^{'})\,,\nonumber \\
w&=&2r_1\;.
\end{eqnarray}
 The mapped
region is shown with solid boundaries in Figure \ref{map2}.
\subsection{Divergences in \boldmath{$(M-M_{C})_{q\bar{q}}$} in jet region \boldmath{$J$}}
\begin{figure}[h!]
\[
\psfig{figure=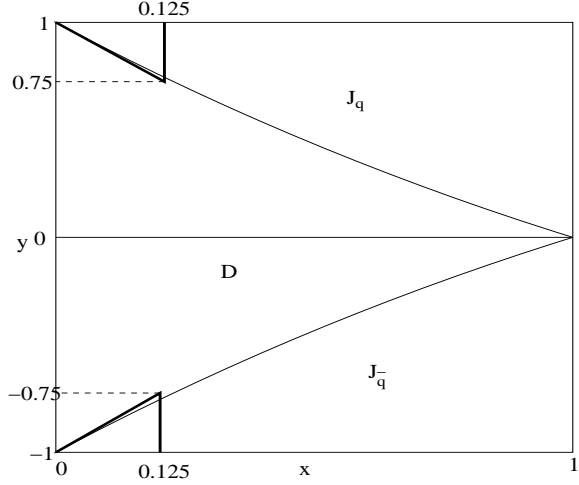,%
width=3in,height=2.5in,angle=0}
\]
\caption{Mapped regions}
\label{map3}
\end{figure}

In regions $J_q$ and $J_{\bar{q}}$ depicted in Figure \ref{map3}, the relevant integral to be
evaluated is $M-M_C$ which in region $J_q$ is given by
\begin{eqnarray}
(M-M_{C})_{q\bar{q}}=\frac{D_{q}(x_1)D_{\bar{q}}(x_2)}{D_{q}(x_q)D_{\bar{q}}(x_{\bar{q}})}\frac{\alpha_S}{2
  \pi}C_F\frac{(x-1)((x-1)y^2+2y)+(3x+5)(x+1)}{4x(1+y)} \;.
\end{eqnarray}
This diverges as $x\rightarrow 0$. There are no singularities at the points $x=1,y=\pm1$
because the divergences at these points exactly cancel between $M$ and $M_{C}$. Points in the region $x<\frac{1}{8},y>\frac{3}{4}$ and
between the lines $y=0$ and $y=1-2x$ are mapped and re-weighted according to
(\ref{map}).
\begin{eqnarray}
\label{map}
x^{'}&=&\frac{1}{8}r_1\,, \nonumber \\
y^{'}&=&1-2x^{'}r_2\,,\nonumber \\
w&=&2r_1 \;.
\end{eqnarray}
Interchange $y \leftrightarrow -y$ for the region $J_{\bar{q}}$. 
\subsection{Divergences in \boldmath{$(M-M_{C})_{qg}$} in jet region \boldmath{$J$}}
$(M-M_{C})_{qg}$ is given by
\begin{equation}
(M-M_{C})_{qg}=\frac{D_{q}(x_1)D_g(x_2)}{D_{q}(x_q)D_{\bar{q}}(x_{\bar{q}})}\frac{\alpha_S}{2
  \pi}T_F\frac{(x-1)[4x^2-3x+1+3xy(x-1)+y^2(x-1)^2]}{4x} 
\end{equation}
which diverges as $x\rightarrow0$ in region $J$. This is regularized by using the same
mapping presented in (\ref{map}).
\clearpage
\bibliography{project}
\bibliographystyle{utphys}
\end{document}